\documentclass[superscriptaddress,showpacs,amssymb,10pt,aps,prd,reprint,longbibliography,footinbib]{revtex4-1}
\usepackage{graphicx,epsfig,amssymb,times}
\usepackage{amsmath,amsfonts}
\usepackage{bm}
\usepackage{epstopdf}
\usepackage[linktocpage,colorlinks]{hyperref}
\usepackage[caption=false]{subfig}
\usepackage[usenames]{color}
\usepackage{float}
\usepackage{natbib}
\usepackage{ulem}
\usepackage{cancel}
\usepackage{verbatim}
\usepackage{enumitem}
\usepackage[utf8]{inputenc}
\definecolor{coolblack}{rgb}{0.0, 0.18, 0.39}
\definecolor{darkred}{rgb}{0.5,0,0}
\definecolor{darkgreen}{rgb}{0,0.5,0}
\definecolor{darkblue}{rgb}{0,0,0.5}
\definecolor{lapislazuli}{rgb}{0.15, 0.38, 0.61}
\definecolor{venetianred}{rgb}{0.78, 0.03, 0.08}
\definecolor{bleudefrance}{rgb}{0.19, 0.55, 0.91}
\definecolor{dogwoodrose}{rgb}{0.84, 0.09, 0.41}
\hypersetup{colorlinks=true, citecolor=darkblue, linkcolor=darkblue,
urlcolor = darkblue}
\def\be{\begin{equation}}
\def\ee{\end{equation}}

\newcommand{\bea}{\begin{eqnarray}}
\newcommand{\eea}{\end{eqnarray}}
\newcommand{\ben}{\begin{enumerate}}
\newcommand{\een}{\end{enumerate}}
\newcommand{\bi}{\begin{itemize}}
\newcommand{\ei}{\end{itemize}}

\def\ga{\mathrel{\raise.3ex\hbox{$>$\kern-.75em\lower1ex\hbox{$\sim$}}}}
\def\la{\mathrel{\raise.3ex\hbox{$<$\kern-.75em\lower1ex\hbox{$\sim$}}}}

\def\be{\begin{equation}}
\def\ee{\end{equation}}

\def\I_M{{I_{\scriptscriptstyle M\times M}}}

\def\be{\begin{equation}}
\def\ee{\end{equation}}
\def\bea{\begin{eqnarray}}
\def\eea{\end{eqnarray}}
\begin{document}

\title{\large Schwarzschild black hole surrounded by quintessential matter field as an accelerator for spinning particles}

\author{Pankaj Sheoran}\email{hukmipankaj@gmail.com}
\affiliation{Instituto de F\'{\i}sica y Matem\'{a}ticas, Universidad Michoacana de San Nicol\'{a}s de Hidalgo,\\
Edificio C-3, 58040 Morelia, Michoac\'{a}n, M\'{e}xico.}

\author{Hemwati Nandan}\email{hnandan@associates.iucaa.in}
\affiliation{Department of Physics, Gurukul Kangri Vishwavidyalaya, Haridwar-249 407, India.}
\affiliation{Center for Space Research, North-West University, Mafikeng, South Africa.}

\author{Eva Hackmann}\email{eva.hackmann@zarm.uni-bremen.de}
\affiliation{ZARM, University of Bremen, Am Fallturm 2, 28359 Bremen, Germany.}

\author{Ulises Nucamendi}\email{unucamendi@gmail.com}
\affiliation{Instituto de F\'{\i}sica y Matem\'{a}ticas, Universidad Michoacana de San Nicol\'{a}s de Hidalgo,\\
Edificio C-3, 58040 Morelia, Michoac\'{a}n, M\'{e}xico.}
\affiliation{Mesoamerican Centre for Theoretical Physics, Universidad Aut\'{o}noma de Chiapas, Ciudad Universitaria, Carretera Zapata Km. 4, Real del Bosque (Ter\'{a}n), 29040 Tuxtla Guti\'{e}rrez, Chiapas, M\'{e}xico}
\affiliation{Departamento de F\'isica, Cinvestav, Avenida Instituto Politecnico Nacional 2508, San Pedro Zacatenco, 07360 Gustavo A. Madero, Ciudad de Mexico, Mexico.}

\author{Amare Abebe}\email{amare.abbebe@gmail.com}
\affiliation{Center for Space Research, North-West University, Mafikeng, South Africa.}
%\affiliation{Department of Physics, Sultan Qaboos University, Al Khodh 123 Muscat, Oman}
\date{\today}

%%%%%%%%%%%%%%%%%%%%%%%%%%%%%%%%%%%%%%%%%%%%%%
\begin{abstract}
We study the collision of two massive particles with non-zero intrinsic spin moving in the equatorial plane in the background of a Schwarzschild black hole surrounded by quintessential matter field (SBHQ). For the quintessential matter equation of state (EOS) parameter, we assume three different values. It is shown that for collisions outside the event horizon, but very close to it, the centre-of-mass energy ($E_{\rm CM}$) can grow without bound if exactly one of the colliding particles is what we call near-critical, i.e., if its constants of motion are fine tuned such that the time component of its four-momentum becomes very small at the horizon. In all other cases, $E_{\rm CM}$ only diverges behind the horizon if we respect the M{\o}ller limit on the spin of the particles. We also discuss radial turning points and constraints resulting from the requirement of subluminal motion of the spinning particles.
\end{abstract}

\maketitle
%%%%%%%%%%%%%%%%%%%%%%%%%%%%%%%%%%%%%%%%%%%%%%
\section{Introduction}
%%%%%%%%%%%%%%%%%%%%%%%%%%%%%%%%%%%%%%%%%%%%%%
The first simplest black hole (BH) solution of Einstein's field equations was obtained by Schwarzschild in 1916 \cite{Schwarzschild:1916uq} immediately after the discovery of general relativity (GR) by Einstein. The BH solution found by Schwarzschild is the simplest in the sense that it has only one observable parameter (i.e., mass). Black Holes (BHs) are one of the most interesting topics in gravity research, and it took almost a century to confirm that these mysterious objects do exist in our universe. Recently the LIGO and VIRGO collaborations have detected the first ever gravitational waves signals from BH merger \cite{Abbott:2016blz}. Even more recently, the first ever direct image of a BH observed by the Event Horizon Telescope (EHT) suggests to us to strongly believe in the presence of BHs in our universe \cite{Akiyama:2019cqa}.

The appearance of BHs is not only limited to GR or alternative theories of gravity (ATG) like string theory \cite{Maldacena:1996ky}, but they have also played a crucial role in understanding %the mysterious behavior of dark energy having negative pressure, which is one of the importantunsolved problems in the
cosmology. There are two major classes of cosmological models for dark energy. One of them is the cosmological constant $\Lambda$ \cite{Padmanabhan:2002ji} having an equation of state (EOS) parameter $\epsilon=-1$. But in this model the fine tuning problem is yet to be resolved \cite{Weinberg:1988cp}. The other class of cosmological model mainly depends on a dynamical scalar field such as, but not confined to, quintessence \cite{Carroll:1998zi}, chameleon fields \cite{Khoury:2003aq}, K-essence \cite{ArmendarizPicon:2000dh}, tachyons \cite{Padmanabhan:2002cp}, phantom \cite{Caldwell:1999ew} and dilatons \cite{Gasperini:2001pc}. In these models, the main difference is the EOS parameter $\epsilon$ which varies from -1 to -1/3 for quintessence like models and less than $-1$ for phantom like models. A comprehensive study of various dark energy models is presented in \cite{Copeland:2006wr}.
%As we know that the existence of dark energy having negative pressure is not well studied in the context of BHs. Hence, we try to understand the consequences of BHs in this paper surrounded by the quintessence like models.
In this paper, we restrict to three different equation of state parameters, including the cosmological constant case and two quintessence like models. In particular, we focus on particle collisions in the background of a static BH solution surrounded by quintessence like matter obtained by Kiselev in \cite{Kiselev:2002dx}. The geodesic motion and geodesic deviation around this BH spacetime is investigated in detail in \cite{Uniyal:2014paa}.

A rotating BH under some specific conditions can act as a particle accelerator for two spinless particles which start from rest at infinity and collide near the event horizon of a rotating BH (Kerr BH) pointed out by Ba\~nados, Silk, and West (BSW) ~\cite{Banados:2009pr}. They showed that the collisional energy (i.e., center-of-mass (CM) energies) of these spinless particles will be infinitely high if the BH is rotating in addition to the condition that one of the particle must have attained a critical value (a very fine-tuned value) of the angular momentum. They also mentioned that if the BH is non-rotating (i.e., Schwarzschild), it is not possible to obtain an infinite amount of CM energy. After this pioneering work by BSW \citep{Banados:2009pr}, a number of studies have been performed on the particle acceleration by all sorts of BHs in GR \cite{Jacobson:2009zg,Grib:2010dz,Lake:2010bq,Wei:2010vca,Grib:2010zs,Zaslavskii:2010aw,Harada:2010yv,Grib:2010xj,Banados:2010kn,Williams:2011uz,Harada:2011xz,Patil:2011yb,Zhu:2011ae,Liu:2011wv,Zaslavskii:2011dz,Zaslavskii:2011ex,Grib:2011ph,Yao:2011ai,Gao:2011sv,Zaslavskii:2011uu,Patil:2011uf,Zhu:2011ja,Harada:2011pg,Zaslavsky:2011zz,Zaslavskii:2011tj,Frolov:2011ea,Zaslavskii:2012ua,Grib:2012iq,Hussain:2012zza,Harada:2012ap,Tanatarov:2012xj,Nemoto:2012cq,Grib:2013vc,Galajinsky:2013as,Zaslavsky:2013dra,Stuchlik:2013yca,Abdujabbarov:2013qka,Tsukamoto:2013dna,Zaslavskii:2014dxa,Yumisaki:2016biz} and in different ATG models \cite{Sadeghi:2013gmf,Pradhan:2014oaa,Pradhan:2014eza,Ghosh:2014mea,Hussain:2014aea,Zakria:2015eua,Amir:2015pja,Halilsoy:2015rna,Pourhassan:2015lfa,Ghosh:2015pra,Saadat:2013iba,Abdujabbarov:2011af,Sadeghi:2011qu,Wei:2010gq,Mao:2010di,Li:2010ej,Liu:2010ja,Halilsoy:2015qta,Abdujabbarov:2015rqa,Debnath:2015bna,Toshmatov:2015gna,Sultana:2015avz,Zaslavskii:2016stw,Amir:2016nti,Oteev:2016fbp,Jawad:2016ccw,Fernando:2017kut,Fernando:2017qrq,Majeed:2017txa,Sharif:2017owq,Tsukamoto:2017rrl,An:2017tlp,An:2017hby,Becar:2017aag,Gonzalez:2018lfs,Ahmed:2018fge,Shaymatov:2018azq,Ogasawara:2018gni,Gonzalez:2018zuu,Saha:2019xql,Rudra:2019ssz,Rahim:2019lip}. These studies conclude in their individual works that the conditions obtained by BSW to get infinite amount of high CM energy are universal and these results were also generalized by Harada in \cite{Harada:2014vka}. It is worth noting here that the conditions mentioned by BSW such as the BH must be extremal and one of the colliding particles should have a critical angular momentum are very rare to observe in nature. In turn, the BSW process is a very rare event to observe in nature which needs careful attention in diverse context.

The BSW mechanism is so far mainly studied for spinless test particles (i.e. particles that follow geodesics) only. However, in general a particle moving in the vicinity of a BH is an extended object having self interaction such as the case of a spinning particle. It has been shown by Matisson, Papapetrou and Dixon (MPD) \cite{Mathisson:1937zz,Papapetrou:1951pa,Dixon:1964} that the trajectory followed by a spinning particle is non-geodesic due to the coupling between the spin of the particle and curvature of the spacetime around a massive central object like a BH.

In 2016, it was shown by Armaza et al. \cite{Armaza:2015eha} that it is still possible to obtain an infinite amount of CM energy for the Schwarzschild BH if one considers the collision of spinning particles instead of a collision of spinless particles. The study of BHs as a particle accelerator for spinning particles is further extended to the case of charged and spinning BHs in \cite{Zhang:2016btg}, where it was shown that it is possible to obtain infinitely high CM energy outside the event horizon of a nonextremal Reissner-Nordstrom (RN) BH. Zhang et al \cite{Zhang:2016btg} also concluded that the area belonging to the infinitely high CM energy in spin and total orbital angular momentum ($s,l$) plane of the spinning particles is very sensitive to the BH charge as it decreases as the charge of the black hole increases. They further showed that for a non-extremal Kerr BH case, we can also obtain infinitely high CM away from the event horizon and the corresponding area in the ($s,l$) plane increases with an increase in the spin of the BH.
%In the same work, they also discussed the CM energy of the colliding spinning particles in the background of Kerr-Newman BH as well.
Combining charge and rotation in the Kerr-Newman background they finally concluded that the spin parameter and the charge of the BH affect the CM energy of the colliding particles in a completely opposite way. Recently, the universality of BSW mechanism for spinning particles, for a class of stationary axisymmetric BH, is also discussed in \cite{Jiang:2019cuc}. However, in \cite{Jiang:2019cuc}, the calculations of $E_{CM}$ in terms of 4-momentum which is a conserved quantity are not performed. Also, the timelike condition is not verified explicitly for 4-velocity. Hence, the results are not conclusive for two spinning particles colliding in the vicinity of a  BH and it is therefore worthy to discuss the collision of such particles in the vicinity of more BHs to draw the definite conclusions in this regard.

%However, the study of BHs as  particle accelerators for spinning particles has been performed for only a few BH models. Hence, it is very interesting to explore more about BH particle acceleration processes in the context of spinning particles. Based on such standpoints,
In this work, we extent the study of BH spinning particle acceleration processes and investigate two spinning particles colliding outside the event horizon of the non-extremal Schwarzschild BH which is surrounded by the quintessence like matter, which we will abbreviate as SBHQ henceforth \cite{Uniyal:2014paa}. We have observed that the CM energy of the colliding particles might be infinitely high for the collisions of the spinning particles, but the collisions must take place inside the cosmological horizon of the SBHQ. The CM energy in our case is found to be very sensitive to the value of normalization constant ($\lambda$) and the EOS parameter ($\epsilon$) which, for quintessential matter, varies from $-1$ to $-1/3$.

Our paper is organized as follows. We start in Sec. \ref{sec:spin_particles} with a brief overview of the equations of motion for  spinning  particles in Einstein's theory of general relativity (GR). In Sec. \ref{sec:metric}, we discuss the spacetime geometry of the SBHQ and and its event and cosmological horizons. Following Refs. \cite{Hojman:1976kn,Zalaquett:2014eia,Saijo:1998mn} we also derived the expressions for the four-momentum of a spinning particle. In Sec. \ref{sec:CM}, we obtain the expression for the CM energy of the colliding spinning particles in the vicinity of SBHQs and show that it reduces to the Schwarzschild black hole case \cite{Armaza:2015eha} if the normalization constant $\lambda$ vanishes and the energy $e$ per unit mass becomes unity. We then discuss the possible scenarios where arbitrarily high $E_{\rm CM}$ is possible.  Section \ref{sec:potential} is devoted to the study of the effective potential ($V_{\rm eff}$) and radial turning points for the trajectories of the spinning particles. We have divided this section into two parts: in the first part, we find the expression for $V_{\rm eff}$, as it helps to characterize the path of the spinning particle moving in the background of SBHQ. Based on this, in the second part we classify the spinning particles and their trajectories according to \cite{Zaslavskii:2016dfh} into three sub-classes: usual particle, critical particle and near-critical particle, respectively.
%Further, we also classify the trajectories of a spinning particle depending on its behavior (i.e., usual, critical and near-critical).
In Sec. \ref{super_lum}, we study the superluminal constraint and the conditions to avoid the superluminal region for the spinning particles.  Finally, Sec. \ref{sec:final_remarks} is devoted to the summary and conclusions of our results and to future prospects.

Throughout our work in this paper, we set the fundamental constants to unity (i.e., $c=G=1$), the signature of spacetime as $(-,+,+,+)$, Greek indices (i.e., $\alpha$, $\beta$,\, \ldots) run from $0$ to $3$ and Latin indices runs from $1$ to $3$ unless otherwise stated. Also, in the following sections, we chose the spin $s$ per unit mass of the colliding particles within the M\o ller limit (i.e., $r_{p}>s$) \cite{Moller:1949,PhysRevD.6.406}, where $r_{p}$ is the size of the spinning particle. It is important to note that size of the spinning particle is very less than the size of the BH (i.e. $r_{p}\ll r_{0(1)}$), therefore we have $s\ll M$ \cite{Zhang:2018gpn}.

%%%%%%%%%%%%%%%%%%%%%%%%%%%%%%%%%%%%%%%%%%%%%%%%%%%%%%
\section{Equations of motion of spinning particles in curved spacetime}\label{sec:spin_particles}
%%%%%%%%%%%%%%%%%%%%%%%%%%%%%%%%%%%%%%%%%%%%%%%%%%%%%%
The study of the chargeless spinning particles in GR started with the pioneering work of MPD \cite{Mathisson:1937zz,Papapetrou:1951pa,Dixon:1964} on spinning tops in curved spacetime. In their formulation, they showed that the trajectories followed by the chargeless spinning tops were not in accordance with the equivalence principle i.e. the above massive particles follow the non-geodesic paths. Further, Hojman \cite{Hojman:1976kn, Hojman:2016gep} extensively studied and extended the formulation by MPD. In this section, we will present a brief overview of the equations of motion developed by Hojman with the help of Lagrangian formulation.
The aforesaid equations of the motion read as
\begin{equation}
 \label{hojeq1}
 \frac{dx^{\alpha}}{d\tau}=u^{\alpha}\;,
\end{equation}
\begin{equation}
 \label{hojeq2}
 \frac{DP^{\alpha}}{D\tau}=-\frac{1}{2}R^{\alpha}_{\beta\gamma\delta}u^{\beta}S^{\gamma\delta}\;,
\end{equation}
\begin{equation}
 \label{hojeq3}
 \frac{DS^{\alpha\beta}}{D\tau}=S^{\alpha\gamma}\sigma_{\gamma}^{\beta}-\sigma^{\alpha\gamma}S_{\gamma}^{\beta}=P^{\alpha}u^{\beta}-P^{\beta}u^{\alpha}\;,
\end{equation}
where $\tau$, $u^{\alpha}$, $P^{\alpha}$, $S^{\alpha\beta}$ and $\sigma^{\alpha\beta}$ are an affine parameter, the 4-velocity, the 4-momentum vector, the spin tensor, and the antisymmetric angular velocity tensor, respectively. The antisymmetric angular velocity tensor is in turn defined as
\begin{equation}
 \sigma^{\alpha\beta}\equiv\eta^{(\gamma\delta)}
e_{(\gamma)}^{\alpha}\frac{De_{(\delta)}^{\beta}}{D\tau}=-\sigma^{\beta\alpha}\;.
\end{equation}
Here, $e_{(\gamma)}^{\alpha}$ is an orthonormal tetrad which is used to define the orientation of the top, $De_{(\delta)}^{\beta}/D\tau$ is the usual covariant derivative of the orthonormal tetrad and $\eta_{(\gamma\delta)}\equiv$ diag$(-1, 1, 1, 1)=\eta^{(\gamma\delta)}$.

As the Eqs. (\ref{hojeq1})-(\ref{hojeq3}) does not form a closed set of equations (i.e., they are insufficient to determine the complete trajectory of spinning particles  in a curved spacetime) and hence, spin supplementary conditions are needed. For simplicity purposes,  we choose the Tulczyjew spin supplementary condition (TSSC) $S^{\alpha\beta}P_{\beta}=0$ which conserves the dynamical mass of the spinning particle and choose a particular frame of the spinning particles for which only 3-components of $S^{\alpha\beta}$ are non-vanishing (i.e., $S^{0i}=0$) \citep{Armaza:2015eha}.

Additionally, the 4-momentum $P^{\alpha}$ is not parallel to the four velocity $u^{\alpha}$ for the case of a spinning particle and a relation between $P^\alpha$ and $u^\alpha$ is essential and can be written as \cite{Suzuki:1998}
\begin{equation}
 u^{\alpha}=\frac{\kappa}{m}\left[{P^{\alpha}}+\frac{2S^{\alpha\beta}P^{\gamma}R_{\beta\gamma\rho\epsilon}S^{\rho\epsilon}}{4m^{2}+R_{\mu\nu\kappa\lambda}S^{\mu\nu}S^{\kappa\lambda}}\right]\;.
\end{equation}
%with the help of the condition $S^{\alpha\beta}P_{\beta}=0$ (also known as Tulczyjew spin supplementary condition (TSSC)) which gives rise to rotations only \cite{Tulczyjew:1959}.
Here, $\kappa$ is a normalization constant. It is worth mentioning here that the above condition on the spin tensor comes naturally from the theory if one suitably chooses the corresponding Lagrangian (for detailed analysis see \cite{Hojman:2012me}).

We now define the conserved quantities \cite{Hojman:2012me} related to the spinning top and these are the mass $(m)$ of the spinning top
\begin{equation}
\label{mt}
 m^{2}=-P^{\alpha}P_{\alpha}\;,
\end{equation}
and its spin $(S)$,
\begin{equation}
\label{st}
S^{2}=\frac{1}{2}S^{\alpha\beta}S_{\alpha\beta}\;.
\end{equation}
In addition to the above-mentioned conserved quantities, we have an extra conserved quantity $D_{\xi}$ defined as below,
\begin{equation}
 D_\xi \equiv P^{\alpha}\xi_{\alpha}-\frac{1}{2}S^{\alpha\beta}\xi_{\alpha;\beta}\;,
\end{equation}
which is independent of the choice of the background metric as shown in \cite{Dixon:1964}. Here, $\xi_{\alpha}$ is a Killing vector associated with the spacetime metric. The motion of the tops in the background of SBHQ is presented in the next section.

%%%%%%%%%%%%%%%%%%%%%%%%%%%%%%%%%%%%%%%%%%%%%%%%%%
\section{Spinning particles in SBHQ background}\label{sec:metric}
\begin{figure*}
\begin{tabular}{c c c}
\hspace{-0.5cm}
\includegraphics[scale=0.45]{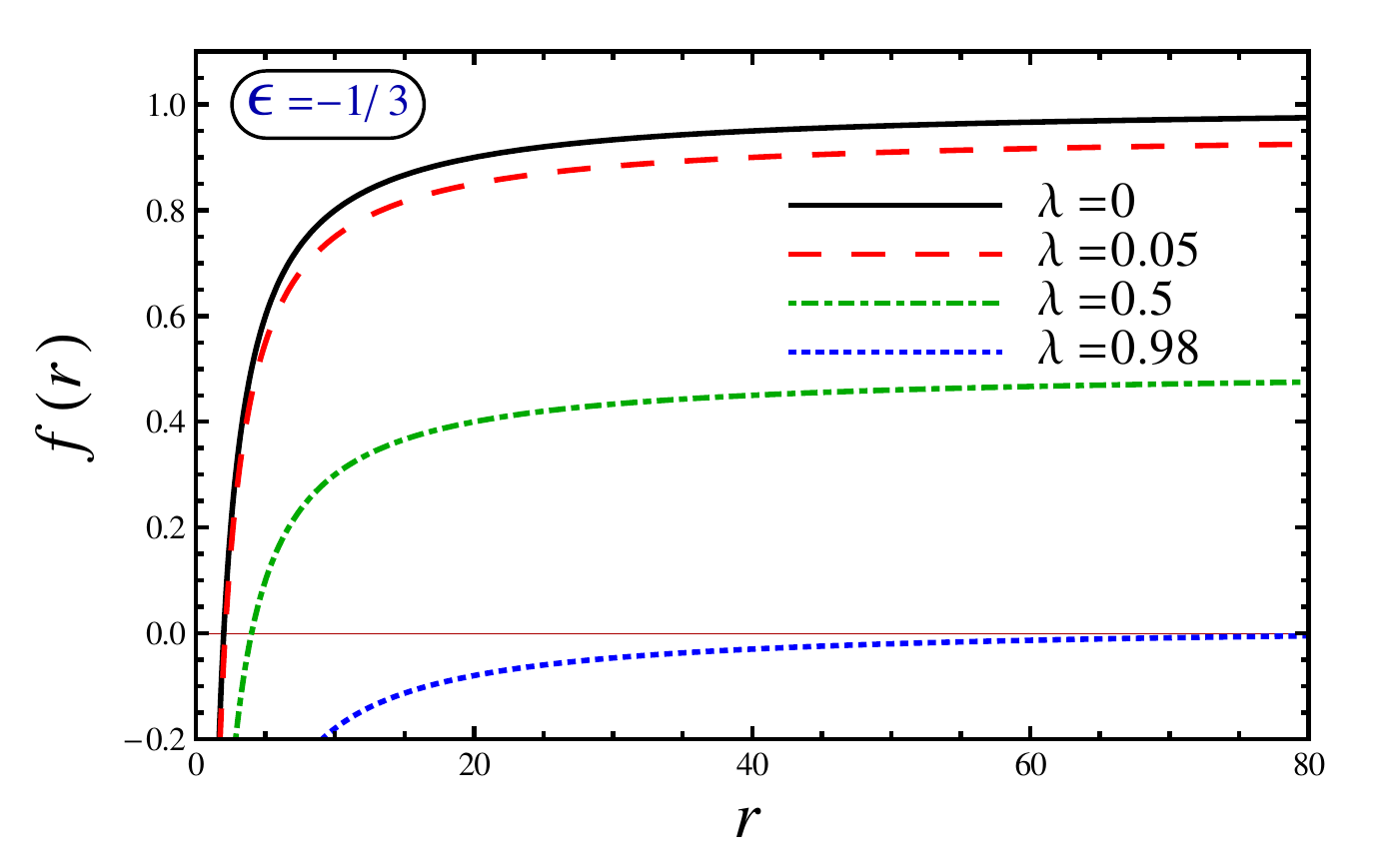}\hspace{-0.5cm}
&\includegraphics[scale=0.45]{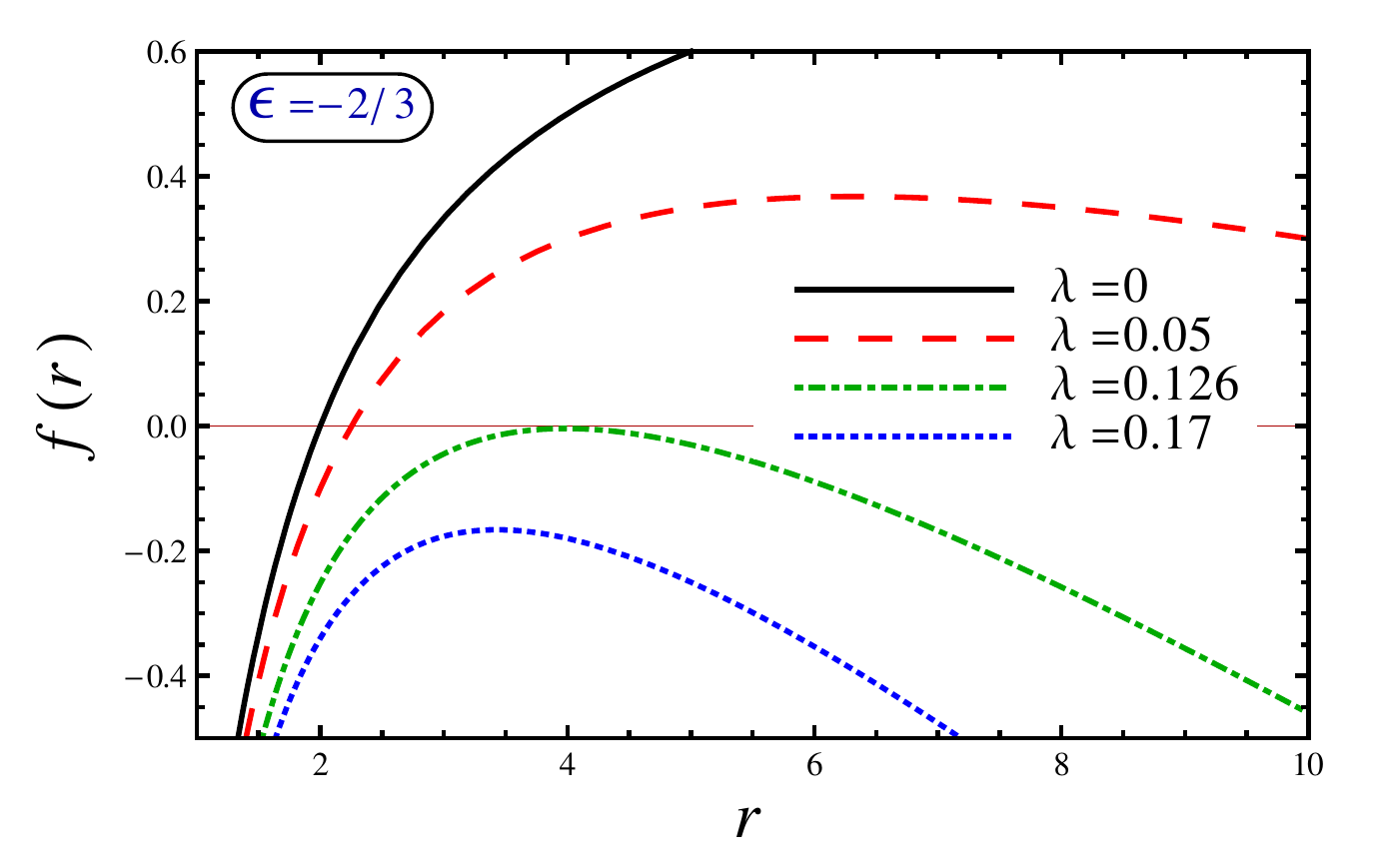}
 \hspace{-0.5cm}
 &\includegraphics[scale=0.45]{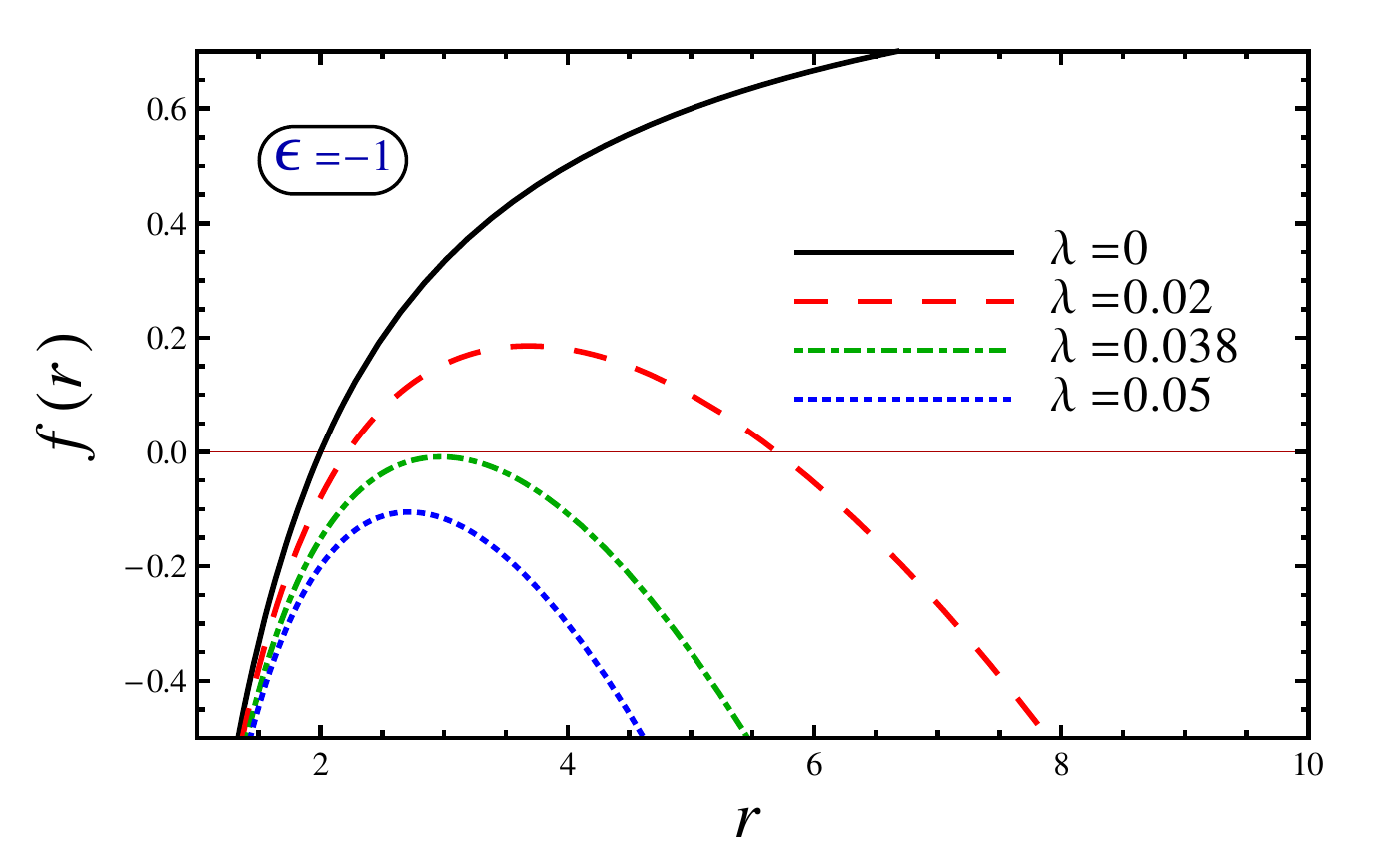}
 \end{tabular}
 \caption{The behavior of $f(r)$ with $r$ for different values of normalization constant $\lambda$, for fixed values of $\epsilon$ ($M=1$) (Color-online).}
\label{fig1}
\end{figure*}

%%%%%%%%%%%%%%%%%%%%%%%%%%%%%%%%%%%%%%%%%%%%%%%%%%%
The metric for SBHQ in the Schwarzschild coordinate system ($t,r,\theta,\phi$) reads as
\begin{equation}
\label{metric}
 ds^{2}=g_{\alpha\beta}dx^{\alpha}dx^{\beta}=-f(r)dt^{2}+\frac{1}{f(r)}dr^{2}+r^{2}d\Omega^{2}_{2}\;,
\end{equation}
where
\begin{eqnarray}
 && \label{fr} f(r)=\left(1-\frac{2M}{r}-\frac{\lambda}{r^{3\epsilon+1}}\right)\;,\\
 &&\label{dO} d\Omega^{2}_{2}=d\theta^{2}+\text{sin}^{2}d\phi^{2}\;.
\end{eqnarray}
Here, $\lambda$ is a normalization constant whose physical interpretation depends on the specific EOS parameter value $\epsilon$. The behaviour of $f(r)$ is shown in Fig. \ref{fig1} for different combinations of $\lambda$ and $\epsilon$.

In order to analyze the properties of SBHQ, we study the structure of horizon which has a two-sphere topology
(except in the case $\epsilon = -1/3$ and $0<\lambda<1$ which has the topology of a two-sphere but a deficit solid angle \cite{Nucamendi:1996ac}-\cite{Nucamendi:2000af}) and is calculated by the equation $g^{rr}=0$ of the above metric.
Now, using Eqs. (\ref{metric}), (\ref{fr}) and the above definition, the horizon satisfies the following condition
\begin{equation}
\Delta_{0}\equiv r^{3\epsilon+1}-2M r^{3\epsilon}-\lambda=0\;.
  \label{EH}
\end{equation}
From Eq. (\ref{EH}), we find that the horizon of SBHQ depends upon two extra parameters, i.e. $\lambda$ and $\epsilon$ respectively, besides the usual mass $M$ of a static spherical BH as in general relativity (i.e. SBH). We consider in this work three different choices of the EOS parameter, namely $\epsilon=-1/3, -2/3, -1$. For these choices, we now analyse the possible horizons of the spacetime:
% Depending on the  equation of state parameter $\epsilon (-1/3\;, -2/3\;, -1)$, we have following three different cases:
\begin{itemize}
    \item When $\epsilon=-1/3$ and $0<\lambda<1$, the Eq. (\ref{EH}) becomes linear in r and has only one root at $r=r_{0(1)}=2M/(1-\lambda)$, known as event horizon.
    \item For $\epsilon=-2/3$, the Eq. (\ref{EH}) becomes quadratic in r and has two roots $r_{0(1)}$ and $r_{0(2)}$, known as event and cosmological horizons, located at
    \begin{equation}
        r=r_{0(1,2)}=\frac{1\pm\sqrt{1-8 M\lambda}}{2\lambda}.
    \end{equation}
    It is clear from above equation that for $\lambda=1/8M$ both horizons coincide at the position $r=4M$.
    \item For $\epsilon=-1$, Eq. (\ref{EH}) becomes a depressed cubic equation in r whose discriminant and roots are as follows:
    \begin{eqnarray}
     \Box&=&\frac{1}{27\lambda^{3}}\left(-1+27M^{2}\lambda \right)\;, \label{delta}\\
     \tilde{r}_{1}&=& Y_{1}+Y_{2}\;,\\
     \tilde{r}_{2,3}&=& -\left(\frac{Y_{1}+Y_{2}}{2}\right)\pm\frac{i\sqrt{3}}{2}\left(Y_{1}-Y_{2}\right)\;,
    \end{eqnarray}
    where
    \begin{equation}
        Y_{1,2} = \sqrt[3]{\frac{-M}{\lambda}\pm\sqrt{\Box}}\;.
    \end{equation}
   Depending on the values of $\lambda$ we have following three sub-cases:
    \begin{enumerate}[label=\roman*)]
        \item If $\lambda=1/27M^{2} \implies \Box=0$, then all roots are real, and at least two are equal (i.e. $\tilde{r}_{1}<0$, $\tilde{r}_{2}\equiv r_{0{(2)}}=3M$ and $\tilde{r}_{3}\equiv r_{0{(1)}}=3M$). This means both the event $r_{0{(1)}}$ and the cosmological $r_{0{(2)}}$ horizons coincide.
        \item If $0<\lambda<1/27M^{2} \implies \Box<0$, then all roots are real and unequal (i.e. $\tilde{r}_{1}<0$, $\tilde{r}_{2}\equiv r_{0{(2)}}>0$ and $0<\tilde{r}_{3}\equiv r_{0{(1)}}<r_{0{(2)}}$). This means the event $r_{0{(1)}}$ and the cosmological $r_{0{(2)}}$ horizons do not coincide.
         \item If $\lambda>1/27M^{2} \implies \Box>0$, then one root is real and two are complex conjugates (i.e. $\tilde{r}_{1}<0$, $\tilde{r}_{2}\equiv r_{0{(2)}}=$ imaginary and $\tilde{r}_{3}\equiv r_{0{(1)}}=$ imaginary). This means for the case $\lambda>1/27M^{2}$, there are no horizons and hence corresponds to no BH spacetime. In fact, it corresponds to a naked singularity as evident from the expression for Kretschmann scalar ($\mathit{K}$) give below,
\begin{equation}
\mathit{K} =8\left(\frac {6\,{M}^{2}{r}^{2}+12\,M\lambda\,r+7\,{
\lambda}^{2}}{{r}^{8}}\right).
\end{equation}      
%This means$\lambda>1/27M^{2}$ is nonphysical.
   \end{enumerate}
\end{itemize}
Numerical values of horizon for different combination of normalization constant $\lambda$ and the EOS parameter $\epsilon$ are shown in Table. \ref{table1}.

%%%%%%%%%%%%%FIGURE%%%%%%%%%%%%%%
%%%%%%%%%%%%%%%%%%%%%%

Now, we study the motion of spinning particles in the background of the spacetime defined by Eq. (\ref{metric}). We restrict here to the case that the motion is planar. Due to spherical symmetry, we may assume that the particle is initially in the equatorial plane $\theta=\pi/2$. To ensure that $P^\theta=0$, $u^\theta=0$ we then assume that the spin vector is perpendicular to the equatorial plane \cite{Hackmann2014}. We find the constants of motion with the help of Eqs. (\ref{mt}) - (\ref{st}) which in the equatorial plane read as
\begin{eqnarray}
\label{mt1}
 m^{2}&=&-\frac{(P^{r})^{2}}{f(r)}+f(r)(P^{t})^{2}-r^{2}(P^{\phi})^{2}\;,\\
 \label{st1}
 S^{2}&=&-(S^{tr})^{2}+\frac{r^{2}(S^{r\phi})^{2}}{f(r)}-f(r)r^{2}(S^{t\phi})^{2}\;.
\end{eqnarray}
By using the symmetries of the background metric (\ref{metric}) and the Killing vectors, the energy ($E$) of the particles and the total angular momentum ($J$) orthogonal to the plane of motion as the conserved quantities are defined below,
\begin{eqnarray}
\label{E1}
 E&=&f(r)P^{t}-\frac{f(r)'S^{tr}}{2}\;,\\
 \label{j1}
 J&=&r(rP^{\phi}+S^{r\phi})\;,
\end{eqnarray}
where the ($'$) denotes the derivative with respect to the radial coordinate.
%%%%%%%%%%%%%%%%%%%%%%begin comment%%%%%%%%%%%%%%%%%%%%%%%%%%%%
%\begin{comment}
%%%%%%%%%%%%%%%%%%%%%%%HORIZONS%%%%%%%%%%%%%%%%%%%%%%%%%%%%%%%
\begin{table}
\begin{center}
\caption{Numerical values of horizons for SBHQ with $M=1$.}\label{table1}
\resizebox{\linewidth}{!}
{
\begin{tabular}{l l l l l l l}
 \hline \hline
 & & \multicolumn{1}{c}{$\epsilon=-1/3$} & \multicolumn{2}{c}{$\epsilon=-2/3$} &\multicolumn{2}{c}{$\epsilon=-1$}\\
 \cline{3-3}\cline{4-5}\cline{6-7}
& $\lambda$    & ${r_{0_{(1)}}}$ & \;\;\;${r_{0_{(1)}}}$ & ${r_{0_{(2)}}}$ & \;\;\;${r_{0_{(1)}}}$ & ${r_{0_{(2)}}}$\\
\hline
  & 0.0   & 2.0  &\;\; 2.0 & &\;\; 2.0 &  \\
& 0.00001    & 2.00002 &\;\; 2.00004 & 99997.99 &\;\; 2.00008 & 315.22  \\
  & 0.0001     & 2.0002  &\;\; 2.0004 & 9997.99 &\;\; 2.0008 & 98.98 \\
 & 0.001     & 2.002  &\;\; 2.004  & 997.99 &\;\; 2.008 & 30.57 \\
 & 0.01     & 2.02  &\;\; 2.04 & 97.96 &\;\; 2.09 & 8.78 \\
 & 0.1     & 2.22 &\;\; 2.76 & 7.23  \\
 & 0.2     & 2.50    \\
  & 0.3     & 2.85   \\
 & 0.4     & 3.33    \\
    & 0.5    & 4.0   \\
 \hline \hline
\end{tabular}
}
\end{center}
\end{table}
%%%%%%%%%%%%%%%%%%%%%%%%%%%%%%%%%%%%%%%%%%%%%%%%%%%%%%
%\end{comment}
%%%%%%%%%%%%%%%%%%%%%%%%%%%%end comment%%%%%%%%%%%%%%%%%%%%%%%
Now, by using the Eqs. (\ref{mt1}), (\ref{st1}) and the TSSC $S^{\alpha \beta}P_{\beta}=0$, the components $S^{t\phi}$ and $S^{r\phi}$ come out as
\begin{equation}
 S^{tr}=srP^{\phi}\;\;,\;\;S^{t\phi}=\frac{sP^{r}}{rf(r)}\;\text{and}\; S^{r\phi}=\frac{sf(r)P^{t}}{r}.
 \label{S_comps}
 \end{equation}
 It is worth to note here that $s=\pm S/m$ is the spin per unit mass; the $\pm$ signs are related to (anti) parallel spin of the particle with respect to the total angular momentum, respectively. The component of spin perpendicular to the equatorial plane may then reads as
 \begin{equation}
  S_{z}=rS^{r\phi}=s\left(\frac{2 e r-jsf(r)'}{2r-s^{2}f(r)'}\right)\;.
\label{Sz}
 \end{equation}
Further, all the non-zero components of the 4-momentum vector $P^{\alpha}$ calculated with the help of Eqs.\,(\ref{mt1}),\,(\ref{st1}),\,(\ref{E1}), \,(\ref{j1}),\,(\ref{S_comps}),\,and\,(\ref{Sz}) as follows,
\begin{eqnarray}
\label{pt}
 P^{t}&=& m\left(\frac{r^{3\epsilon+1}}{\Delta_{0}}\right)\mathcal{K}\;,\\
 \label{ph}
 P^{\phi}&=& m\left(\frac{2}{r}\right)\mathcal{L}\;,\\
 \label{pr}
 (P^{r})^{2}&=& m^{2}\left[\mathcal{K}^{2}-f(r)\left(1+4\mathcal{L}^{2}\right)\right]\;.
\end{eqnarray}
where
\begin{eqnarray}
%\Delta_{0}&=&r^{3\epsilon+1}-2 M r^{3\epsilon}-\lambda\;,\nonumber\\
\mathcal{K}&=&\frac{2 e r-j s f(r)'}{2r-s^{2}f(r)'}\;,\nonumber\\
\mathcal{L}&=&\frac{j-e s}{2r-s^{2}f(r)'}\;. \label{defL}
\end{eqnarray}
Here, $e=E/m$ is energy per unit mass and $j=J/m$ is the total angular momentum per unit mass. Hereafter, we normalize $m$ to unity for simplicity.

Finally, one can write the expression for $\dot{\phi}$ and $\dot{r}$ as follows:
\begin{eqnarray}
\label{dotph}
 \dot{\phi}&=&\frac{u^{\phi}}{u^{t}}=\frac{[2r-rs^{2}f(r)'']P^{\phi}}{[2r-s^{2}f(r)']P^{t}}\;,\\
 \dot{r}&=&\frac{u^{r}}{u^{t}}=\frac{P^{r}}{P^{t}}\;.
 \label{dotr}
\end{eqnarray}
It is worth mentioning here that the parameter corresponding to the proper time ($\tau$) has to be fixed in order to obtain the velocity components $u^{t}$, $u^{\phi}$, and $u^{r}$. However, for the above discussed relativistic invariants, one does not need to make any such specific choices.

%%%%%%%%%%%%%%%%%%%%%%%%%%%%%%%%%%%%%%%%%%%%%%%%%%%%%%%%%
\section{Centre-of-mass energy of the spinning particles}\label{sec:CM}
%%%%%%%%%%%%%%%%%%%%%%%%%%%%%%%%%%%%%%%%%%%%%%%%%%%%%%%%%
Let us consider two spinning massive particles ($m_{1}$ and $m_{2}$) colliding near to the horizon of the BH. The centre-of-mass energy ($E_{\rm CM}$) of these two particles can with the help of the formula derived as in \cite{Banados:2009pr} be written as
\begin{eqnarray}
\label{ecm1}
 E_{\rm CM}^{2}&=&-g_{\alpha\beta}\left(P_{1}^{\alpha}+P_{2}^{\alpha}\right)\left(P_{1}^{\beta}+P_{2}^{\beta}\right)\;,\\
 \label{ecm2}
 &=& m_{1}^{2}+m_{2}^{2}-2g_{\alpha\beta}P_{1}^{\alpha}P_{2}^{\beta}\;.
\end{eqnarray}
Here, with a constraint $m_{1}+m_{2}=constant$ along with the condition $R(r;,e,l,s)>1$ which follow under the fixed parameters $(e,j,s)$, the  and the condition $R(r;,e,l,s)>1$ (see appendix for details including the definition of the function $R$), the particles (spinning as well spinless) with equal masses acquire the maximum $E_{\rm CM}$ in comparison to the particles with unequal masses. This $E_{\rm CM}$ increases as the BH spin increases and diverges for the extremal rotating BH under specific conditions on the angular momentum of one of the particles. Hence, to have the maximum collisional energy, it is assumed that both the spinning particles have the same mass (i.e., $m_{1}=m_{2}=m$) and for simplicity we consider $m=1$. Therefore, the Eq. (\ref{ecm2}) with these assumptions in the equatorial plane becomes
\begin{equation}
 \label{ecm3}
 E_{\rm CM}^{2}=2\left[1-\left(g_{tt}P_{1}^{t}P_{2}^{t}+g_{rr}P_{1}^{r}P_{2}^{r}+g_{\phi\phi}P_{1}^{\phi}P_{2}^{\phi}\right)\right]\;,
\end{equation}
which after substituting the values of $P^{t}$, $P^{r}$ and $P^{\phi}$ from Eqs. (\ref{pt}), (\ref{ph}) and (\ref{pr}) respectively, reduces to
\begin{eqnarray}
\label{ecmf}
 E_{\rm CM}^{2}&=&\frac{2}{\Delta_{0}C_{1}C_{2}}\Bigg[r^{3\epsilon+1}D_{1}D_{2}+\Delta_{0}\Big[C_{1}C_{2}\nonumber\\
 &-&4r^{6\epsilon+4}(j_{1}-e_{1}s_{1})(j_{2}-e_{2}s_{2})\Big]\nonumber\\
 &-&\sqrt{r^{3\epsilon+1}D_{1}^{2}-\Delta_{0}[C_{1}^{2}+4\;r^{6\epsilon+4}(j_{1}-e_{1}s_{1})^{2}]}\nonumber\\&&\sqrt{r^{3\epsilon+1}D_{2}^{2}-\Delta_{0}[C_{2}^{2}+4\; r^{6\epsilon+4}(j_{2}-e_{2}s_{2})^{2}]}\Bigg],\nonumber\\
\end{eqnarray}
where
\begin{eqnarray}
%&& \Delta_{0}=r^{3\epsilon+1}-2Mr^{3\epsilon}-\lambda\;,\nonumber\\
 &&C_{1,2}=2r(r^{3\epsilon+2})-s_{1,2}^{2}\Big[2Mr^{3\epsilon}+\lambda(3\epsilon+1)\Big]\;,\nonumber\\
 &&D_{1,2}=2r(r^{3\epsilon+2})e_{1,2}-j_{1,2}s_{1,2}\Big[2Mr^{3\epsilon}+\lambda(3\epsilon+1)\Big]\;.\nonumber\\
 \label{C12}
 \end{eqnarray}
One can easily verify from Eq. (\ref{ecmf}) that $E_{\rm CM}$ could possibly diverge not only for $\Delta_{0}=0$ but also for $C_{1,2}=0$. In case one substitutes $\lambda=0$ and $e=1$ in Eq. (\ref{ecmf}), the expression for $E_{\rm CM}$ reduces to
\begin{eqnarray}
\label{ecmschw}
 E_{\rm CM}^{2}&=&\frac{2}{\Delta\Delta_{1}\Delta_{2}}\Bigg[r(r^{3}-Mj_{1}s_{1})(r^{3}-Mj_{2}s_{2})\nonumber\\
 &+&\Delta\big[\Delta_{1}\Delta_{2}-r^{4}(j_{1}-s_{1})(j_{2}-s_{2})\big]\nonumber\\
 &-&\sqrt{r(r^{3}-Mj_{1}s_{1})^{2}-\Delta[\Delta_{1}^{2}+r^{4}(j_{1}-s_{1})^{2}]}\nonumber\\
 &&\sqrt{r(r^{3}-Mj_{2}s_{2})^{2}-\Delta[\Delta_{2}^{2}+r^{4}(j_{2}-s_{2})^{2}]}
 \Bigg]\;.\nonumber\\
\end{eqnarray}
Here $\Delta=r-2M$ and $\Delta_{1,2}=r^{3}-Ms_{1,2}^{2}$.  Eq. (\ref{ecmschw}) matches with $E_{\rm CM}$ of two spinning test particles colliding near the Schwarzschild BH \citep{Armaza:2015eha}.

%%%%%%%%%%%%%%Divergence Radius $\epsilon=-1/3$%%%%%%%%%%%%%%%%%%%%%%%%%%
\begin{table}
\begin{center}
\caption{Numerical values of divergence radius $r_{d}$ for SBHQ with $\epsilon=-1/3$ and $M=1$.}\label{table2}
%\resizebox{\linewidth}{!}
%{
\begin{tabular}{l l l l }
 \hline \hline
& $s$ & ${r_{d}}$ \\
\hline
 & 0.2   & 0.341995  \\
  & 0.4   & 0.542884 \\
 & 0.6   & 0.711379 \\
 & 0.8   & 0.861774  \\
 & 0.99  & 0.993322  \\
\hline \hline
\end{tabular}
%}
\end{center}
\end{table}
%%%%%%%%%%%%%%%%%%%%%%%%%%%%%%%%%%%%%%%%%%%%%%%%%%%%%%
%%%%%%%%%%%%%%Divergence Radius $\epsilon=-2/3$%%%%%%%%%%%%%%%%%%%%%%%%%%
\begin{table}
\begin{center}
\caption{Numerical values of divergence radius $r_{d}$ for SBHQ with $\epsilon=-2/3$ and $M=1$.}\label{table3}
%\resizebox{\linewidth}{!}
%{
\begin{tabular}{c c c c c c}
 \hline \hline
  & & \multicolumn{1}{c}{$\lambda=0.00001$} & \multicolumn{1}{c}{$\lambda=0.0001$} &\multicolumn{1}{c}{$\lambda=0.001$}&\multicolumn{1}{c}{$\lambda=0.01$}\\
 \cline{3-3}\cline{4-4}\cline{5-5}\cline{6-6}
& $s$ & ${r_{d}}$ & ${r_{d}}$ & ${r_{d}}$ & ${r_{d}}$  \\
\hline
& 0.2   & 0.341995 & 0.341994 & 0.341989 & 0.341929\\
  & 0.4   & 0.542883 & 0.542881 & 0.542857 & 0.542617\\
  & 0.6   & 0.711378 & 0.711373 & 0.711319 & 0.710779\\
  & 0.8   & 0.861773 & 0.861763 & 0.861667 & 0.860709\\
  & 0.99  & 0.993321 & 0.993306 & 0.993159 & 0.991691\\
\hline \hline
\end{tabular}
%}
\end{center}
\end{table}
%%%%%%%%%%%%%%%%%%%%%%%%%%%%%%%%%%%%%%%%%%%%%%%%%%%%%%
%%%%%%%%%%%%%%Divergence Radius $\epsilon=-1$%%%%%%%%%%%%%%%%%%%%%%%%%%
\begin{table}
\begin{center}
\caption{Numerical values of divergence radius $r_{d}$ for SBHQ with $\epsilon=-1$ and $M=1$.}\label{table4}
%\resizebox{\linewidth}{!}
%{
\begin{tabular}{c c c c c c}
 \hline \hline
  & & \multicolumn{1}{c}{$\lambda=0.00001$} & \multicolumn{1}{c}{$\lambda=0.0001$} &\multicolumn{1}{c}{$\lambda=0.001$}&\multicolumn{1}{c}{$\lambda=0.01$}\\
 \cline{3-3}\cline{4-4}\cline{5-5}\cline{6-6}
& $s$ & ${r_{d}}$ & ${r_{d}}$ & ${r_{d}}$ & ${r_{d}}$  \\
\hline
& 0.2   & 0.341995 & 0.341994 & 0.341991 & 0.341949\\
 & 0.4   & 0.542883 & 0.542881 & 0.542855 & 0.542594\\
  & 0.6   & 0.711378 & 0.711370 & 0.711293 & 0.710527\\
 & 0.8   & 0.861772 & 0.861755 & 0.861590 & 0.859943\\
& 0.99  & 0.993319 & 0.993289 & 0.992998 & 0.990098\\
\hline \hline
\end{tabular}
%}
\end{center}
\end{table}
%%%%%%%%%%%%%%%%%%%%%%%%%%%%%%%%%%%%%%%%%%%%%%%%%%%%%%

For Eq. (\ref{ecmf}), the case when $\Delta_{0}=0$ is not of much interest because both numerator and denominator vanish at the horizon, and the energy in this limit becomes finite. It can be generally shown from \eqref{ecm3} that in the limit $\Delta_0=0$ or, equivalently, $f=0$ we find
\begin{align}
\frac{1}{2} E_{\rm CM}^2 & = 1 + \frac{1}{2} \left( \frac{\mathcal{K}_1}{\mathcal{K}_2} + \frac{\mathcal{K}_2}{\mathcal{K}_1}\right) + 2 \frac{(\mathcal{K}_1\mathcal{L}_2-\mathcal{K}_2\mathcal{L}_1)^2}{\mathcal{K}_1\mathcal{K}_2}\,, \label{limitECM}
\end{align}
where $\mathcal{K}_i$ and $\mathcal{L}_i$ refer to particle $i$. In the limit $\lambda=0$, $s=0$, $e=1$ this reduces to the result in \cite{Banados:2009pr}, and for $\lambda=0$, $e=1$ we recover the result in \cite{Armaza:2015eha}.

Also, the case $C_{1,2}=0$ is not of significant interest in contrast with \cite{Armaza:2015eha}, because the radius $r_d$, where the divergence occurs, always is behind the horizon, when the restriction on the particle's spin is taken into consideration. This can be seen as follows: $C_{i}$ is zero exactly if the denominator $(2r-s_i^2f')$ in \eqref{defL} vanishes. Note that this may only happen in the region where $f'>0$, and then $s_i^2 = 2r/f'$. In that region, for all the cases $\epsilon=-1/3,-2/3,-1$, the right hand side $2r/f'$ is a monotonically increasing function of $r$ and, if applicable, also of $\lambda$. This implies that at or outside the horizon we have $2r/f'\geq (2r/f')|_{(r=r_H,\lambda=0)} = 8M^2$, where $r_H$ is the horizon. As $s$ is smaller than the particle radius due to the M\o{}ller bound, and the particle radius is much smaller than $M$, $C_{1,2}$ can therefore not vanish at or outside the horizon. The numerical values of the divergence radius $r_d$ are shown in Tables \ref{table2}, \ref{table3} and \ref{table4} for different combinations of $\lambda$ and $\epsilon$.

Let us return to the case that the collision happens at or close to the horizon $f=0$. Similar to the arguments in \cite{Zaslavskii:2010aw}, from equation \eqref{limitECM} we observe that the center of mass energy may still diverge if $\mathcal{K}_1$ or $\mathcal{K}_2$ vanishes, and that $E_{\rm CM}$ may become arbitrarily large if at least one of the $\mathcal{K}_i$ becomes arbitrarily small. It might however turn out that particles with small or vanishing $\mathcal{K}$ may not be able to reach the near horizon region. Therefore, we will now study particle motion with a particular emphasize on particles that may enable arbitrarily large center of mass energy.

%%%%%%%%%%%%%%%%%%%%%%%%%%%%%%%%%%%%%%%%%%%%%%%%%%%%%%%
\section{Effective potential and radial
turning points}\label{sec:potential}
%%%%%%%%%%%%%%%%%%%%%%%%%%%%%%%%%%%%%%%%%%%%%%%%%%%%%%%
The study of the effective potential and the radial turning points are very important as this help us to characterize the different trajectories of the spinning particles moving around the BHs.

\subsection{Effective potential}
The radial velocity $u^r$ is proportional to the radial component of the conjugate momenta $P^r$ and, therefore, we can determine the radial turning points from $P^r=0$. We rewrite $P^r$  in the form of an effective potential,
\begin{eqnarray}
    &&\left(\frac{P^{r}}{m}\right)^{2}=A \left[1-\frac{s^2 f(r)'}{2 r}\right]^{-2} (e-V_{\rm eff(+)}(r))(e-V_{\rm eff(-)}(r)),\nonumber\\
     &&V_{\rm eff(\pm)}(r)=\frac{B\pm C^{1/2}}{A},
\end{eqnarray}
where
\begin{eqnarray}
 A&=&1-\frac{f(r) s^{2}}{r^2}\;,\nonumber\\
 B&=&\frac{j s}{r^2} \left(\frac{f(r)' r}{2}-f(r)\right)\;,\nonumber\\
 C&=& f(r)\left(1-\frac{s^2 f(r)'}{2 r}\right)^2\left[1+\frac{j^2}{r^2}-f(r)\frac{s^2}{r^2}\right]\;.
\end{eqnarray}
One needs to restrict the values of $r$ such that $e>V_{\rm eff(+)}(r)$ or $e<V_{\rm eff(-)}(r)$ whenever $A>0$, in order to have $u^{r}$ to be real for the motion of the spinning particle. We can easily check for the cases $\epsilon=-1,-\frac13,-\frac23$ that $A$ has some minimum outside the event horizon. For $\epsilon=-1$ we find a minimum at $r=3M$ with $A=1-s^2\left( \frac{1}{27M^2}+\lambda \right)$ implying that $A$ is positive between event and cosmological horizon for $s \ll M$. Analogously, for $\epsilon=-\frac13$ the minimum $A=1-s^2 \frac{(1-\lambda)^3}{27M^2}$ is at $r=\frac{3M}{1-\lambda}$, so again $A$ is positive for $s\ll M$. Finally, for $\epsilon=-\frac23$ we have a minimum between event and cosmological horizon at $r=(1-\sqrt{1-6M\lambda})/\lambda$, and $A$ is always positive for $s \ll M$.

%%%%%%%%%%%%%FIGURE%%%%%%%%%%%%%%
\begin{figure}
\begin{tabular}{c c c}
\includegraphics[scale=0.34]{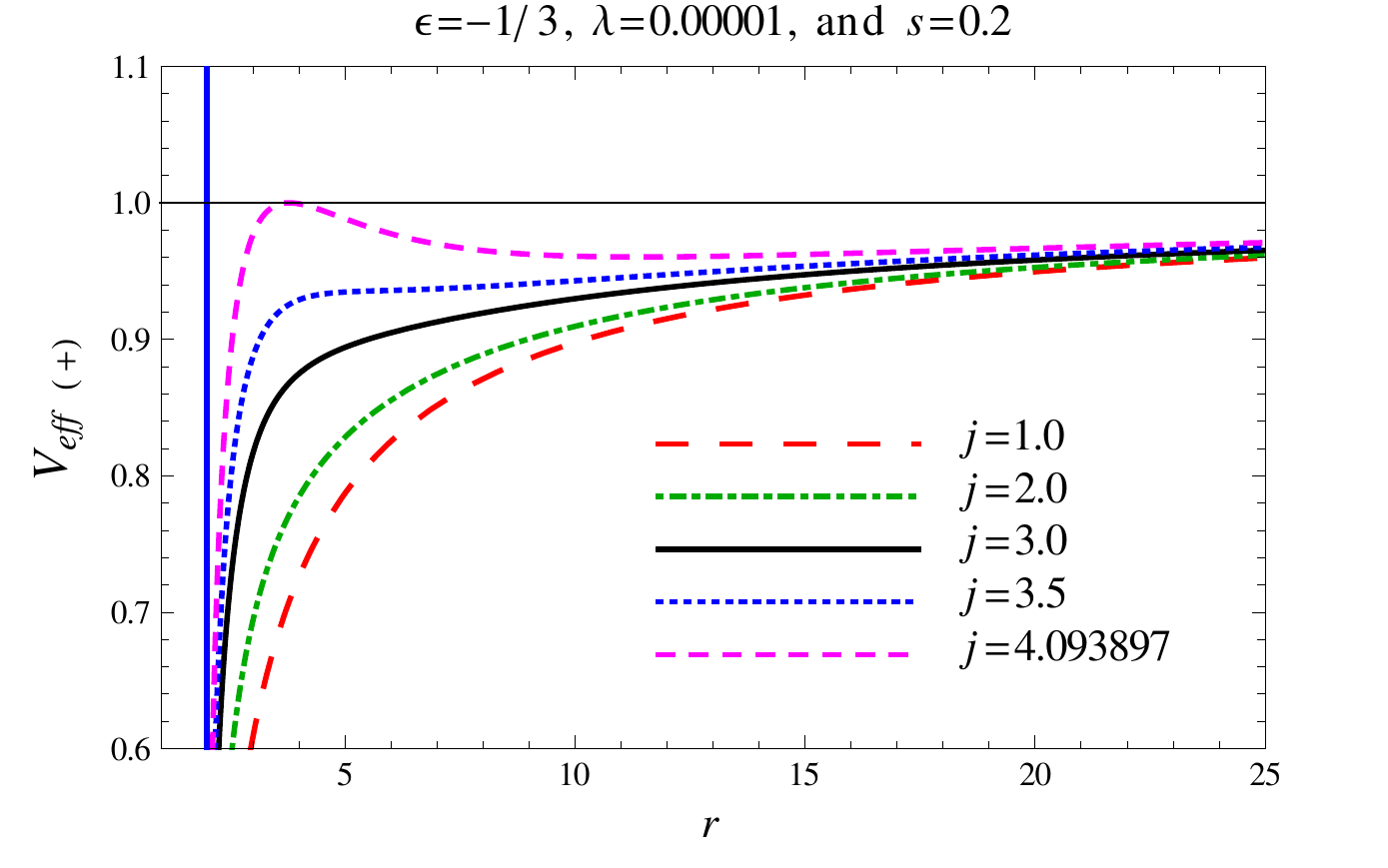}\hspace{-0.6cm}
&\includegraphics[scale=0.34]{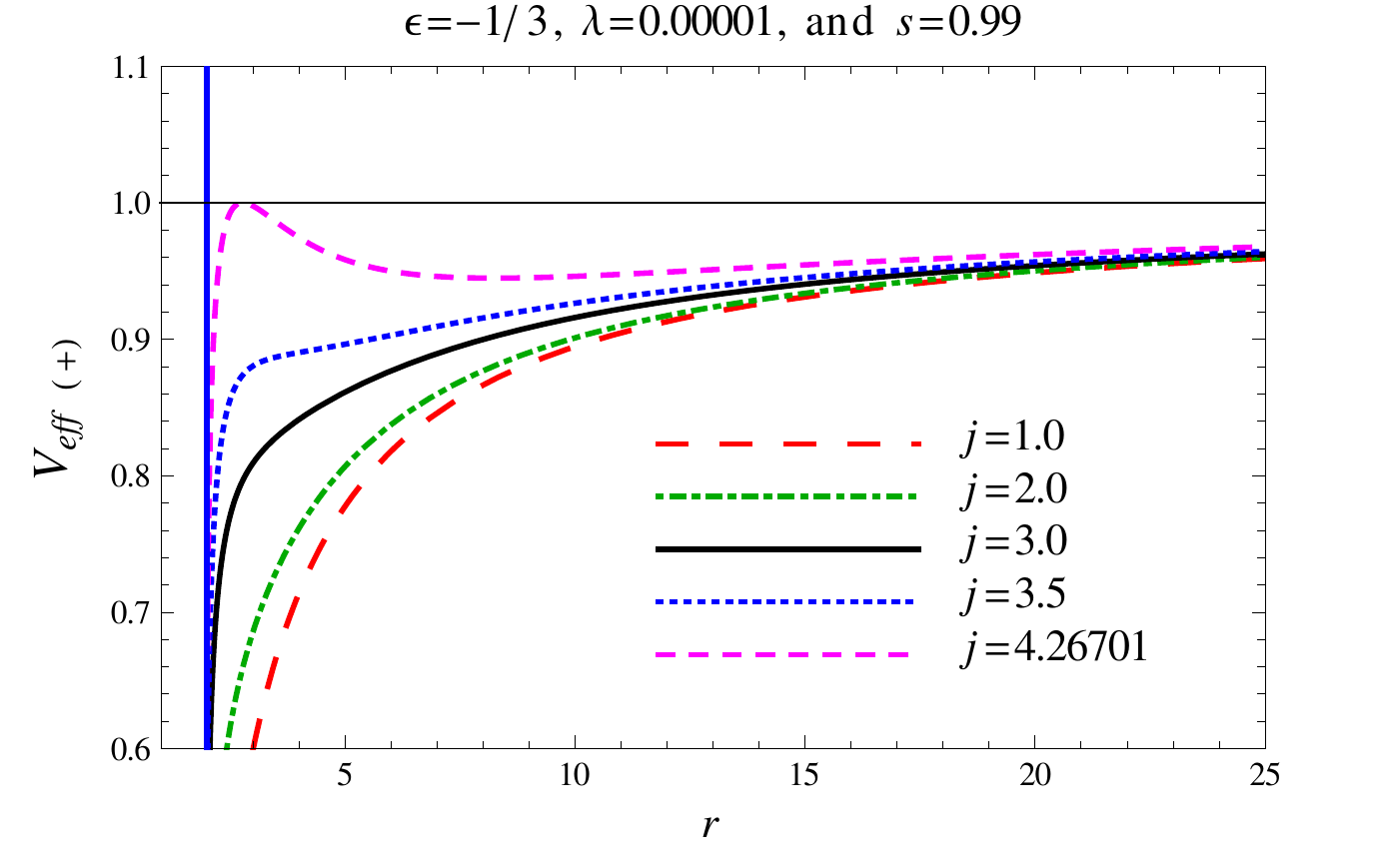}\\
\includegraphics[scale=0.34]{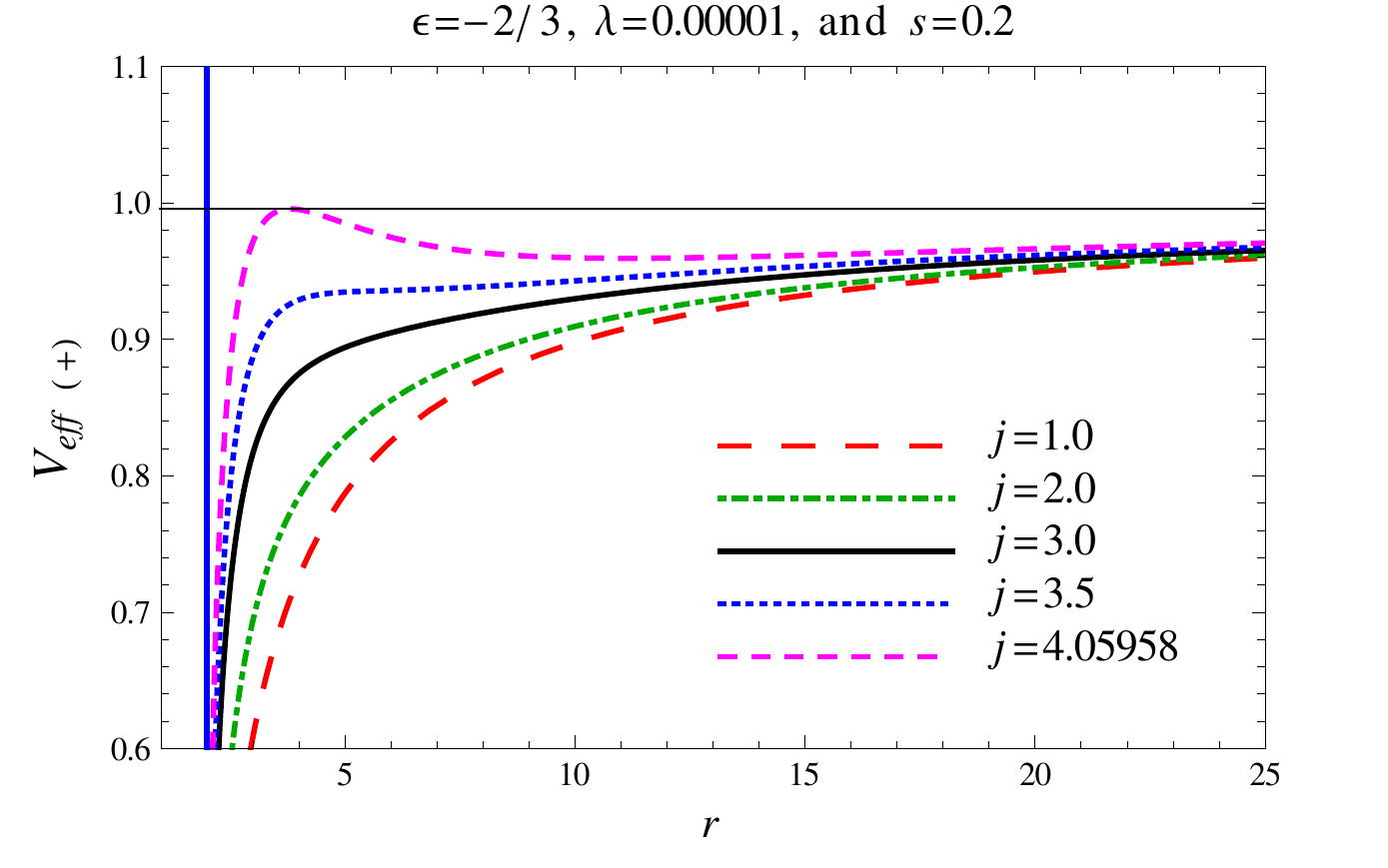}\hspace{-0.6cm}
& \includegraphics[scale=0.34]{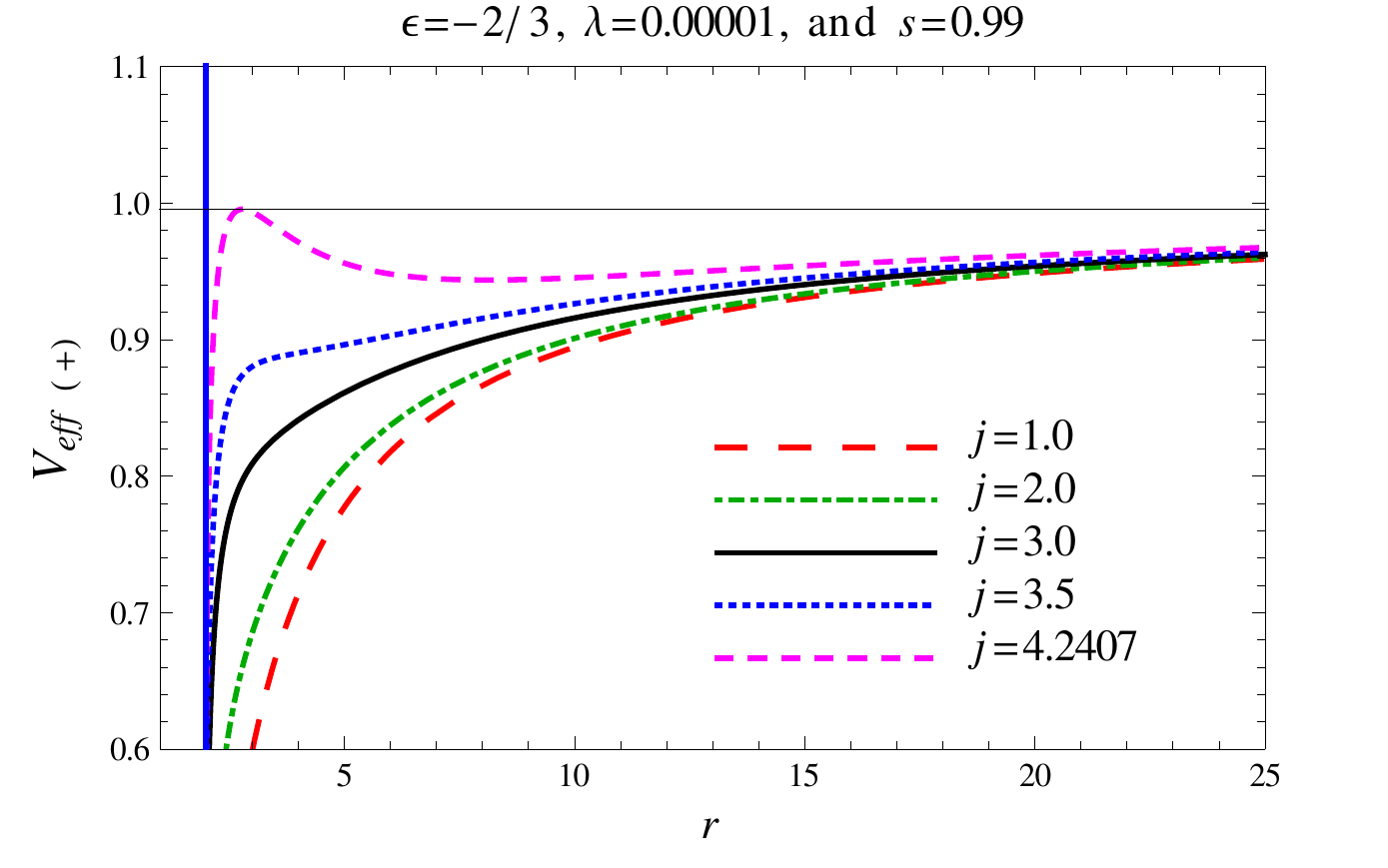}\\
\includegraphics[scale=0.34]{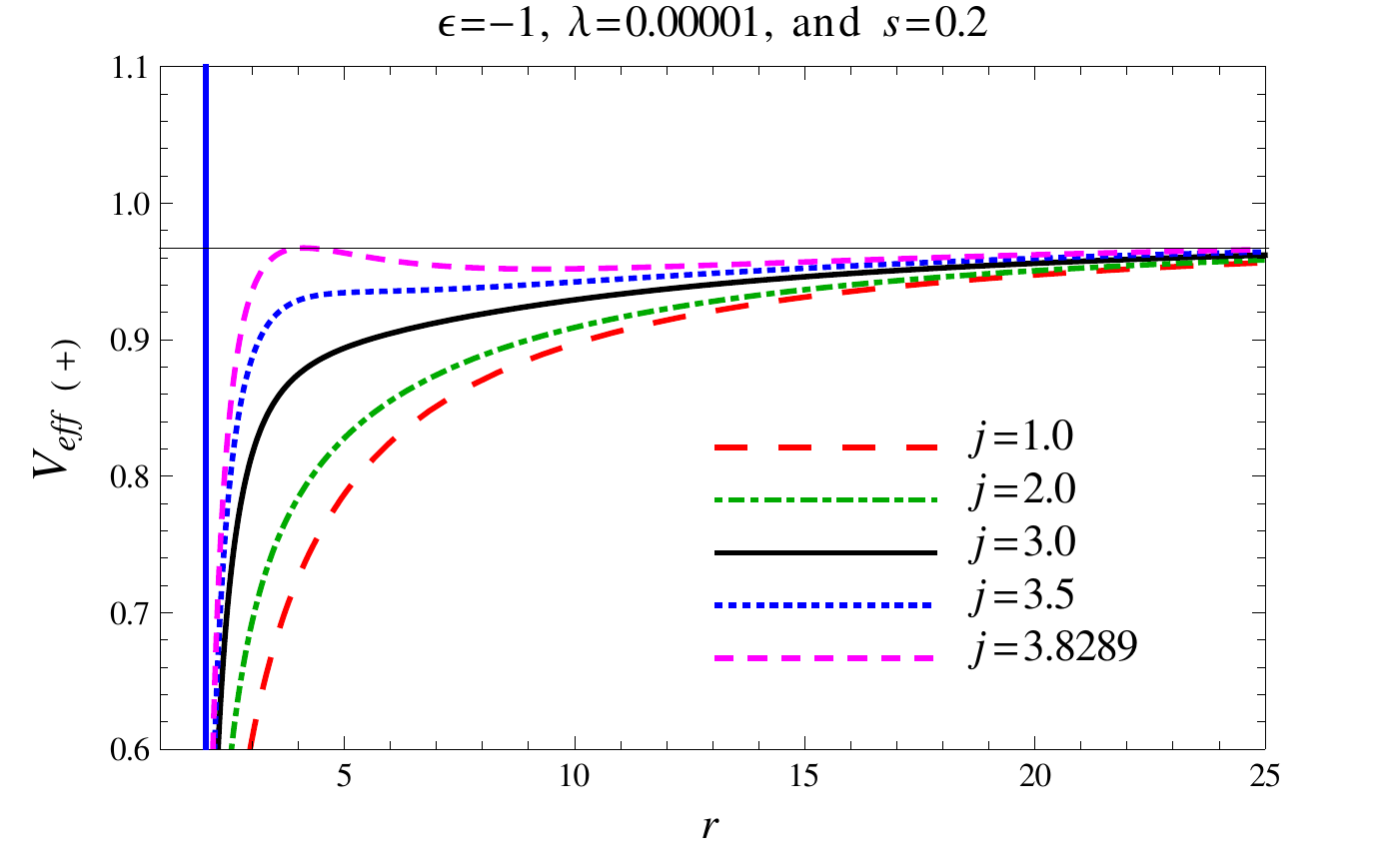}\hspace{-0.6cm}
&\includegraphics[scale=0.34]{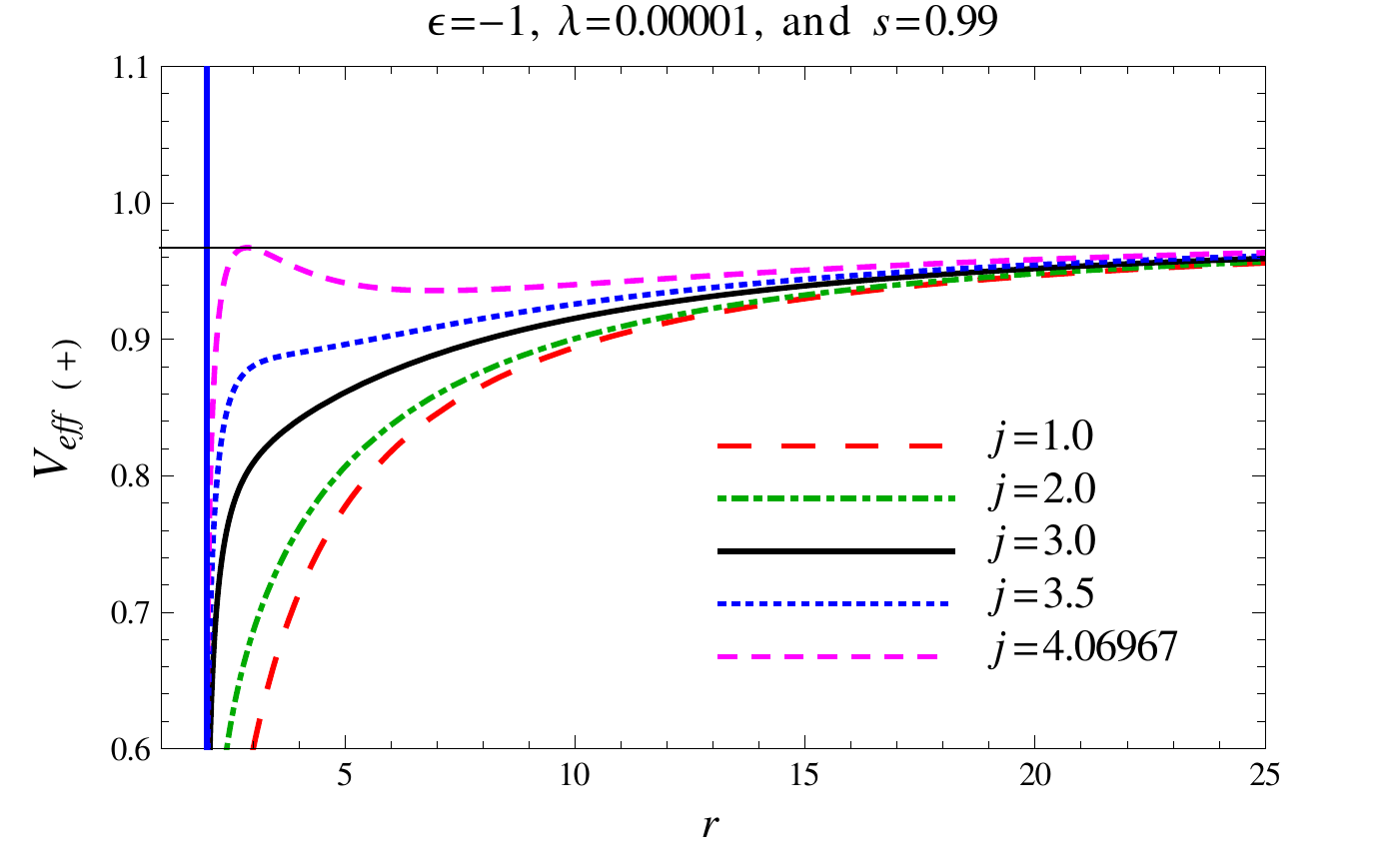}
 \end{tabular}
 \caption{Variation of the effective potential ($V_{\rm eff(+)}$) with respect to $r$, for different values of spin ($s$) corresponding to the constant value of EOS parameter ($\epsilon$), normalization constant ($\lambda$) and total angular momentum ($j$). Here, the solid (blue) vertical line indicate the location of the horizon ($M=1$) (Color-online).}
\label{fig4}
\end{figure}
%%%%%%%%%%%%%%%%%%%%%%

In the original paper by BSW \cite{Banados:2009pr} it is assumed that the colliding particles start from rest at infinity. In our case this is however not generally possible due to the presence of a cosmological horizon. Let us discuss the cases $\epsilon=-1/3,-2/3,-1$ separately: (i) If $\epsilon=-1/3$, we only have an event horizon as explained in section \ref{sec:metric}. We can therefore assume that the particle starts from rest at infinity. In this case, the energy of the particle is given by $e = 1-\lambda$. (ii) For $\epsilon=-2/3$, we have an event and a cosmological horizon if we choose $\lambda<1/(8M)$, see section \ref{sec:metric}. Therefore, it does not make sense to consider a particle starting from infinity. Instead, we could choose to let the particle start from rest from the static radius, see e.g.~\cite{Stuchlik:1983,Toshmatov:2017}, representing an equilibrium between gravitational attraction and cosmological expansion. A particle with $P^r=0$, $P^\phi=0$ can sit at radius $r=\sqrt{2/\lambda}$ with energy $e^2 = 1-2\sqrt{2\lambda}$. (iii) For $\epsilon=-1$ we may again choose the static radius as starting point, giving $r=\lambda^{-\frac13}$ and $e^2=1-3\lambda^\frac13$.

In Fig. \ref{fig4}, the behavior of the positive component of the effective potential $V_{\rm eff(+)}$ is shown as a function of $r$ for two different values of the spin ($s$) and several different $j$. We plotted here $V_{\rm eff(+)}$ only for those values of the particle spin $s$ and the total angular momentum $j$ for which the particle starting from rest from infinity or the static radius, as respectively explained above, will fall into the SBHQ and does not meet the turning point first. It is shown in Fig. \ref{fig4} that the maximum value of $V_{\rm eff(+)}$ decreases with increase in $j$ for each value of $s$ (i.e., $s=0.2$ and $s=0.99$) corresponding to $\epsilon=-1/3, -2/3$ and $-1$, respectively.

\begin{figure*}
\begin{tabular}{c c c}
\hspace{0cm}
\includegraphics[scale=0.46]{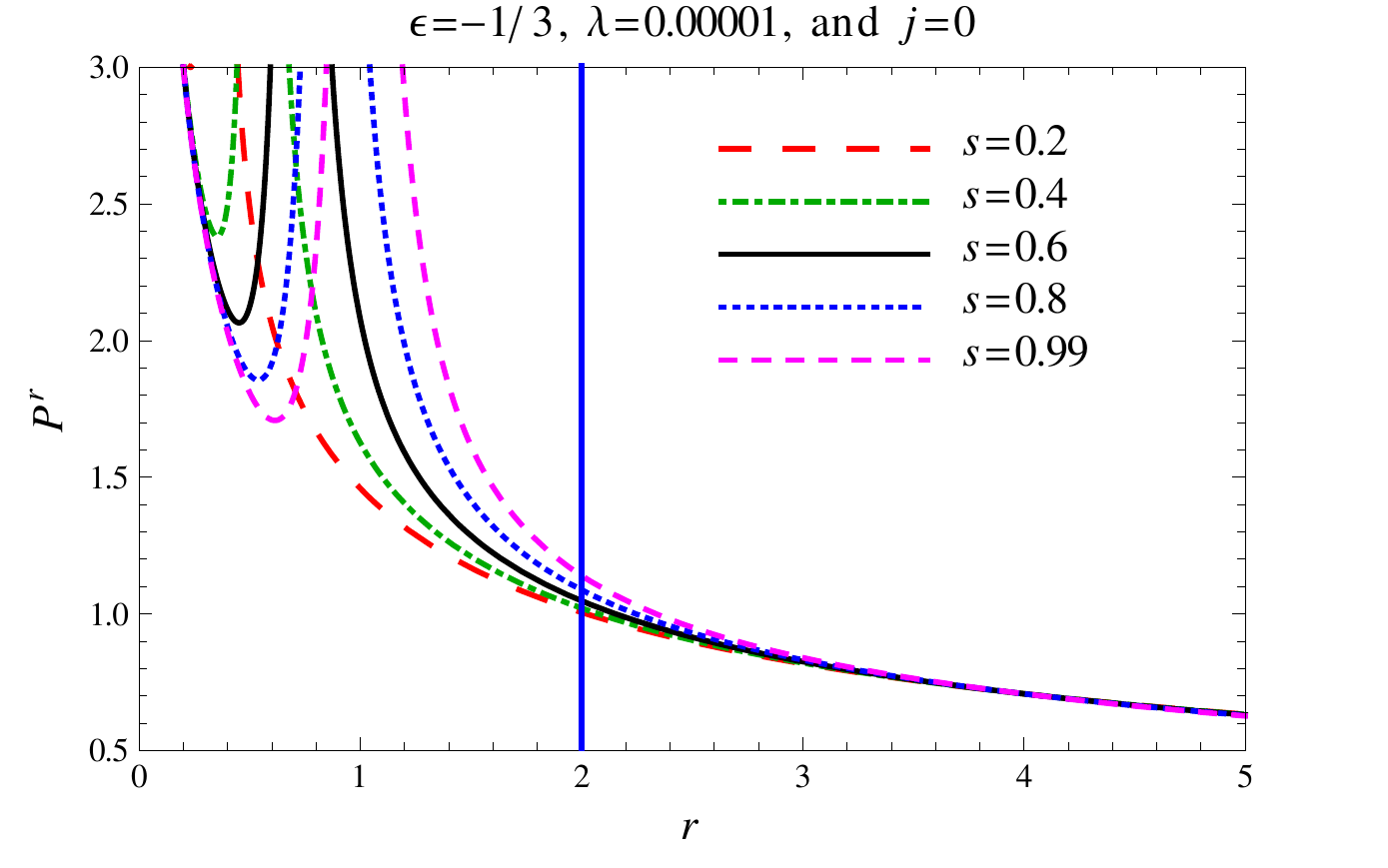}\hspace{-0.7cm}
&\includegraphics[scale=0.46]{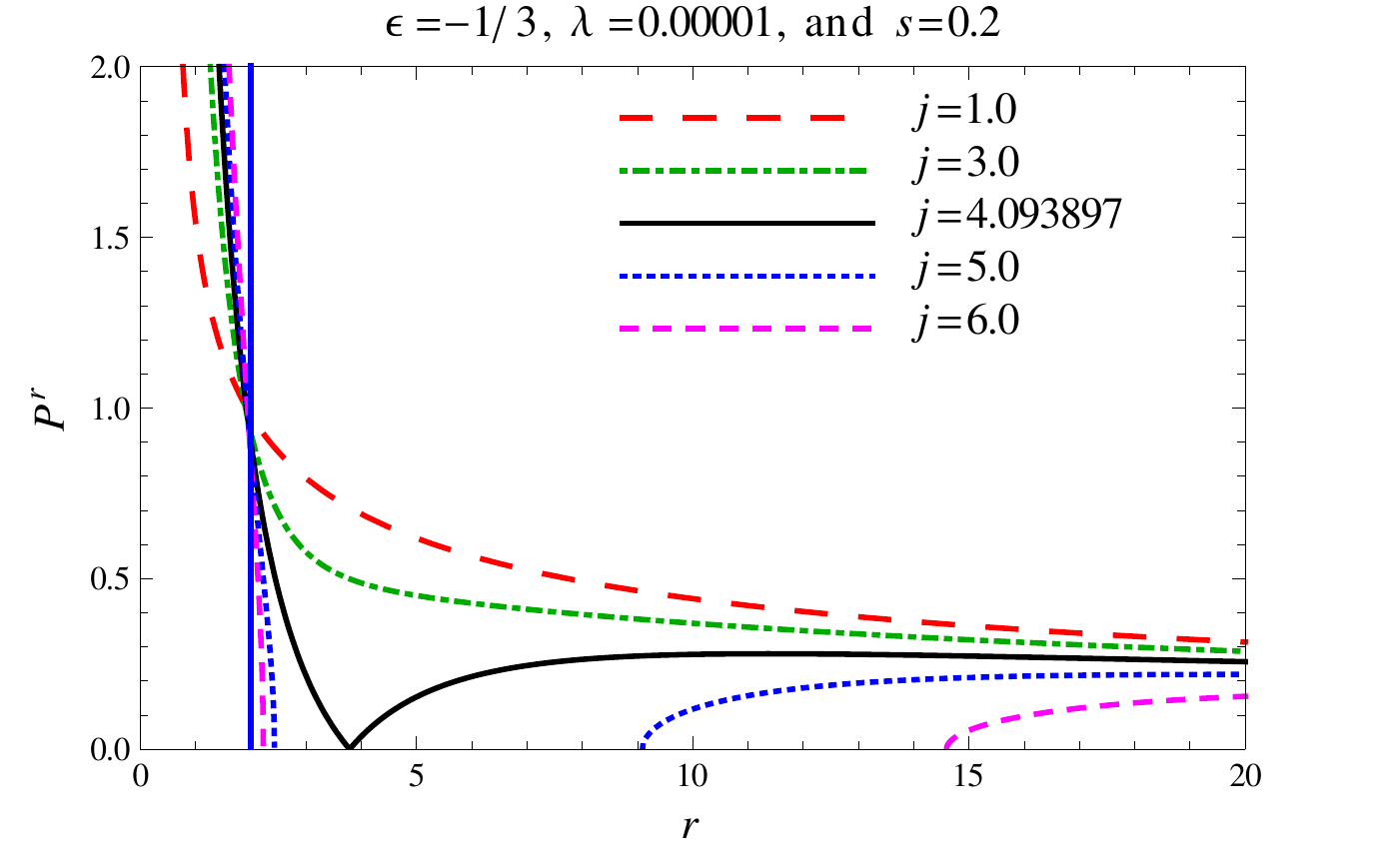}\hspace{-0.7cm}
&\includegraphics[scale=0.46]{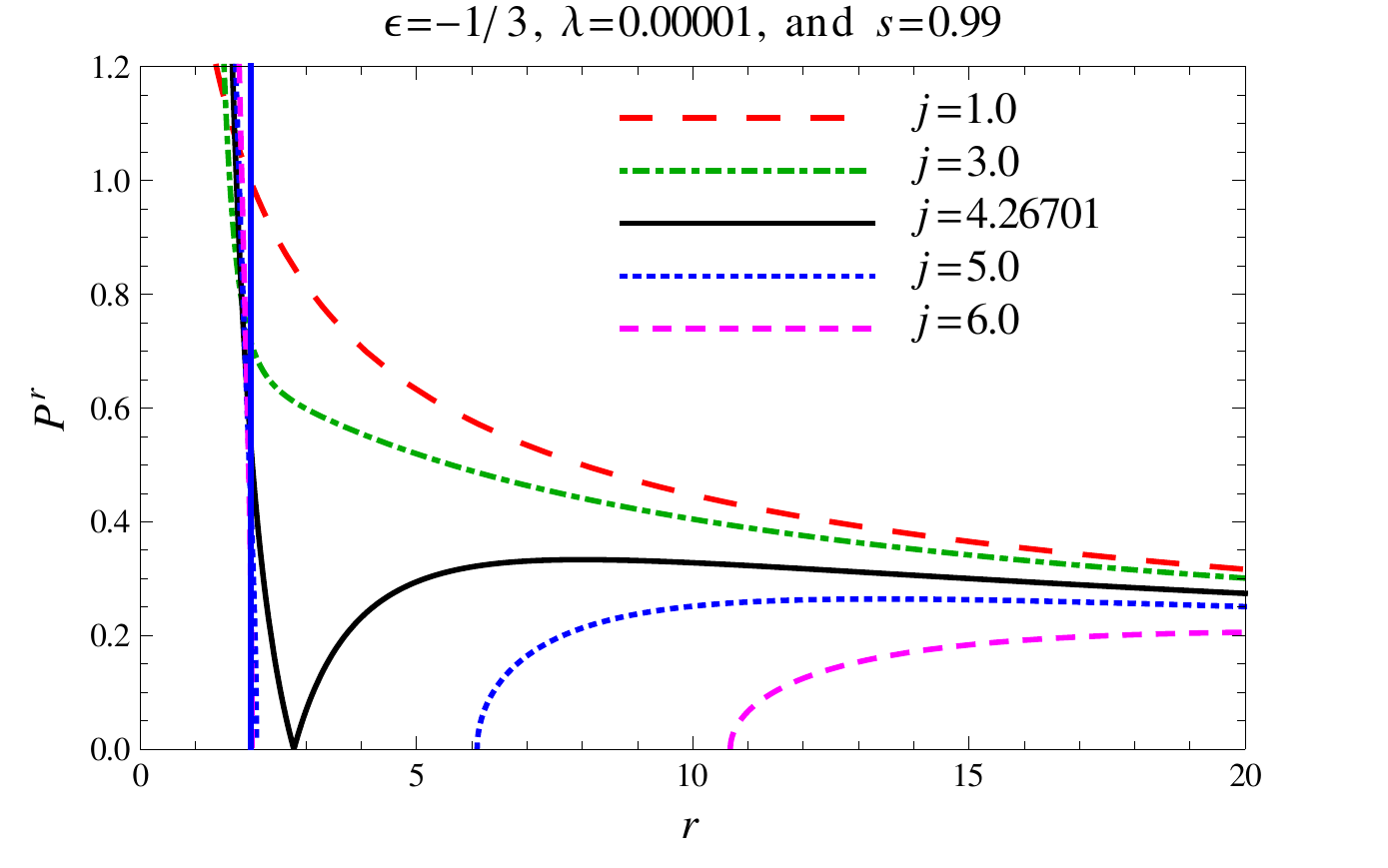}\\
\includegraphics[scale=0.46]{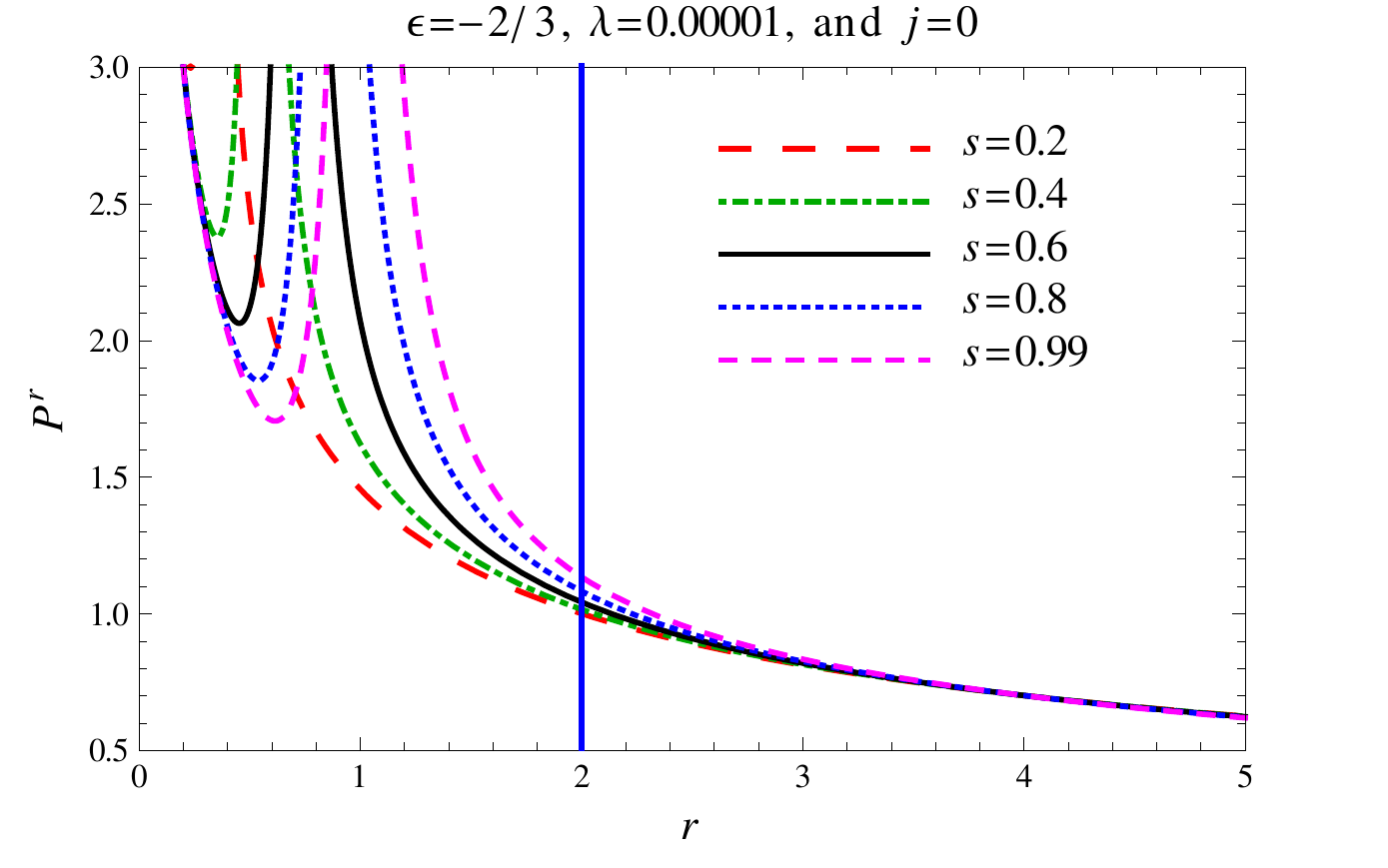}\hspace{-0.7cm}
&\includegraphics[scale=0.46]{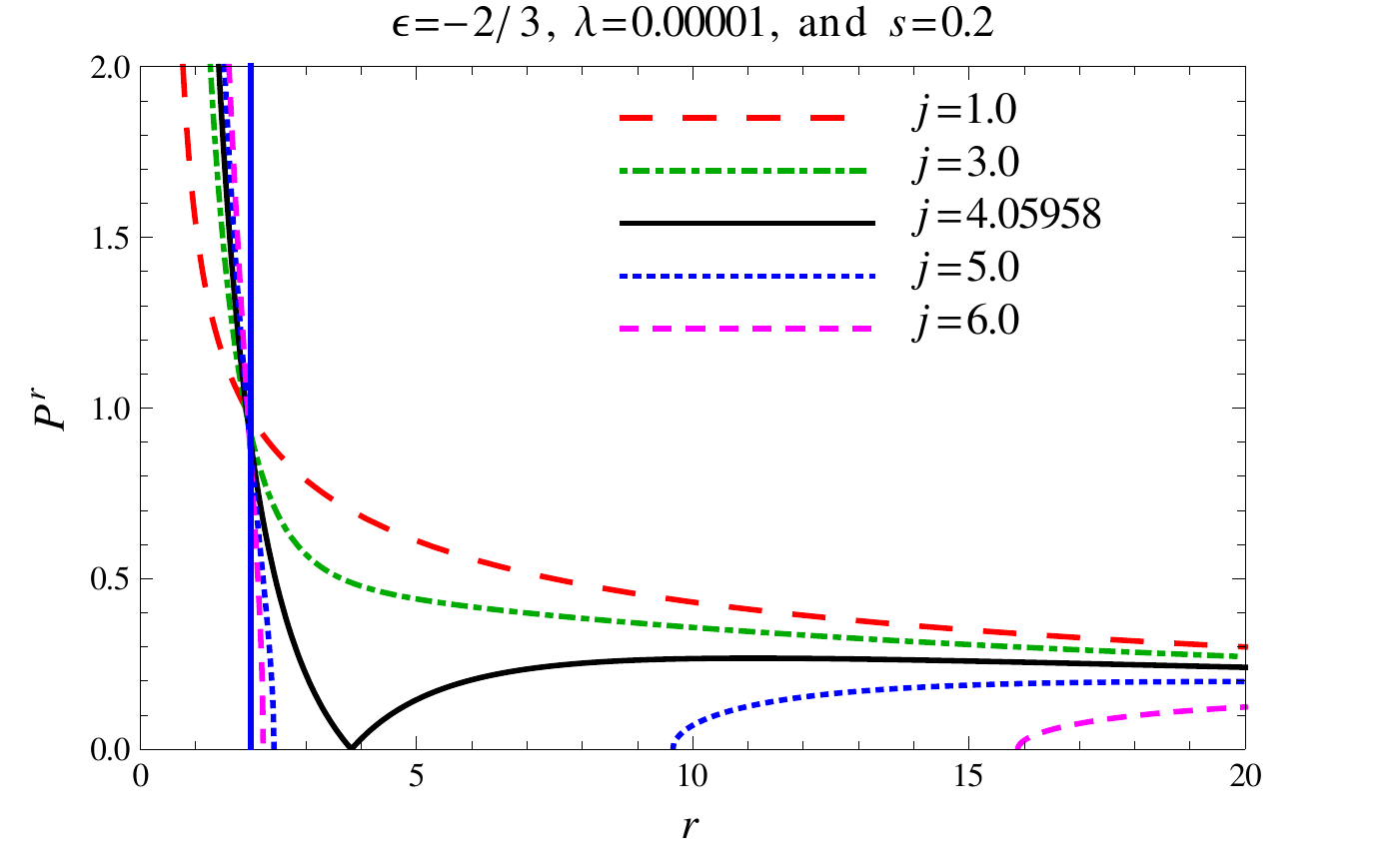}\hspace{-0.7cm}
&\includegraphics[scale=0.46]{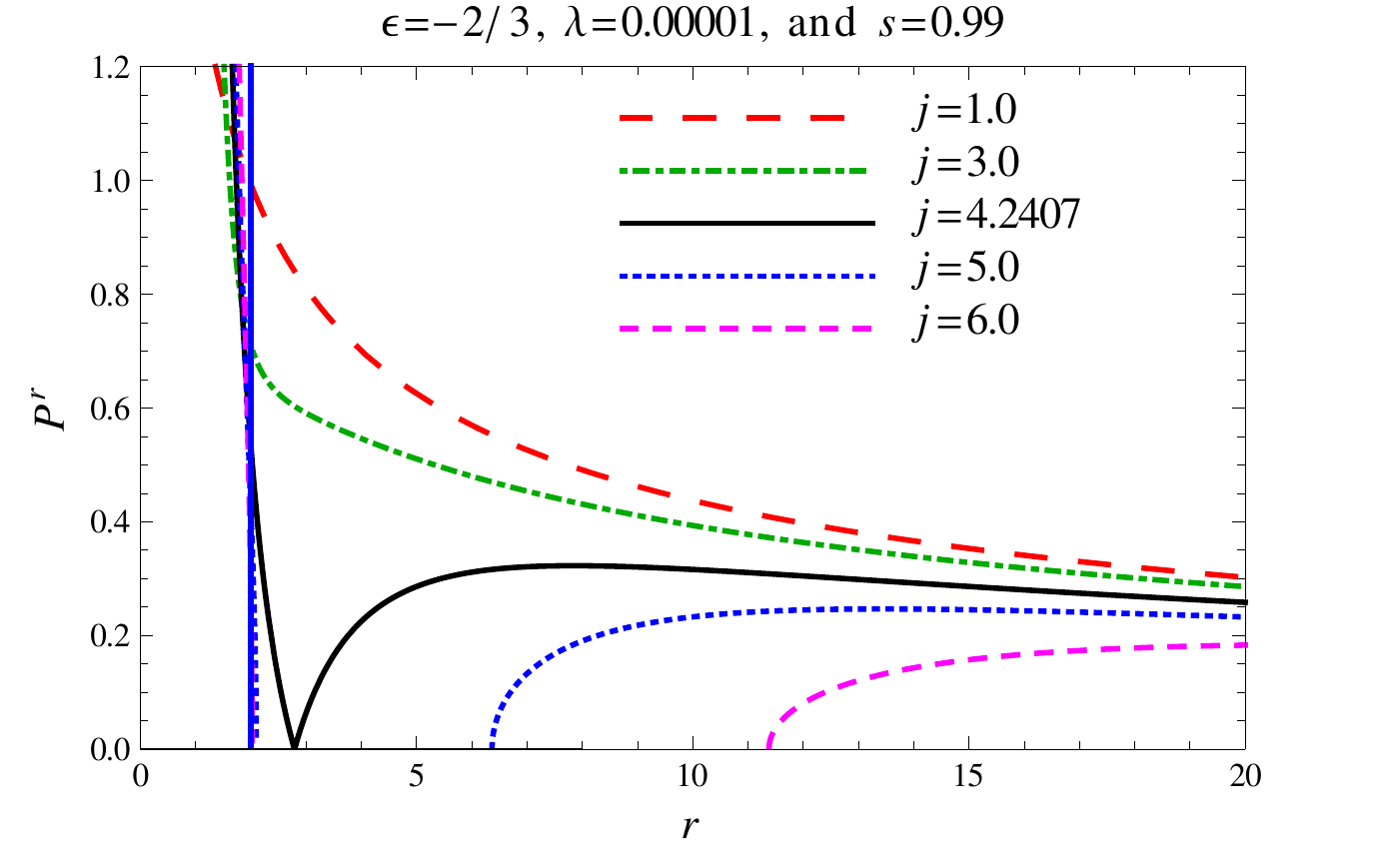}\\
\includegraphics[scale=0.46]{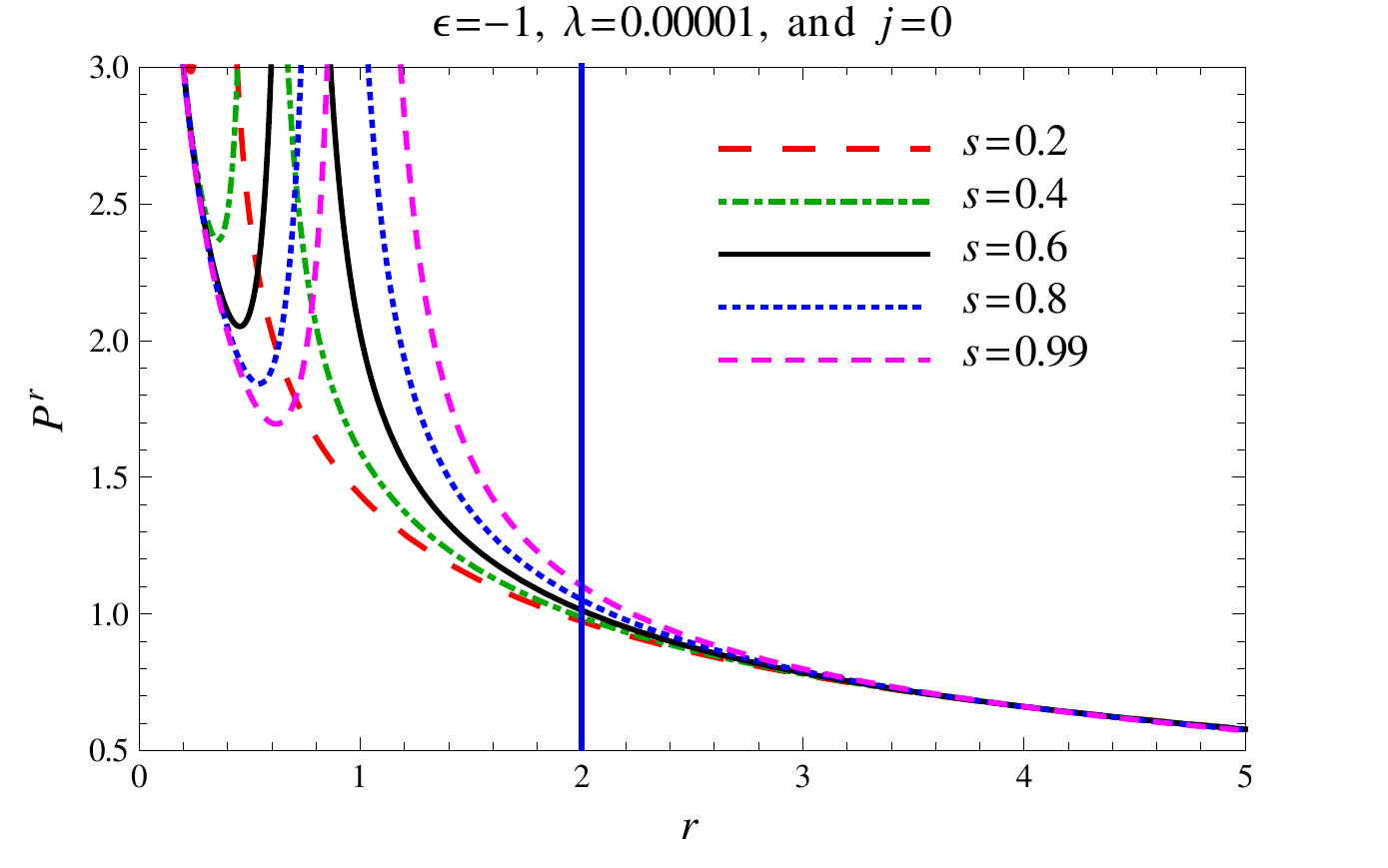}\hspace{-0.7cm}
&\includegraphics[scale=0.46]{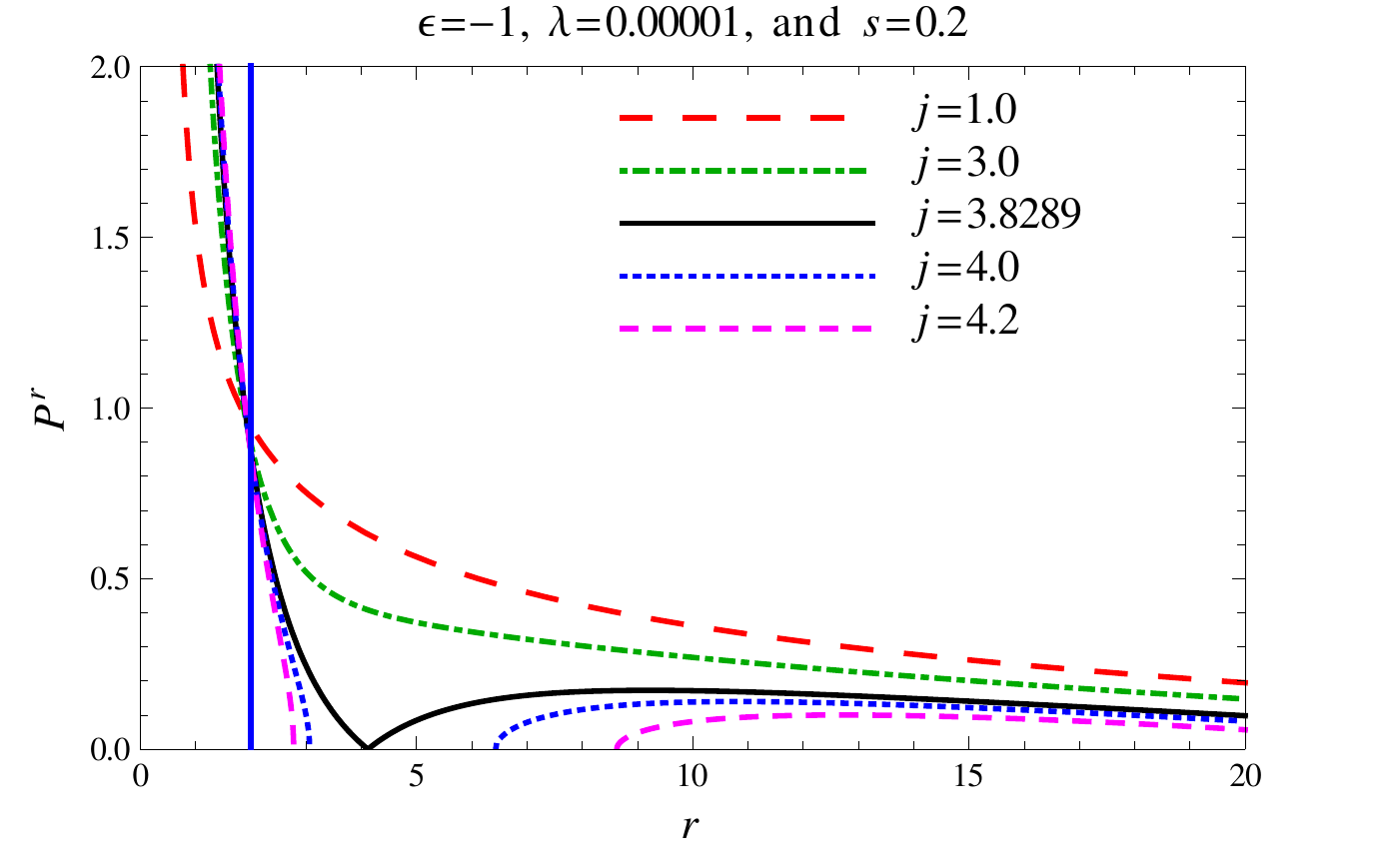}\hspace{-0.7cm}
&\includegraphics[scale=0.46]{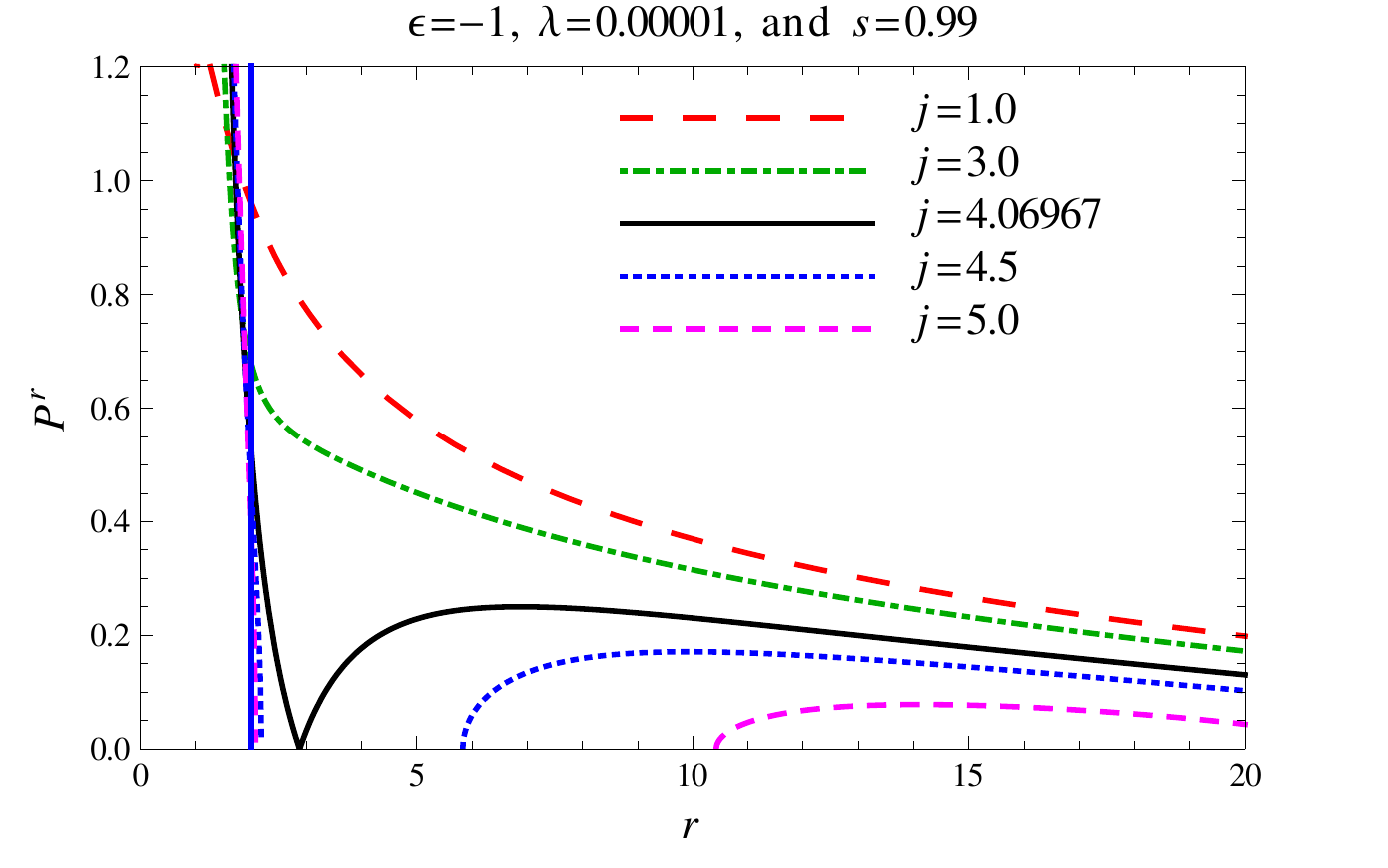}
\end{tabular}
 \caption{The variation of $P^{r}$ with $r$ for a Schwarzschild BH surrounded by quintessential matter. \textit{Left column:} Shows different combinations of spin $s$, keeping $j=0$. \textit{Middle column:} Different combinations of $j$, keeping $s=0.2$. \textit{Right column:} Different combinations of $j$, keeping $s=0.99$. In each of the rows the EOS parameter is fixed to $\epsilon=-1/3, -2/3\; \text{and} -1$, respectively, for the corresponding value of normalization constant (i.e., $\lambda=0.00001$). Here, for the corresponding value of parameter $\epsilon$ $(-1/3, -2/3\; \text{and} -1)$, the value of particle energy per unit mass $e$ is $0.9999, 0.995518$ and $0.967144$. The vertical (blue) solid line represents the location of event horizon ($M=1$) (Color-online).}
\label{fig3}
\end{figure*}

We showed the behavior of the positive component of $P^{r}$ with $r$ in Fig. \ref{fig3} for different combinations of particle spin $s$, total angular momentum $j$, $\lambda$ and $\epsilon$ as it will help in visualizing for which combinations of these parameters the spinning particle will reach the event horizon $r_{0({1})}$ first before meeting the turning point. In the figure, we fixed the normalization parameter $\lambda=0.00001$ and increase the EOS parameter for quintessential matter $\epsilon$ from top to bottom in each column. It is easy to conclude from the Fig. \ref{fig3} (see first column) that all the spinning particles fall into the SBHQ if they obey $s\ll M$, as implied by the M{\o}ller limit, for $j=0$. In the second and third columns the value of $s$ is fixed to $0.2$ and $0.99$, respectively.
%such that these values are well within the MML.
%\eva{Here in each plot the black solid line indicates the smallest value of $j$ such that a particle starting from rest from infinity or the static radius will not fall into the SBHQ but meet a turning point first} \red{ - Is this correct?}\pankaj{yes, you are right.}
It is found from the second and third columns that for each $\epsilon$ value the range of this total angular momentum $j$ increases with increase in $s$. However, the radial distance for which $P^{r}=0$ decreases as $s$ increases.

%%%%%%%%%%%%%FIGURE%%%%%%%%%%%%%%
%%%%%%%%%%%%%%%%%%%%%%

%%%%%%%%%%%%%%%%%%%%%%%%%%%%%%%%%%%%%%%%%%%%%%%%%%%%%%%%%
\subsection{Classification of the spinning particles and their trajectories}\label{sub:classification}
%%%%%%%%%%%%%%%%%%%%%%%%%%%%%%%%%%%%%%%%%%%%%%%%%%%%%

Let us return now to particles that might produce arbitrarily high center of mass energies. According to equation Eq. \eqref{limitECM} this may happen for collisions near the horizon if $\mathcal{K}$ of at least one of the colliding particles becomes very small.

From now onwards we denote the event horizon $r_{0(1)}\equiv r_{0}$ until and otherwise stated. We start by classifying the spinning particles into three different classes: We call a particle \textit{critical} if $\mathcal{K}|_{r=r_{0}}=0$, \textit{near-critical} if $\mathcal{K}|_{r=r_0}=\mathcal{O}(\sqrt{r_c-r_0})$ with the point of collision $r_c$, and all other particles \textit{usual}.

Let us start with critical particles. The condition $\mathcal{K}|_{r=r_{0}}=0$ implies
\begin{equation}\label{con_near_cric}
e=\frac{j s f(r_{0})'}{2 r_{0}}\;.
\end{equation}
Then, the expression for $\mathcal{K}$ near the event horizon (in the first approximation) reads
\begin{equation}\label{near_ho_con}
\mathcal{K}\approx \frac{3js}{r_{0}}\left[\frac{2Mr^{3\epsilon}_{0}+(\epsilon+1)(3\epsilon+1)\lambda}{2r_0^{3\epsilon+3}-s^{2}(2Mr_0^{3\epsilon}+(3\epsilon+1)\lambda)}\right](r-r_{0})\;.
\end{equation}

Thus, the second term in Eq. (\ref{pr}) becomes larger than $\mathcal{K}^{2}$ close to the horizon, where the collision should take place. Hence $(P^{r})^{2}$ is negative there which in turn means that the spinning particle cannot reach the event horizon and meets the turning point first.

For a near critical particle, to have $\mathcal{K}|_{r=r_0}=\mathcal{O}(\sqrt{r_c-r_0})$, we may for instance choose the energy as
\begin{align}
 e = \frac{jsf'(r_0)+a\sqrt{r_c-r_0}(2r_0-s^2f'(r_0))}{2r_0}\,, \label{e_nearcrit}
\end{align}
where $a$ is some positive constant. At the point of collision $r_c$ we then find
\begin{align}
 \mathcal{K}|_{r=r_c} & = \mathcal{K}|_{r=r_0} + \mathcal{K}'|_{r=r_0} (r_c-r_0) + \ldots\nonumber\\
 & = a\sqrt{r_c-r_0} + \mathcal{O}(r_c-r_0)\,.
\end{align}

Now consider the case that one particle, say particle 1, is usual and the other particle is near-critical. To calculate the center of mass energy for this case we write $f = (r-r_0) \tilde{f}$ and derive from \eqref{ecm3},
\begin{align}
 \frac{1}{2} E_{\rm CM} & = 1 - 4 \mathcal{L}_1\mathcal{L}_2 + \frac{1}{f} \left( \mathcal{K}_1\mathcal{K}_2 - \sqrt{R_1R_2} \right)
\end{align}
where $R=\mathcal{K}^2-f(1+4\mathcal{L}^2)$. If we evaluate all quantities at $r=r_c$ we find
\begin{widetext}
\begin{align}
 \frac{1}{2} E_{\rm CM} & = 1 - 4 \mathcal{L}_1\mathcal{L}_2 + \frac{\mathcal{K}_1a_2}{\tilde{f}\sqrt{r_c-r_0}} + \mathcal{O}(\sqrt{r_c-r_0})\nonumber\\
 & \qquad - \frac{\sqrt{\left[\mathcal{K}_1^2-(r_c-r_0)\tilde{f}(1+4\mathcal{L}_1^2)\right] \left[a_2^2+\mathcal{O}(\sqrt{r_c-r_0})-\tilde{f}(1+4 \mathcal{L}_2^2)\right]}\sqrt{r_c-r_0}}{\tilde{f}(r_c-r_0)}\\
  & = 1 - 4 \mathcal{L}_1\mathcal{L}_2 + \frac{\mathcal{K}_1 \left[ a_2-\sqrt{a_2^2-\tilde{f}(1+4\mathcal{L}_2^2)}\right]} {\tilde{f}\sqrt{r_c-r_0}} + \mathcal{O}(1)\,. \label{ECMnearcrit}
\end{align}
\end{widetext}
Here, it is important to note that in the limit $\lambda\rightarrow 0$, the Eq. (\ref{ECMnearcrit}) converges to the result found in \citep{Zaslavskii:2016dfh}.
We see that this expression is only valid if $a_2^2 - \tilde{f}(1+4\mathcal{L}_2^2)>0$. In the limit $s=0$ this condition can be fulfilled, and by continuity it should also hold for small $s$. If the point of collision $r_c$ now approaches the horizon $r_0$ the center of mass energy \eqref{ECMnearcrit} can grow without bound.

If both particles are near-critical, we can calculate the center of mass energy analogously,
\begin{widetext}
\begin{align}
  \frac{1}{2} E_{\rm CM} & = 1 - 4 \mathcal{L}_1\mathcal{L}_2 + \frac{a_1 a_2}{\tilde{f}} - \frac{\sqrt{\left[a_1^2-\tilde{f}(1+4\mathcal{L}_1^2)\right] \left[a_2^2-\tilde{f}(1+4\mathcal{L}_2^2)\right]}}{\tilde{f}} + \mathcal{O}(\sqrt{r_c-r_0})\,, \label{both_near_crit}
\end{align}
\end{widetext}
which will remain finite for $r_c \to r_0$. Finally, if both particles are usual, we can directly see that the diverging parts will cancel and the center of mass energy remains finite, too.

%%%%%%%%%%%%%%%%%%%%%%%%%%%%%%%%%%%%%%%%%%%%%%%%%%%
\section{Avoidance of superluminal region}\label{super_lum}
%%%%%%%%%%%%%%%%%%%%%%%%%%%%%%%%%%%%%%%%%%%%%%%%%%%
It is shown in \cite{Hojman_thesis:1975,Hojman:2012me,Armaza:2015eha,Deriglazov:2015wde,Deriglazov:2015zta} that the four-momentum satisfies the relation $P^{\alpha}P_{\alpha}=-1$ and hence is a conserved quantity, in contrast to the four-velocity $u^{\alpha}$, which is not a conserved quantity for the spinning test particles moving in the curved background. Therefore, the $P^{\alpha}$ vector remains timelike throughout the motion of spinning particle around the BH, whereas the $u^{\alpha}$ vector  might change from the subluminal (timelike) to superluminal (spacelike) region depending upon the invariant relation $u^{\alpha}u_{\alpha}<0$ or $u^{\alpha}u_{\alpha}>0$, respectively. As the four-velocity $u^{\alpha}$ of two colliding spinning test particles will not always lie in the subluminal region, it becomes important to examine closely the behavior of the square of the four-velocity in the region where $E_{\rm CM}$ diverges.
The square of the four velocity thus reads as
\begin{eqnarray}\label{U2_1}
U^{2}&=&\frac{u^{\alpha}u_{\alpha}}{(u^{t})^{2}}
=g_{tt}+g_{rr}\left(\frac{u^{r}}{u^{t}}\right)^{2}+g_{\phi\phi}\left(\frac{u^{\phi}}{u^{t}}\right)^{2}\;.
\end{eqnarray}
Using Eqs. (\ref{pt}), (\ref{ph}), (\ref{pr}), (\ref{dotph}) and (\ref{dotr}) in Eq. (\ref{U2_1}) leads to:
\begin{eqnarray}
\label{U2_2}
&&U^{2}=-f(r)^{2}\left(\frac{2r-s^{2}f(r)'}{2er-jsf(r)'}\right)^{2}
\left(1-\Sigma\right)\;,\\
&&\Sigma = \frac{(2(j-es)s)^{2}(\eta_{-})\Big(4r-s^{2}(\eta_{+})\Big)}{(2r-s^{2}f(r)')^{4}}\;,\label{Sigma}
\end{eqnarray}
where $\eta_{\pm}=f(r)'\pm r f(r)''$.

%%%%%%%%%%%%%%%%%%%%%%%%%%%%%%%%%%%%%%%%%%%%%%%%%%%%%%%%%%%
%%%%%%%%%%%%%FIGURE%%%%%%%%%%%%%%
\begin{figure*}
\begin{tabular}{c c c}
\hspace{0cm}
\includegraphics[scale=0.46]{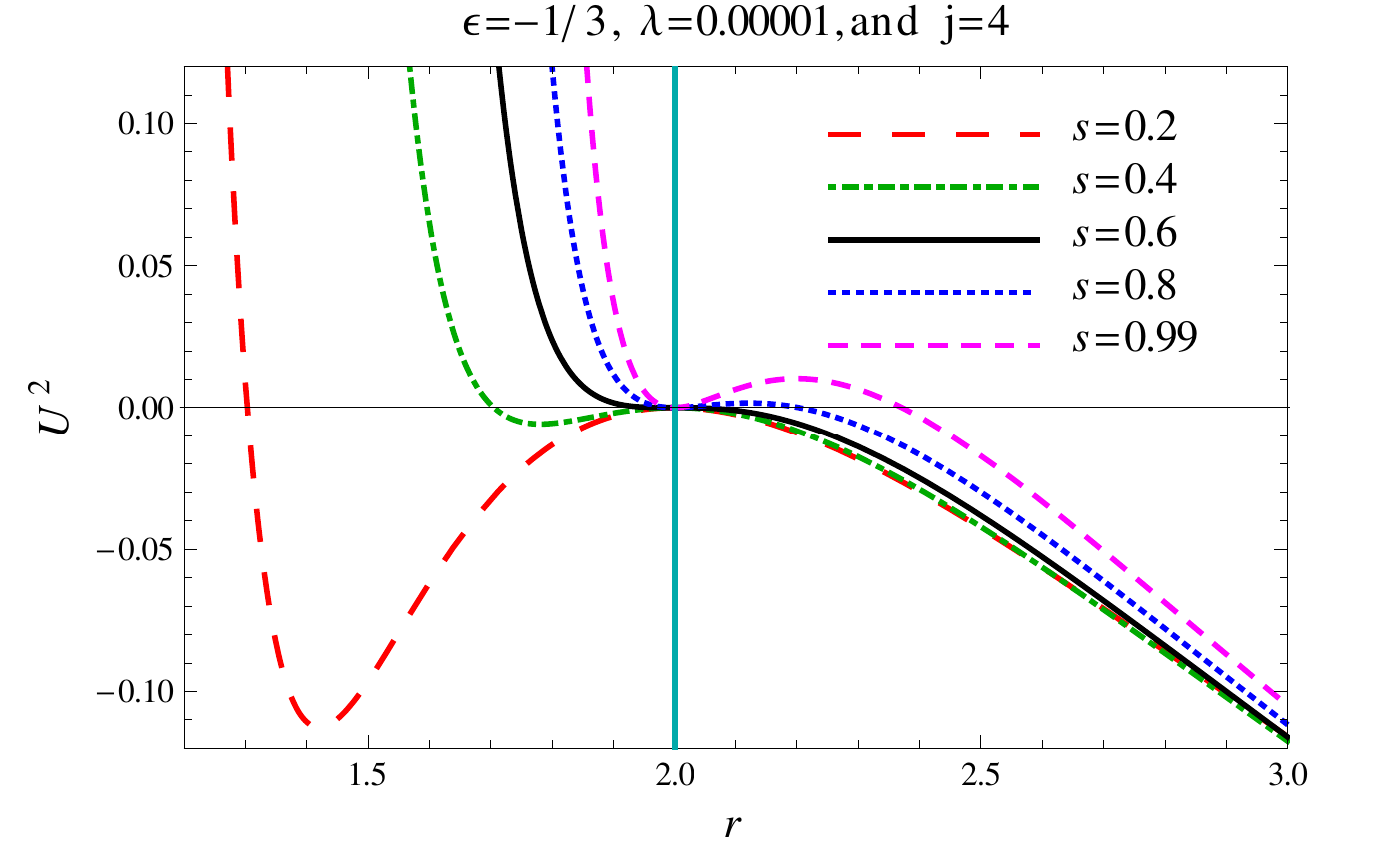}\hspace{-0.7cm}
&\includegraphics[scale=0.46]{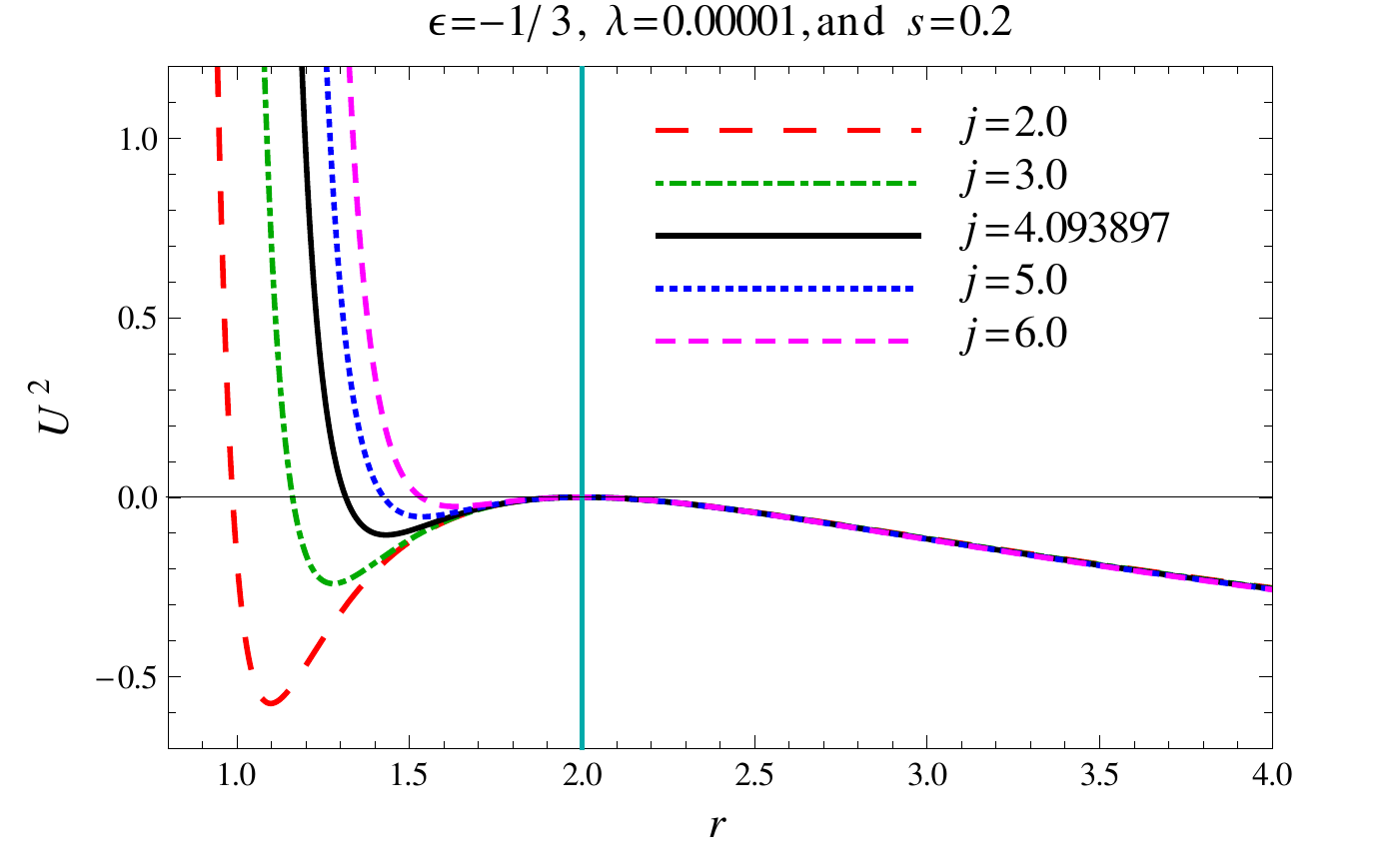}\hspace{-0.7cm}
&\includegraphics[scale=0.46]{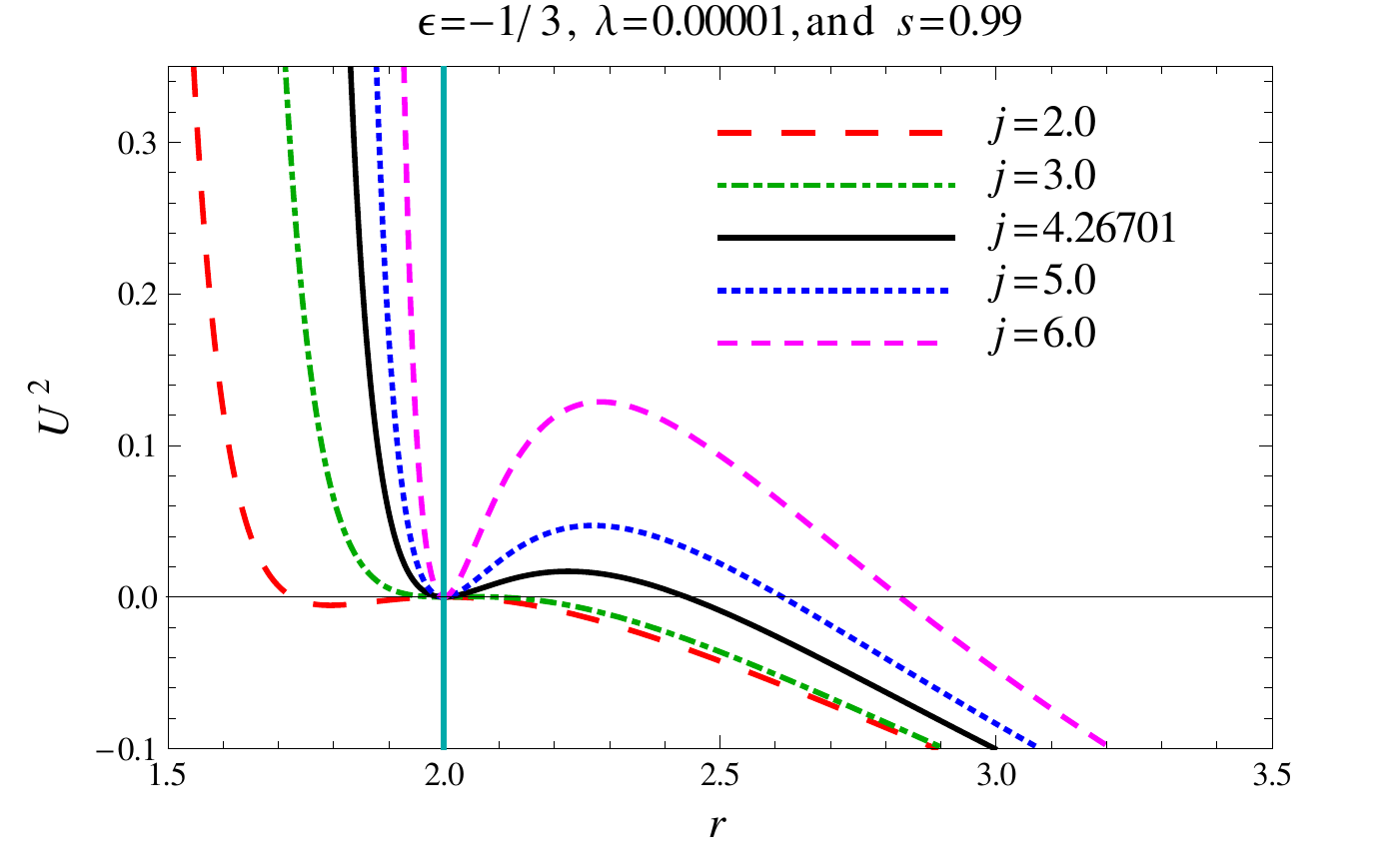}\\
\includegraphics[scale=0.46]{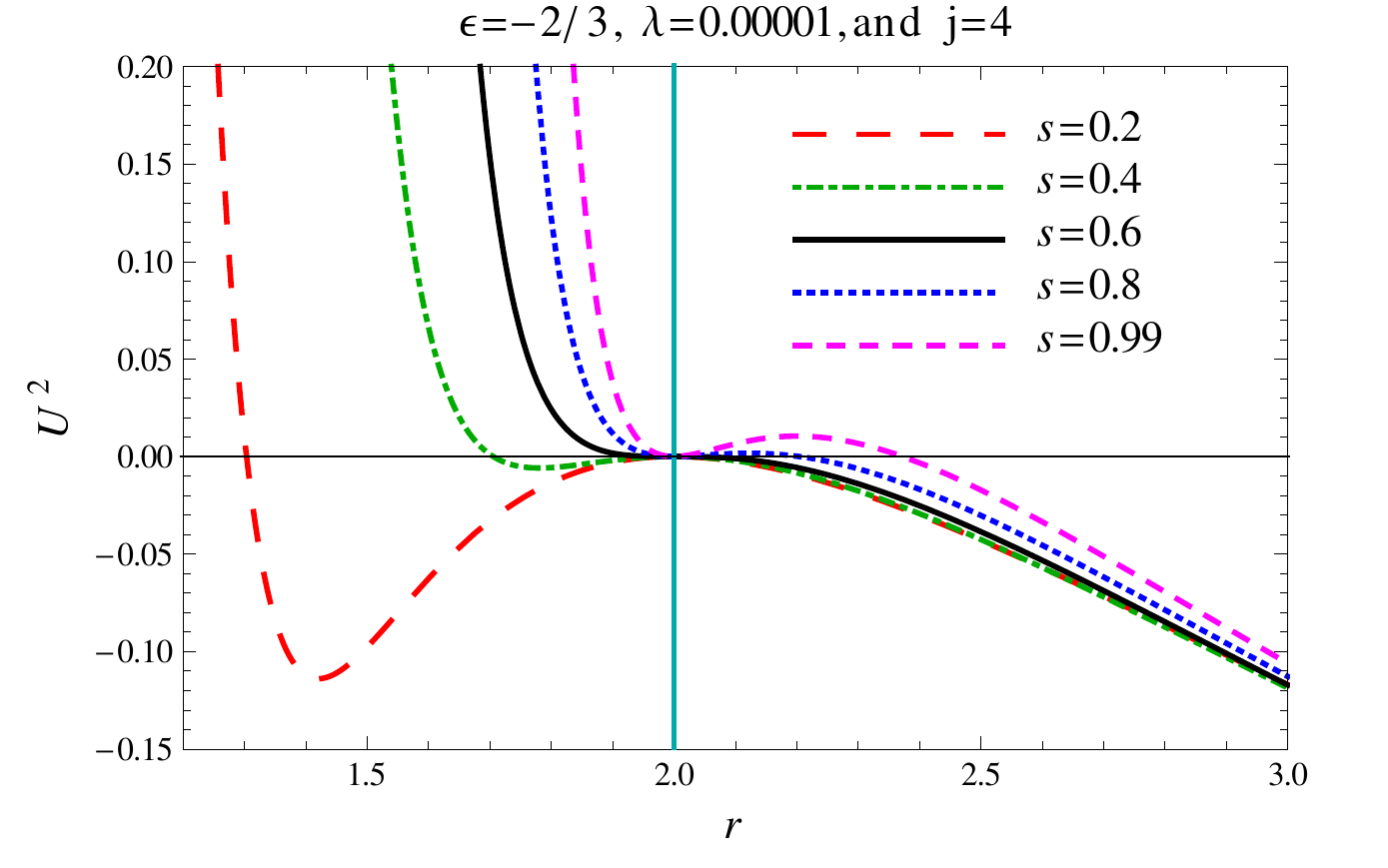}\hspace{-0.7cm}
&\includegraphics[scale=0.46]{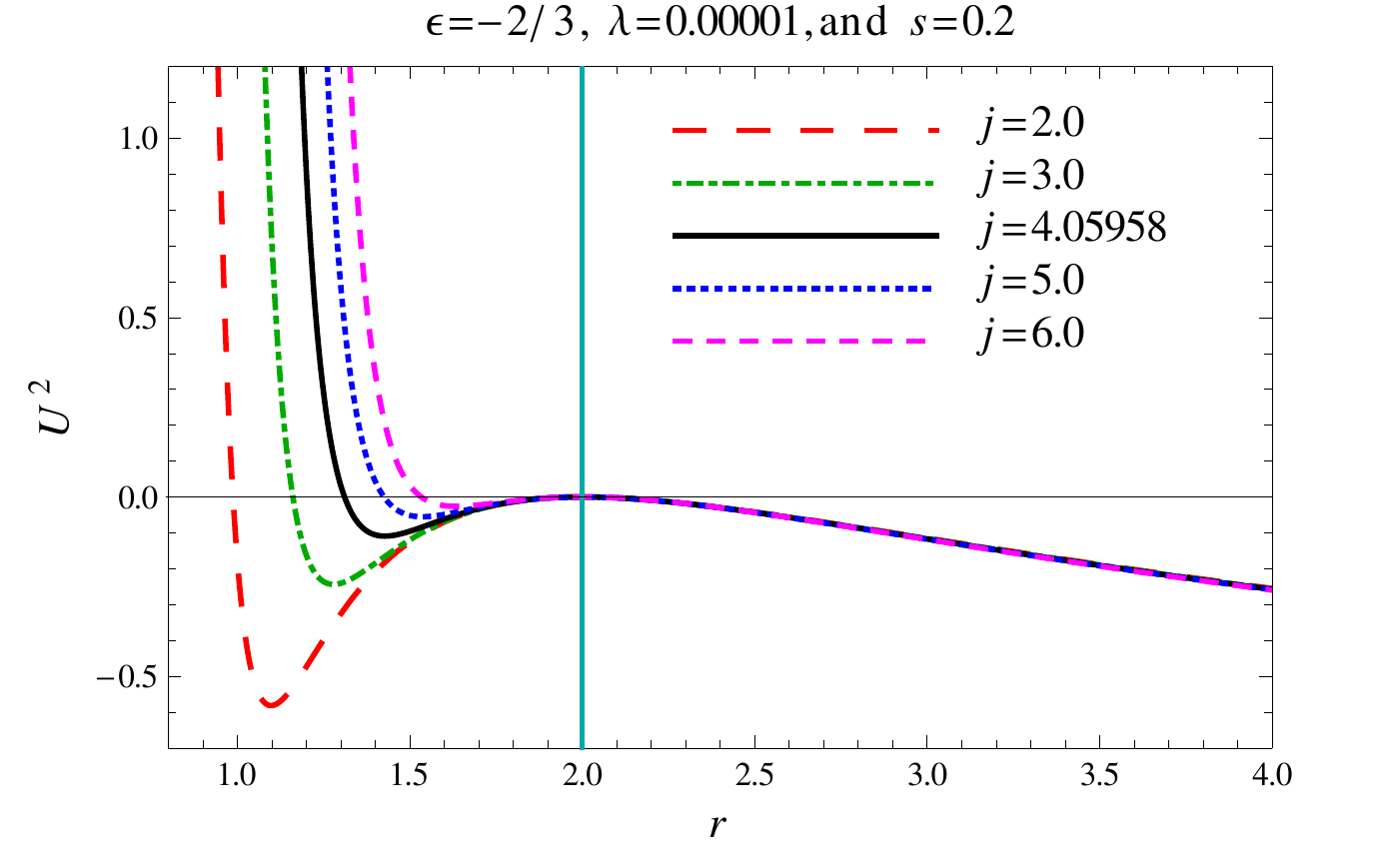}\hspace{-0.7cm}
&\includegraphics[scale=0.46]{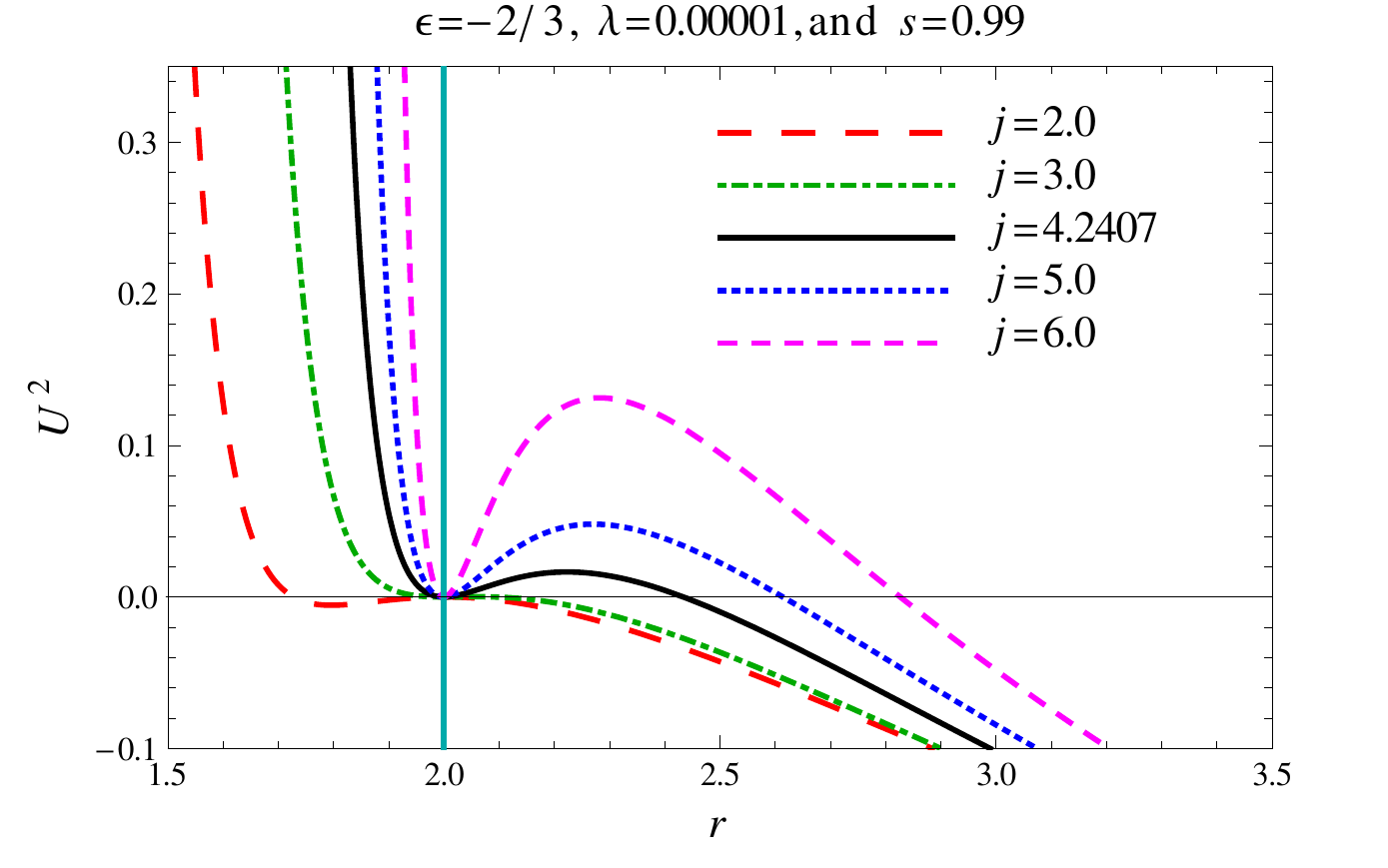}\\
\includegraphics[scale=0.46]{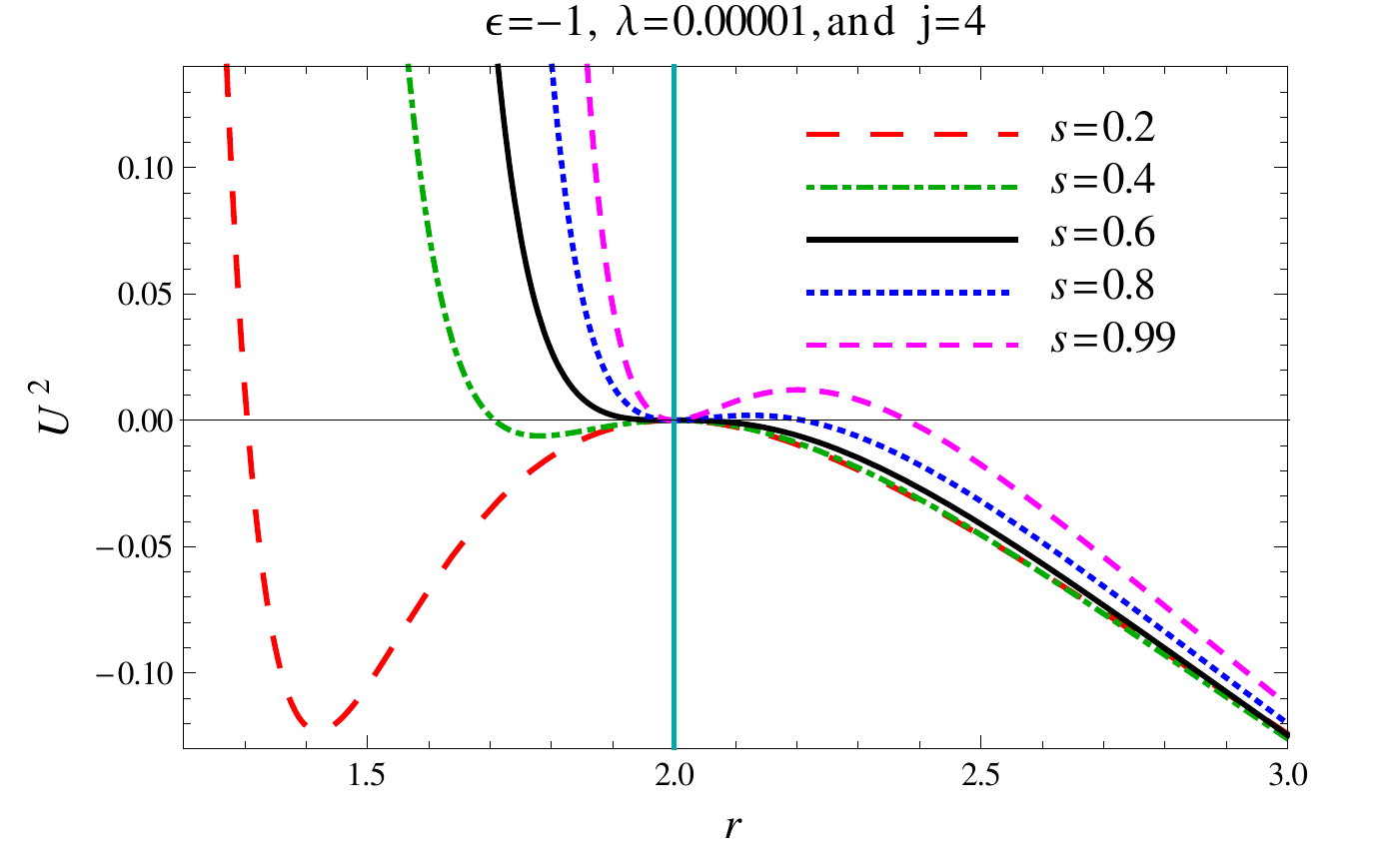}\hspace{-0.7cm}
&\includegraphics[scale=0.46]{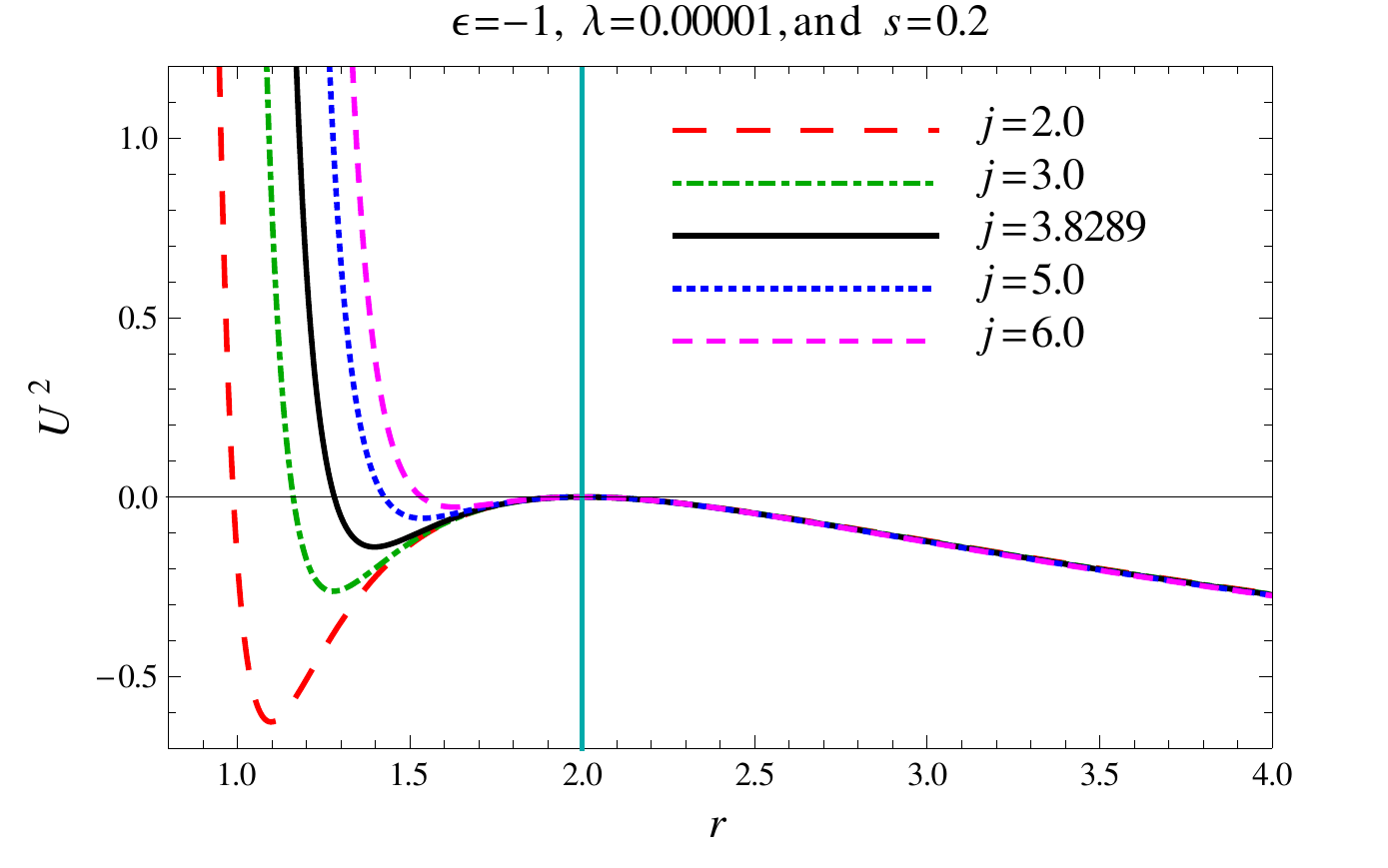}\hspace{-0.7cm}
&\includegraphics[scale=0.46]{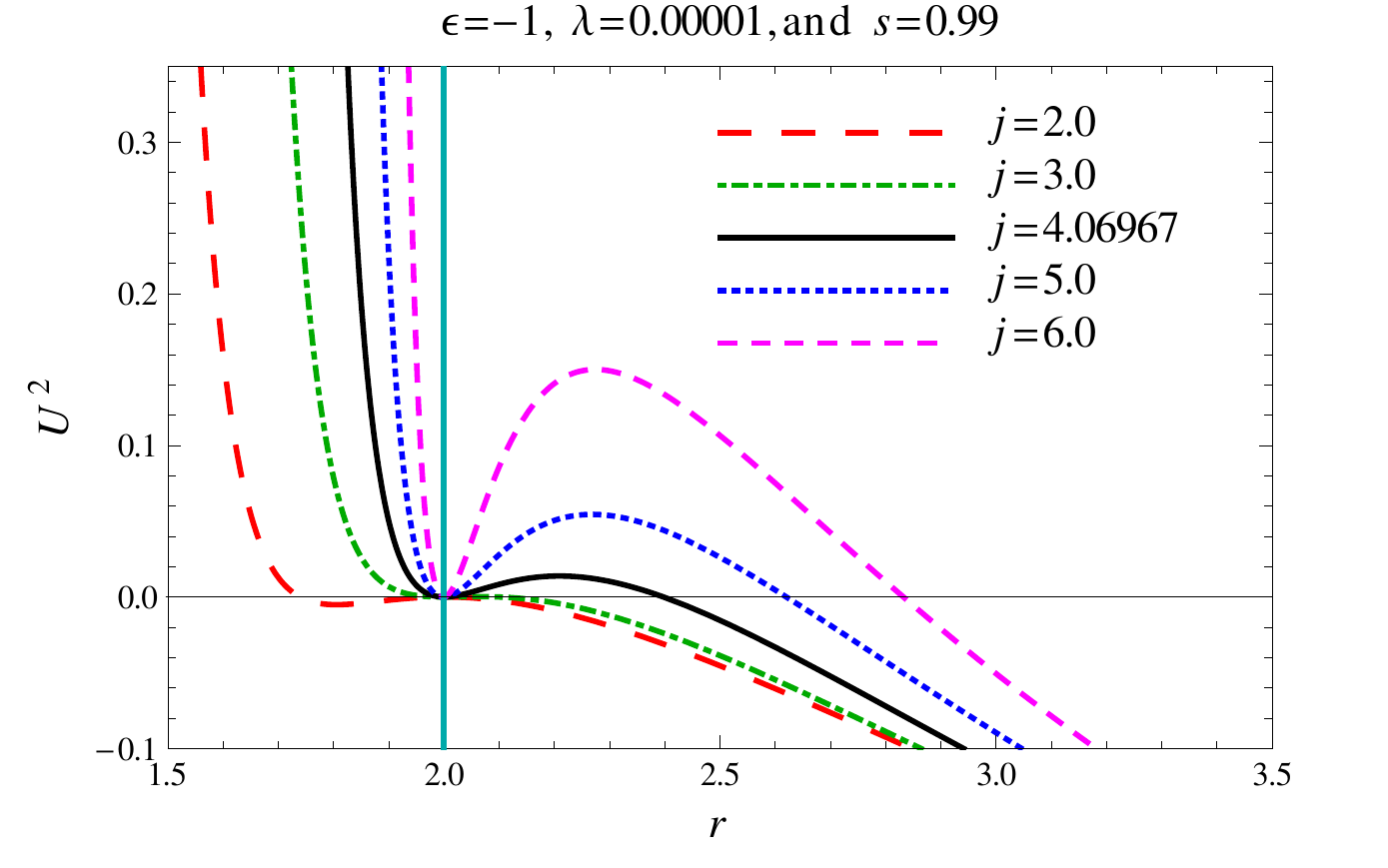}
\end{tabular}
 \caption{Variation of $U^{2}$ as a function of $r$ with different values of $\epsilon, s$ and $j$. The vertical (darker cyan) solid line is the horizon $r_{0}$ ($M=1$). Here, for the corresponding value of parameter $\epsilon$ $(-1/3, -2/3\; \text{and} -1)$, the value of particle energy per unit mass $e$ is $0.9999, 0.995518$ and $0.967144$. (Color-online).}
\label{fig5}
\end{figure*}
%%%%%%%%%%%%%%%%%%%%%%%%%%%%%%%%%%%%%%%%%%%%%%%%%%%%%%%%%%%

The $E_{\rm CM}$ calculated in Eq.~\eqref{ecmf} diverges when either $C_{1}=0$ or $C_{2}=0$ as mentioned in section \ref{sec:CM}. We already showed there that the point where $C_i=0$ always lies behind the horizon and, therefore, is not of importance for our analysis. We note here that, in addition, the condition $C_i=0$ leads to a transition of $U^{2}$ (i.e.~Eq.~\eqref{U2_2}) of the colliding spinning particle from the subluminal region (physical) to the superluminal region (unphysical) as seen in the Fig \ref{fig5}.
We have also concluded earlier from Eq.~\eqref{limitECM}
that the center of mass energy remains finite when the collision takes place at the event horizon.
Hence, in this work we are more interested in finding location outside the event horizon where the square of the four-velocity lies in subluminal region. This leads to the condition $\Sigma<1$ according to Eq.~\eqref{U2_2}.

We may rewrite the expression for $\Sigma$ in Eq.~\eqref{Sigma} as
\begin{align}
\Sigma & = 4 \mathcal{L}^2 (G-1)\,, \\
G & := \left( \frac{2r-rs^2f(r)''}{2r-s^2f(r)'}\right)^2\,.
\end{align}
We first notice that both $G$ and $\mathcal{L}^2$ are monotonically decreasing functions, and that $\Sigma>0$, for $\epsilon=-\frac{1}{3}$ and $\epsilon=-1$. For $\epsilon=-\frac{2}{3}$ this only holds in the vicinity of the horizon. We therefore find that $\Sigma<1$ holds in the vicinity of the horizon $r_0$ if

\begin{equation}\label{cond_Sigma}
\Sigma_{r_{0}}<1.
\end{equation}
In order to have an arbitrarily high collisional $E_{\rm CM}$ outside the event horizon, one of the colliding particles must be the usual particle (i.e. a particle for which $\mathcal{K}_{r_{0}}\neq 0$) and the other must be a near-critical one as shown in the previous section.

For the usual particle with $s=0$, the condition \eqref{cond_Sigma} is satisfied automatically. If $s \neq 0$ we can always choose $(j-es)^{2}$ such that \eqref{cond_Sigma} holds, for instance, one could choose $j=es$. For near-critical particles, we fixed an energy $e$ in \eqref{e_nearcrit}. To achieve the inequality \eqref{cond_Sigma}, we could then for instance choose $j=es$ again. For near-critical particles we can also explicitly solve $\Sigma=1$ for $j$, using the energy $e$ from Eq.~\eqref{e_nearcrit}. We find
\begin{align}
j & = a s \sqrt{r_c-r_0} \pm \frac{r_0}{s}N_\epsilon
\end{align}
with
\begin{align}
N_{-1} & = \frac{Ms^2 -r_0^3(\lambda s^2 -1)}{\sqrt{3M(M s^2 + 2r_0^3(\lambda s^2 +1))}} \,,\\
N_{-1/3} & = \frac{Ms^2-r_0^3}{\sqrt{3M(Ms^2+2r_0^3)}}\,,\\
N_{-2/3} & = \frac{2Ms^2 - r_0^2(\lambda s^2 -2r_0)}{\sqrt{(6M-\lambda r_0^2)(2Ms^2 + r_0^2(\lambda s^2 +4r_0))}}\,.
\end{align}

Hence, we may conclude that the collision of a near-critical particle with a usual particle can produce arbitrarily high center-of-mass energy $E_{\rm CM}$ if we fine tune the parameters. For instance, we could choose a usual particle starting from rest from infinity or the static radius, respectively, with vanishing total angular momentum $j=es$, and a near-critical particle with energy as given in \eqref{e_nearcrit} and vanishing total angular momentum starting from a radius close to the event horizon (but outside of it).

%%%%%%%%%%%%%%%%%%%%%%%%%%%%%%%%%%%%%%%%%%%%%%%%%%%
\section{Summary, Conclusion and Future Prospects }\label{sec:final_remarks}
%%%%%%%%%%%%%%%%%%%%%%%%%%%%%%%%%%%%%%%%%%%%%%%%%%%

We discussed the collision of spinning particles close to the event horizon of a Schwarzschild black hole surrounded by quintessential matter. In particular, we found that the center of mass energy may grow without bound under certain conditions.

After reviewing the equations of motion for spinning particles (under the Tulczyjew spin supplementary condition), we started with an analysis of the horizon structure of the spacetime under discussion. For the equation of state parameter we assumed in this work three different values, $\epsilon=-1/3$, $\epsilon=-2/3$ and $\epsilon=-1$. In addition to the event horizon, for $\epsilon=-2/3$ and $\epsilon=-1/3$ also a cosmological horizon may be present. In the following, we restricted to values of the normalisation constant $\lambda$ which allow for a black hole solution.

We then focused on the center of mass energy of two spinning particles colliding in the vicinity of the event horizon. For generic particles, we showed that the center of mass energy remains finite in the limit that the collision takes place at the horizon. Moreover, potential additional points of divergence given by $C_{1,2}=0$, see Eq. \eqref{ecmf}, are shown to always being located behind the event horizon, if we respect the M{\o}ller limit on the spin of the particles. Therefore, these points are not of further interest.

In order to determine if fine-tuned particles, that might produce unbound center of mass energies, can reach the vicinity of the event horizon, we proceeded with a discussion of the effective potential and radial turning points. The behavior of $P^{r}$ as a function of $r$, plotted in Fig. \ref{fig3}, is used to analyze and distinguish between different trajectories of the spinning particles. From this figure, it seems that the maximum allowed range of total angular momentum $j$, for which a spinning particle reaches the event horizon of the SBHQ, increases with increase in the value of particle's spin $s$. However, it is also observed that the radius at which $P^{r}$ becomes zero, decreases with increase in $s$.

We then identify three different types of spinning particle trajectories dependent on their behavior close to the event horizon: usual, critical, and near-critical particles. From studying collisions between all different combinations of these trajectory types, we concluded that only the collision between a near-critical and a usual particle may produce arbitrarily high center of mass energies.

As for spinning particles with Tulczyjew spin supplementary condition $u_{\alpha}u^{\alpha}$ is not conserved, where $u^{\alpha}$ is the four velocity, the motion may change from subluminal to superluminal. For the combination of interest, namely usual and near-critical particle, we showed that we can always choose the angular momentum of the particles such that the motion is subluminal in the vicinity of the horizon.

Collisions of spinning particles around the Schwarzschild black hole (without surrounding quintessential matter) was studied in \cite{Armaza:2015eha}. In that work, mainly the points where $C_{1,2}=0$, see Eq. \eqref{ecmf}, were analysed. However, these points are always behind the horizon if we take the M{\o}ller limit on the spin into account. The divergence of the center of mass energy discussed in the present work is of a different type. A similar setup for geodesics around the Schwarzschild black hole was shortly discussed in \cite{Grib:2012iq} and around general static and spherically symmetric spacetimes, which includes the present case of SBHQ, was studied recently by some of us in \cite{Hackmann2020}. The physical relevance of these setups, meaning the question if the near-critical particle can be created by a foregoing collision of particles, was studied for the Schwarzschild and extremal Reissner-Nordstr\"om black holes in \cite{Zaslavskii2020}. It is an open question if the near-critical particle becomes physically more relevant if we add spin to the particles or quintessential matter to the spacetime. Another obvious direction for further research is to include the rotation of the black hole.

%%%%%%%%%%%%%%%%%%%%%%%%%%%%%%%%%%%%%%%%%%%%%%%%%%%%%%%
\begin{acknowledgments}
%%%%%%%%%%%%%%%%%%%%%%%%%%%%%%
The authors would like to thank the anonymous referee for the constructive comments and suggestions which helped us to improve the presentation of this paper. HN is thankful to Prof. Philippe Jetzer for invaluable insights and suggestions during the early stage of this work. PS would like to thank \textit{Programa de Desarrollo Profesional Docente} (PRODEP) of the \textit{Secretar\'{\i}a de Educac\'{\i}on P\'{u}blica} (SEP) of the Mexican government, for providing the financial support. HN would like to thank Science and Engineering Research Board (SERB), New Delhi, India for financial support through grant no. EMR/2017/000339. He also thankful to IUCAA, Pune, India (where a part of the work was completed) for support in form of academic visits under its Associateship programme.EH is grateful for support from the research training group RTG 1620 "Models of Gravity" and the center of excellence EXC 2123 "QuantumFrontiers", both funded by the German Research Foundation (DFG). UN acknowledges support from PRODEP-SEP, SNI-CONACYT and CIC-UMSNH. AA acknowledges that this work is based on the research supported in part by the
National Research Foundation (NRF) of South Africa (grant numbers 109257 and 112131). He also acknowledges the hospitality of the High Energy and Astroparticle Physics Group of the Department of Physics of Sultan Qaboos University, where part of this work was completed.
\end{acknowledgments}
%%%%%%%%%%%%%FIGURE%%%%%%%%%%%%%%
\begin{figure*}
\begin{tabular}{c c}
\hspace{0cm}
\includegraphics[scale=0.65]{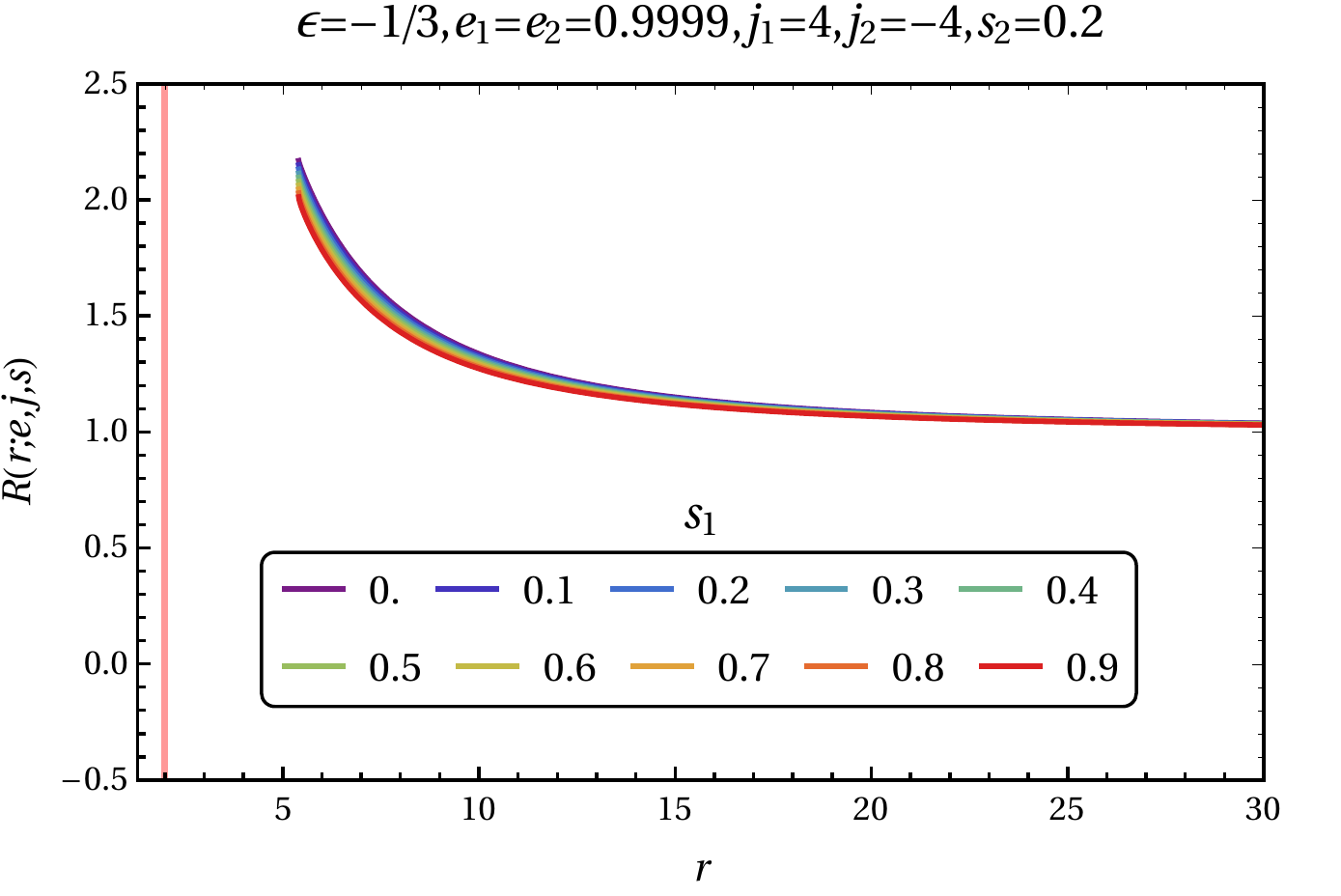}\hspace{-0.5cm}
&\includegraphics[scale=0.65]{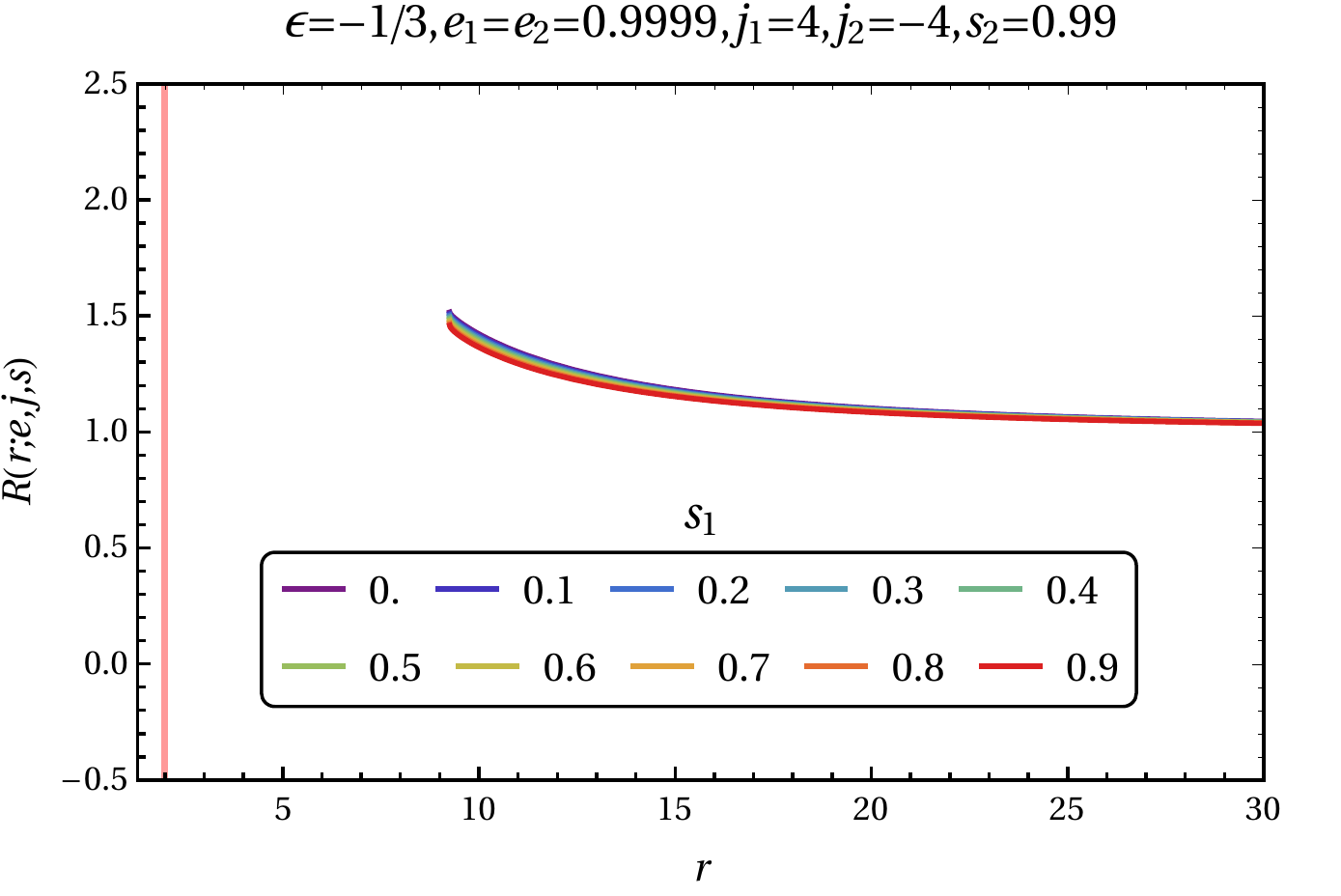}\\
\includegraphics[scale=0.65]{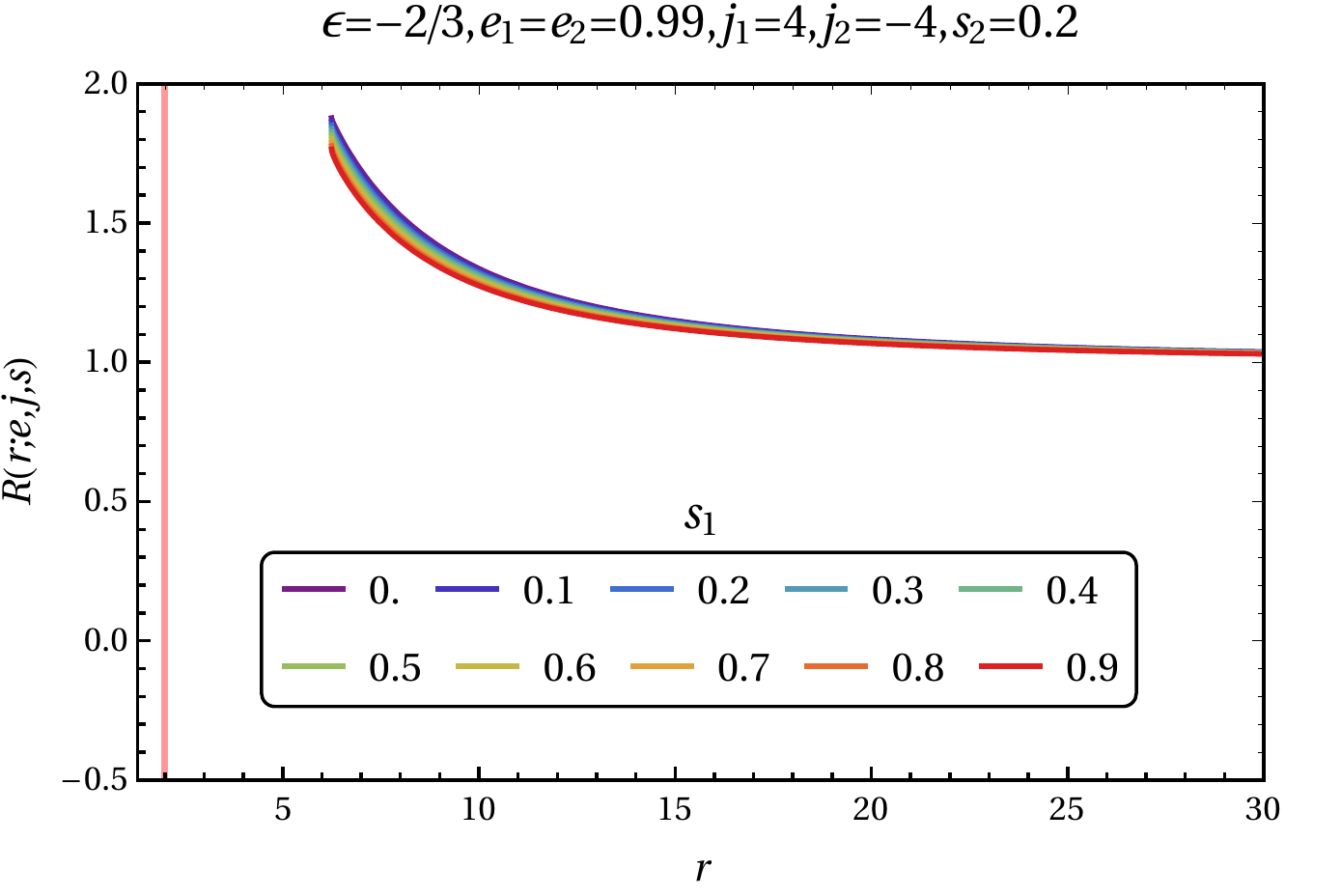}\hspace{-0.5cm}
&\includegraphics[scale=0.65]{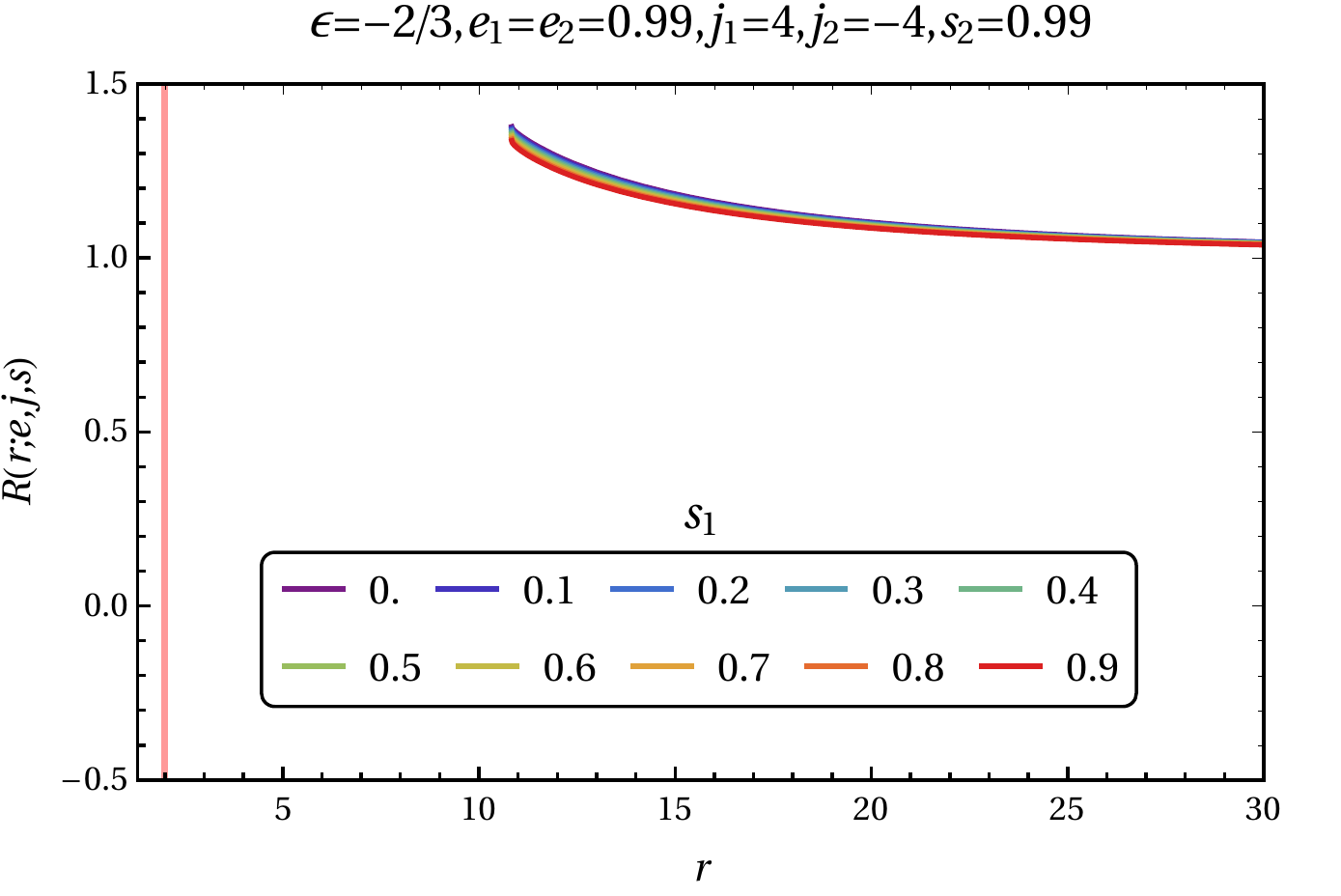}\\
\includegraphics[scale=0.65]{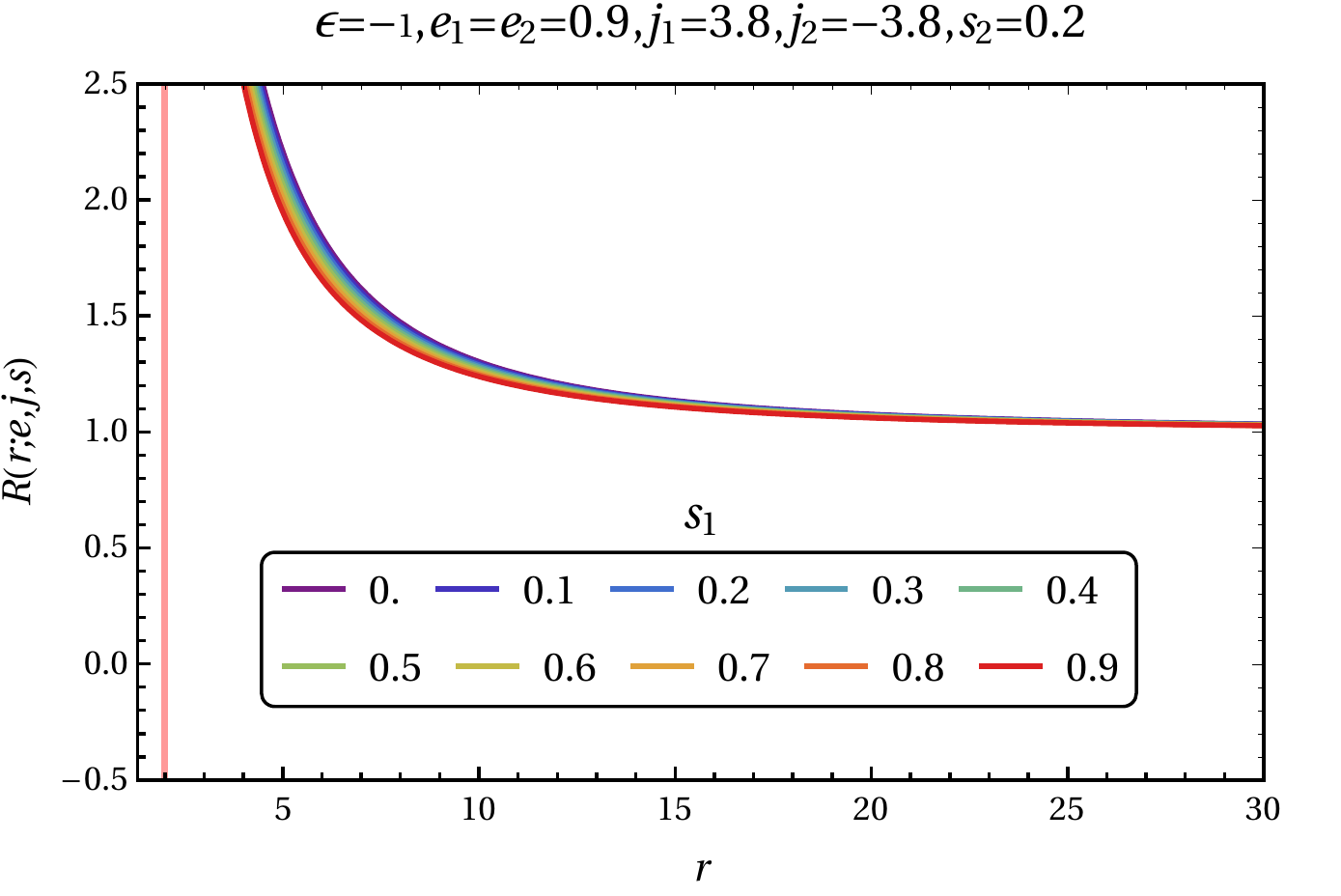}\hspace{-0.5cm}
&\includegraphics[scale=0.65]{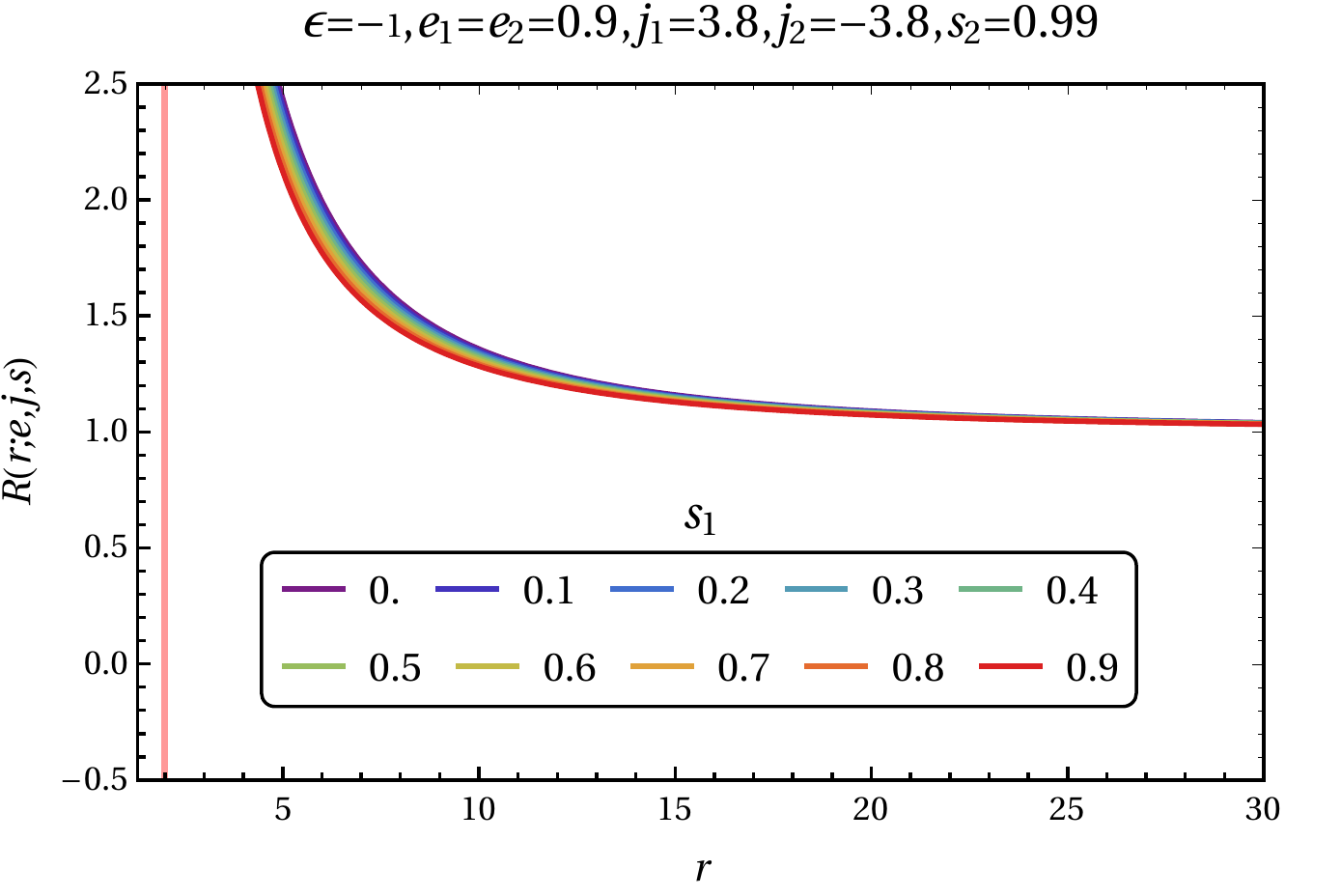}\\
\end{tabular}
 \caption{Variation of $R(r;e,j,s)$ as a function of $r$ with different values of $\epsilon, s$ and $j$. Here, the vertical (red) solid line is the horizon $r_{0}$ ($M=1$) and $\lambda=0.00001$. (Color-online).}
\label{fig6}
\end{figure*}
%%%%%%%%%%%%%%%%%%%%%%%%%%%%%%%%%%%%%%%%%%%%%%%%%%%%%%
\section{Appendix}
For two spinning particles with mass $m_{1}$ and $m_{2}$, the center of mass energy of the collision of both can be written as
\begin{equation}
 \label{A1}
 E_{\rm CM}^{2}= m_{1}^{2} + m_{2}^{2} + 2\left(f(r) P_{1}^{t}P_{2}^{t}-\frac{1}{f(r)}P_{1}^{r}P_{2}^{r}-r^{2}P_{1}^{\phi}P_{2}^{\phi}\right)\;,
\end{equation}
which, after substituting the values of $P^{t}$, $P^{r}$ and $P^{\phi}$ from equations (\ref{pt}), (\ref{ph}) and (\ref{pr}) for every particle respectively, reduces to
\begin{eqnarray}
 \label{A2}
 E_{\rm CM}^{2}= m_{1}^{2} + m_{2}^{2} + 2 m_{1}m_{2} R(r; e, j, s) \;,
\end{eqnarray}
where the function $R(r; e, j, s)$ is given as
\begin{eqnarray}
 \label{A3}
&&R(r; e, j, s)\equiv f(r) \left[\frac{r^{3\epsilon+1}}{\Delta_{0}}\right]^2 \mathcal{K}_{1}\mathcal{K}_{2}  -4\mathcal{L}_1\mathcal{L}_2 \nonumber \\ &&
-\frac{1}{f(r)} \sqrt{\mathcal{K}_{1}^{2}-f(r)(1+4\mathcal{L}_{1}^{2})}
\sqrt{\mathcal{K}_{2}^{2}-f(r)(1+4\mathcal{L}_{2}^{2})} \nonumber \;.\\
&& 
\end{eqnarray}
In order to obtain the maximum of the $E_{CM}$ as given by Eq.(\ref{A2}), we propose the following:  
%for sake of completeness, we will probe the statement: 
\begin{itemize}
\item \textbf{Statement}: 
Assuming the constraint $\mu \equiv m_{1}+m_{2}=$ constant with the fixed parameters $(e, j, s)$, the the maximum of the center of mass energy given by (\ref{A2}) can obtained when both masses are equal $m_{1}=m_{2}$ and $R(r; e, j, s)>1$.
\end{itemize}
%%%%%%%%%%%%%%%%%%%%%FIGURE%%%%%%%%%%%%%%%%%%%%%%%%%%%%%
\begin{figure*}[!t]
\begin{tabular}{c c}
\hspace{0cm}
\includegraphics[scale=0.65]{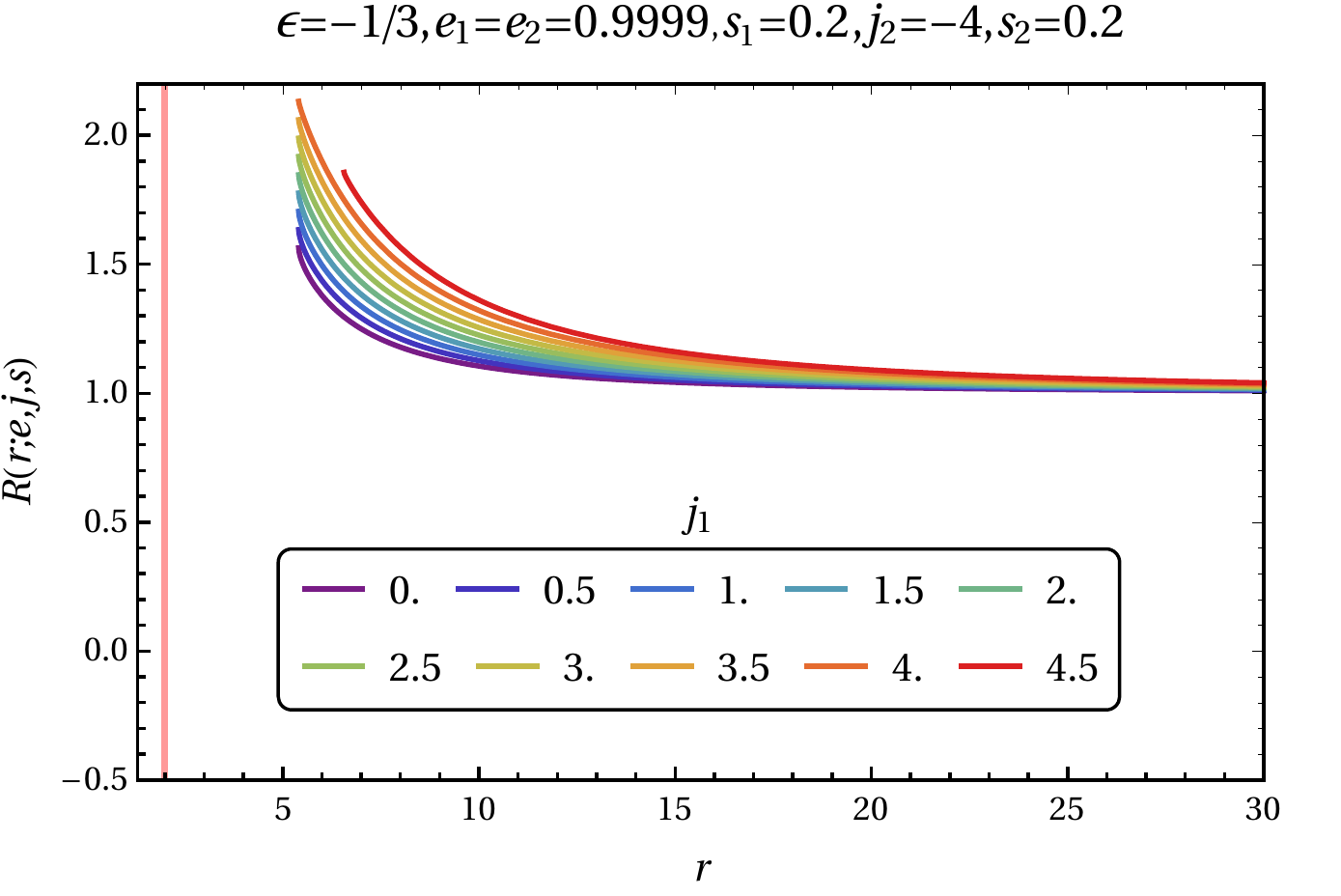}\hspace{-0.5cm}
&\includegraphics[scale=0.65]{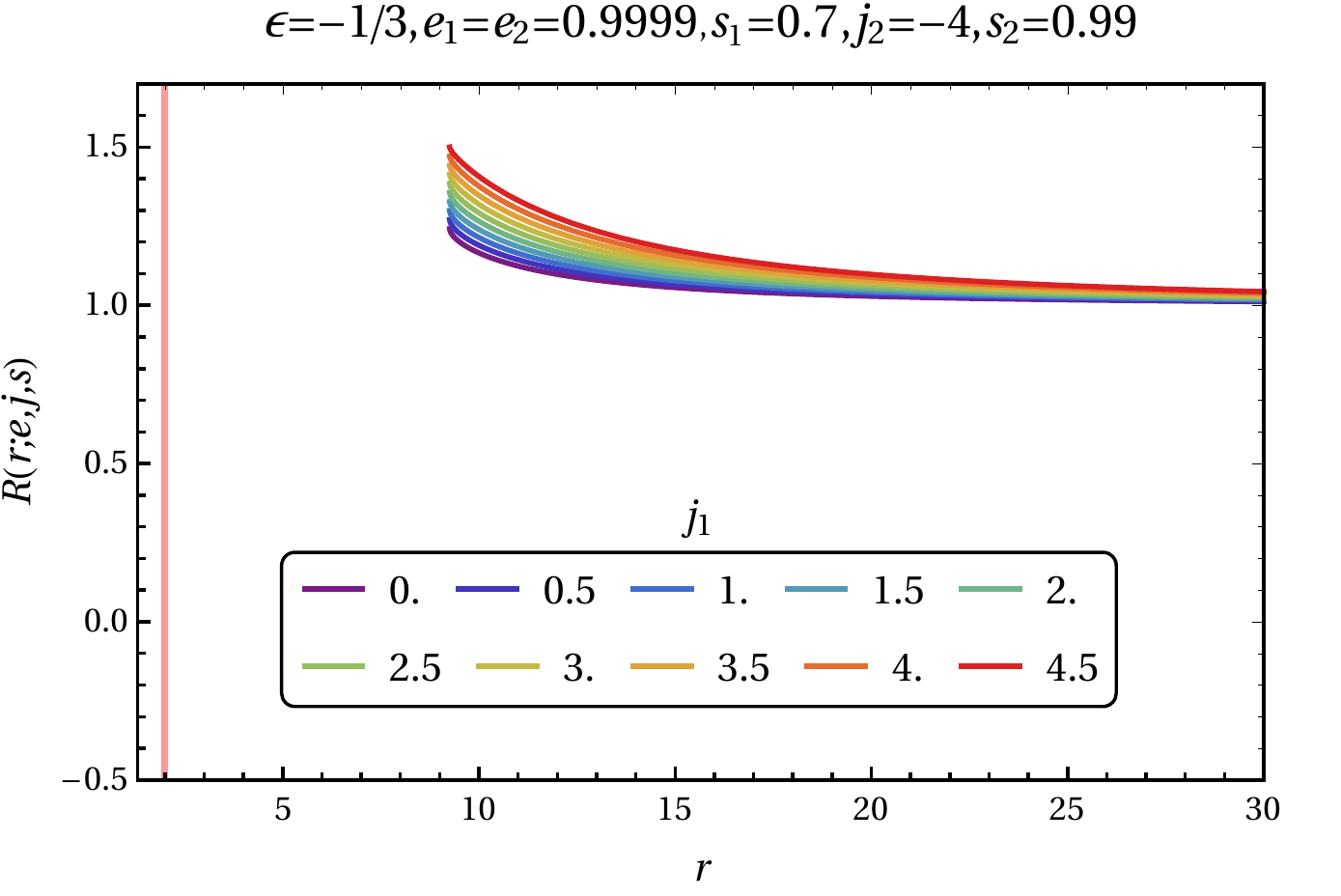}\\
\includegraphics[scale=0.65]{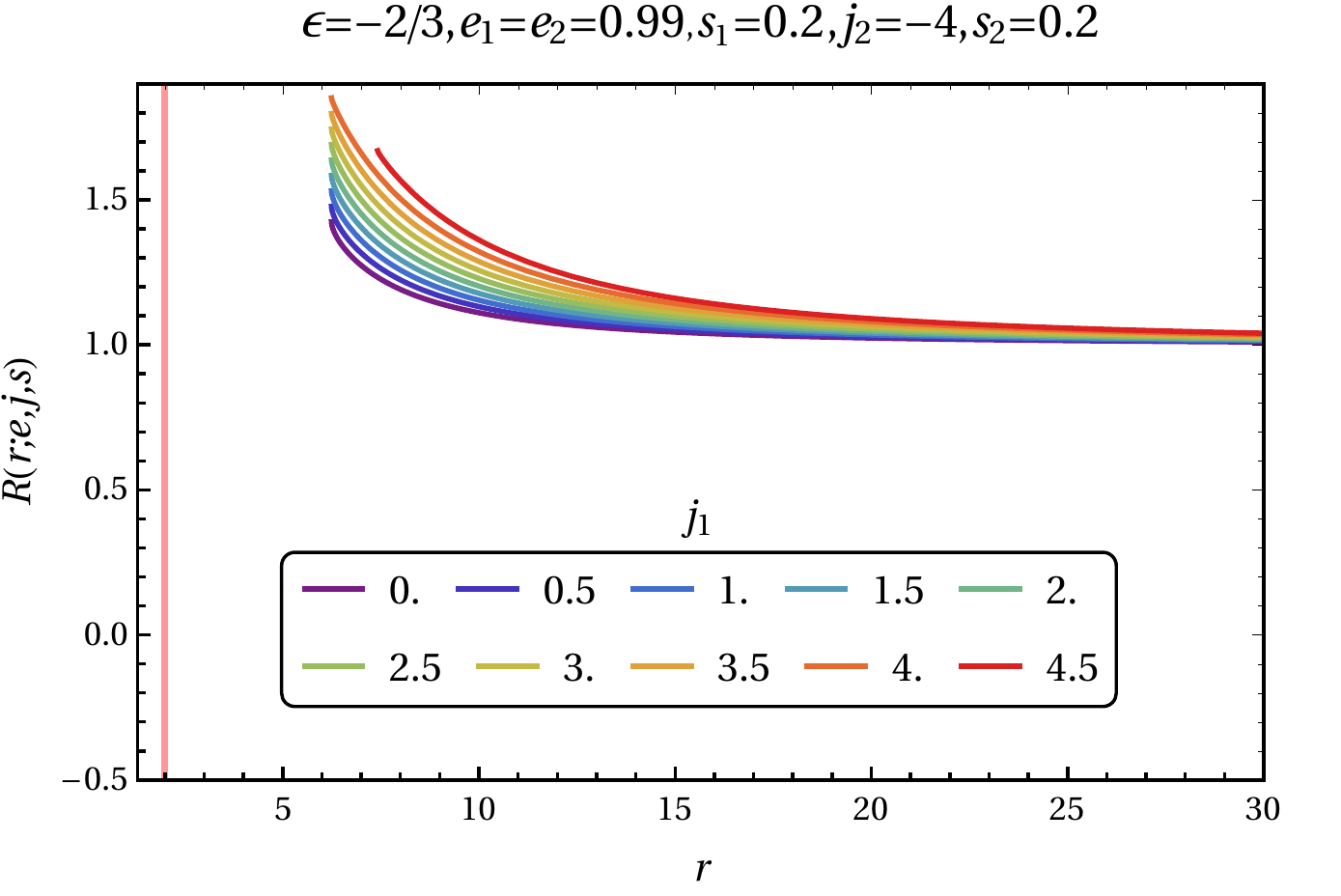}\hspace{-0.5cm}
&\includegraphics[scale=0.65]{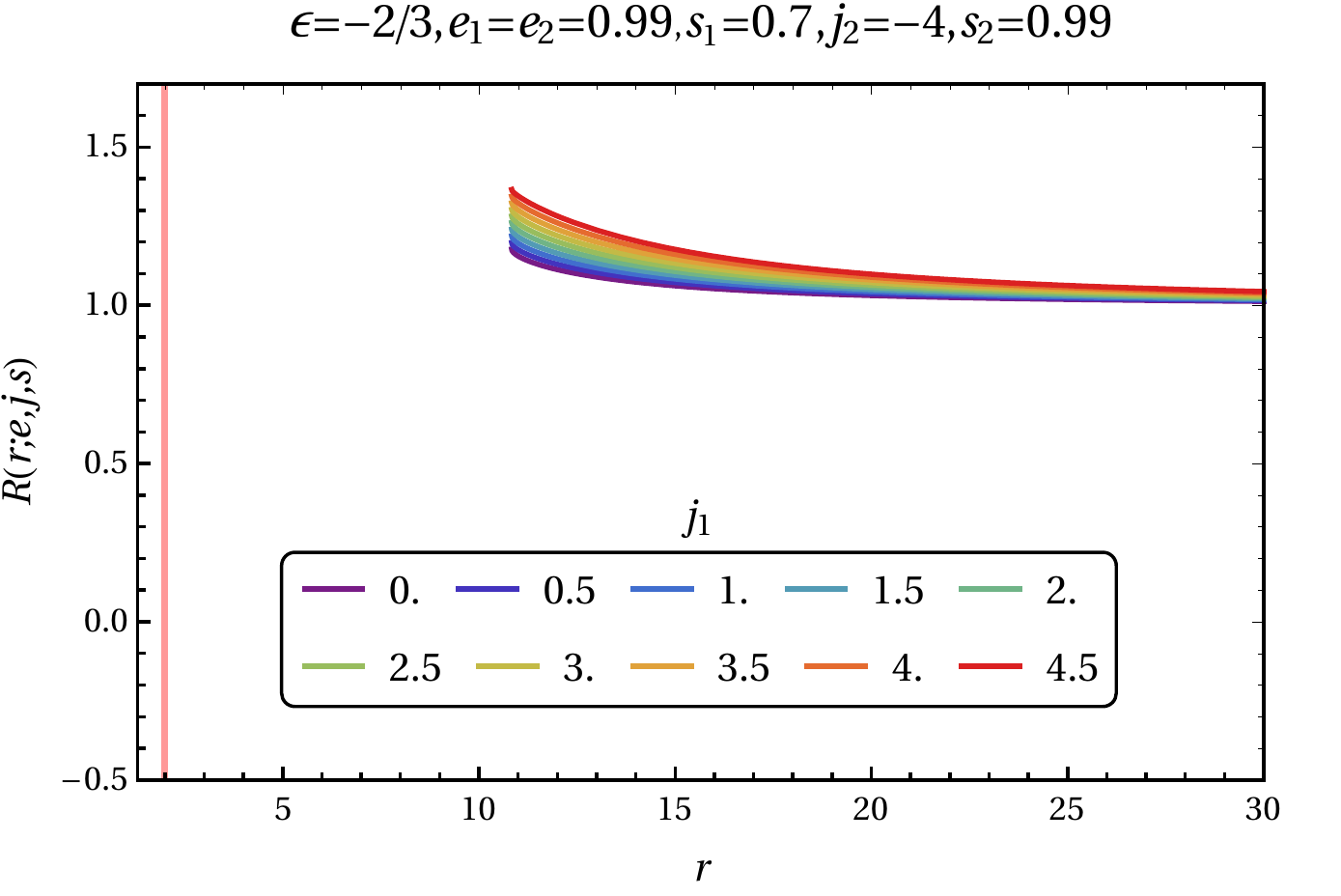}\\
\includegraphics[scale=0.65]{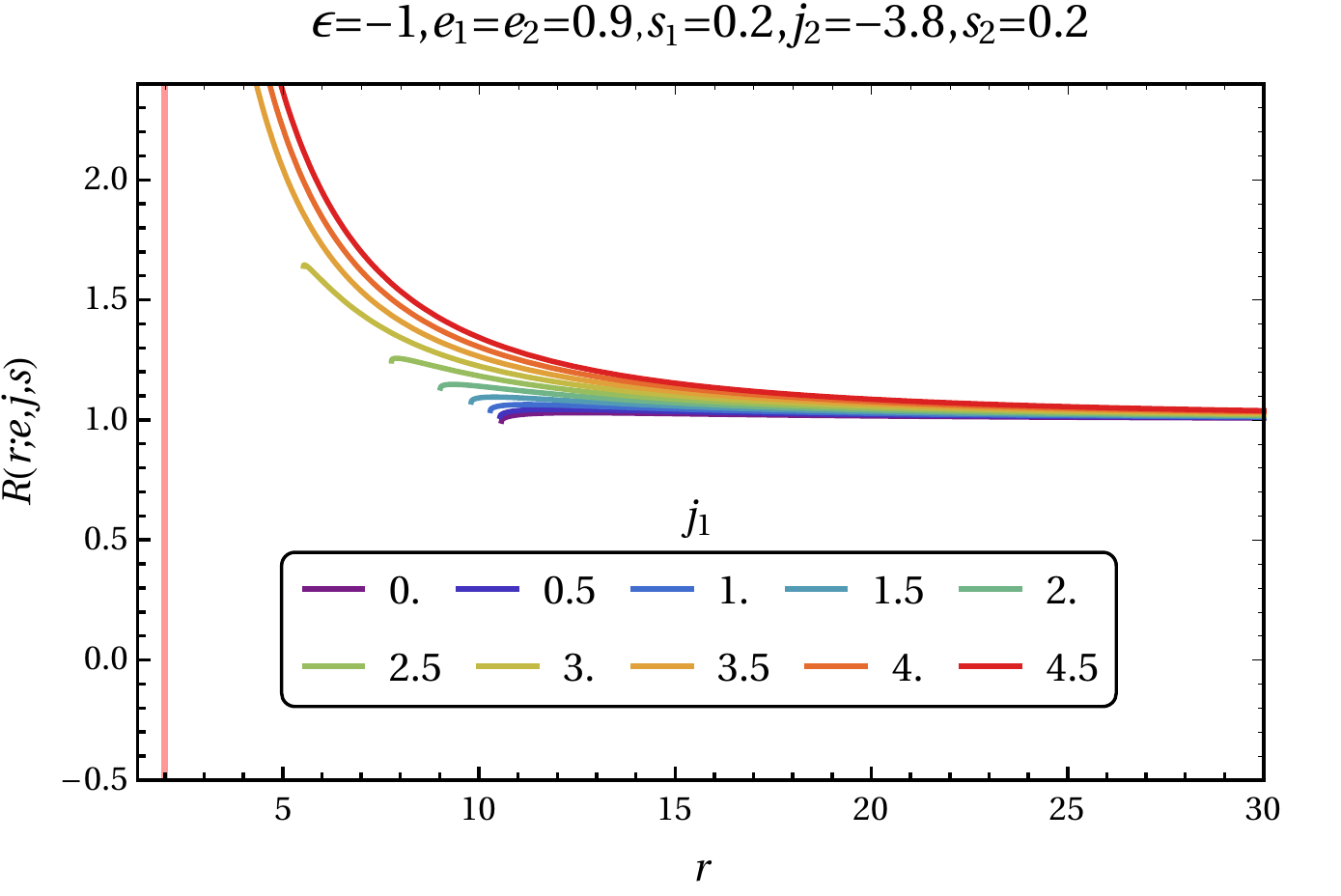}\hspace{-0.5cm}
&\includegraphics[scale=0.65]{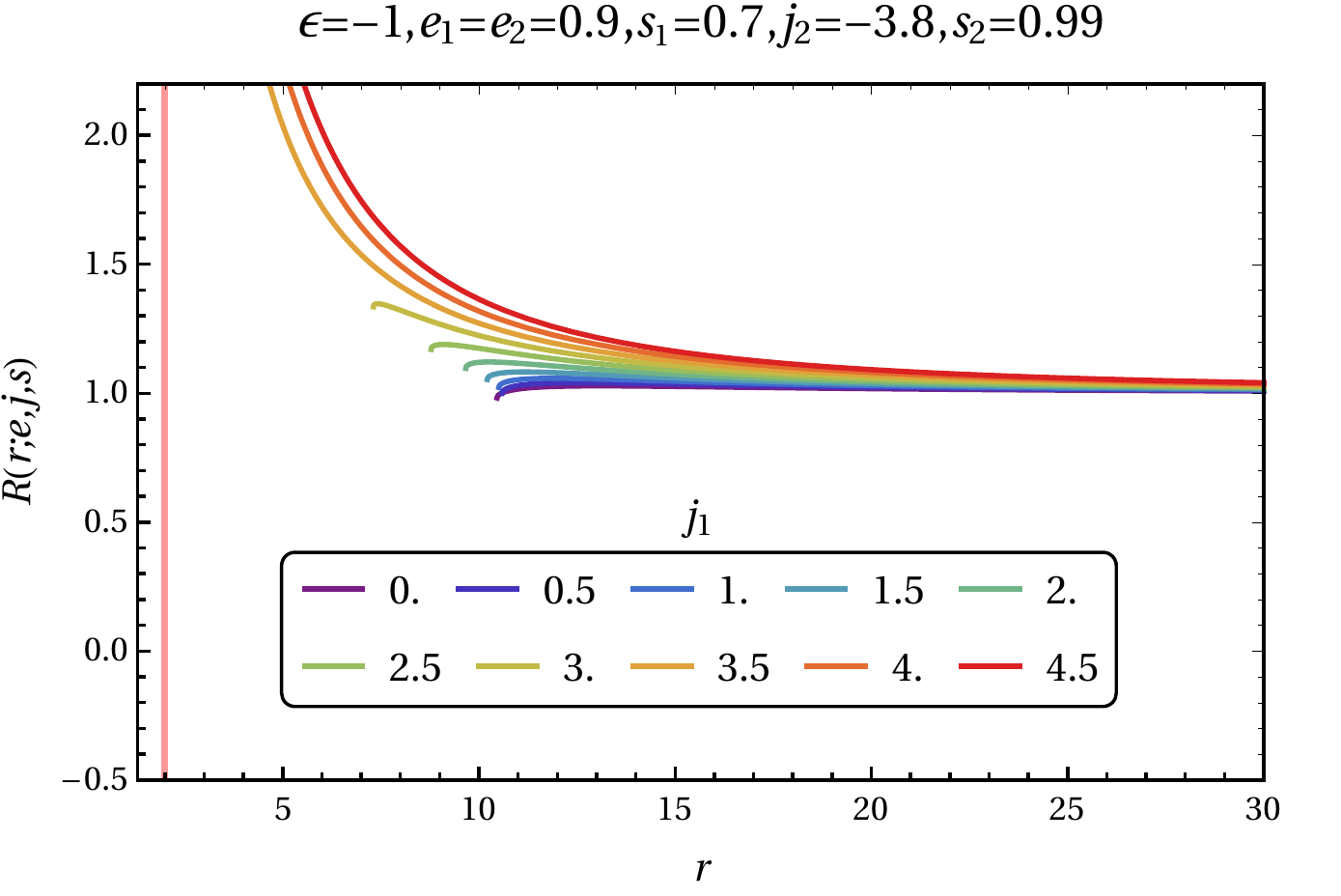}\\
\end{tabular}
 \caption{Variation of $R(r;e,j,s)$ as a function of $r$ with different values of $\epsilon, s$ and $j$. Here, the vertical (red) solid line is the horizon $r_{0}$ ($M=1$) and $\lambda=0.00001$. (Color-online).}
\label{fig7}
\end{figure*}
%%%%%%%%%%%%%%%%%%%%%%%%%%%%%%%%%%%%%%%%%%%%%%%%%%%%%%%%%%%%%
\textbf{Proof}: Once the parameter $m_{2}$ is replaced in terms of $m_{1}$, one can obtain the first and second derivatives of (\ref{A2}) with respect to the parameter $m_{1}$. In order to find the critical points and to obtain the maximum, one need to put them zero and lesser than zero respectively as below, 
\begin{eqnarray}
\label{A4}
  \frac{dE_{\rm CM}^{2}}{dm_{1}} &=& 2(m_{1}-m_{2})[1 - R(r; e, j, s)] = 0  \,,  \\
\label{A5}
  \frac{d^{2}E_{\rm CM}^{2}}{dm_{1}^{2}} &=& 4[1 - R(r; e, j, s)] < 0  \,,
\end{eqnarray} 
from the above Eqns. (\ref{A4}) and (\ref{A5}), it is clear that the maximum of $E_{CM}$ is achievable while both masses are equal 
and $R(r; e, j, s)>1$ (also see Figs. \ref{fig6} and \ref{fig7} for reference). The case for spinless particles can be obtained setting the spin parameter per unit mass $s=0$ in the function $R(r; e, j, s)$.
Further $R(r; e, j, s)>1$ also holds which is evident from Figs. \ref{fig6} and \ref{fig7}.  

One could also conclude the same result when introduce $\mu=m_2/m_1\leq 1$ and then
\begin{align}
\frac{E_{\rm cm}^2}{m_1m_2} &= \mu + \frac{1}{\mu} - 2 g_{ab}v_1^av_2^b\,,
\end{align}
where $v^a_i=P^a_i/m_i$ is independent of the mass. The left hand side of the above equation has a maximum at $\mu=1$, which in turn clearly implies the minimum of $E_{\rm cm}$ at $\mu=1$.

%
%%%%%%%%%%%%%%%%%%%%%%%%%%%%%%%%%%%%%%%%%%%%%%%%%%%%%%
\bibliography{my}

\begin{thebibliography}{125}
\expandafter\ifx\csname natexlab\endcsname\relax\def\natexlab#1{#1}\fi
\expandafter\ifx\csname bibnamefont\endcsname\relax
  \def\bibnamefont#1{#1}\fi
\expandafter\ifx\csname bibfnamefont\endcsname\relax
  \def\bibfnamefont#1{#1}\fi
\expandafter\ifx\csname citenamefont\endcsname\relax
  \def\citenamefont#1{#1}\fi
\expandafter\ifx\csname url\endcsname\relax
  \def\url#1{\texttt{#1}}\fi
\expandafter\ifx\csname urlprefix\endcsname\relax\def\urlprefix{URL }\fi
\providecommand{\bibinfo}[2]{#2}
\providecommand{\eprint}[2][]{\url{#2}}

\bibitem[{\citenamefont{Schwarzschild}(1916)}]{Schwarzschild:1916uq}
\bibinfo{author}{\bibfnamefont{K.}~\bibnamefont{Schwarzschild}},
  \bibinfo{journal}{Sitzungsber. Preuss. Akad. Wiss. Berlin (Math. Phys.)}
  \textbf{\bibinfo{volume}{1916}}, \bibinfo{pages}{189} (\bibinfo{year}{1916}),
  \eprint{physics/9905030}.

\bibitem[{\citenamefont{Abbott et~al.}(2016)}]{Abbott:2016blz}
\bibinfo{author}{\bibfnamefont{B.~P.} \bibnamefont{Abbott}}
  \bibnamefont{et~al.} (\bibinfo{collaboration}{LIGO Scientific, Virgo}),
  \bibinfo{journal}{Phys. Rev. Lett.} \textbf{\bibinfo{volume}{116}},
  \bibinfo{pages}{061102} (\bibinfo{year}{2016}), \eprint{1602.03837}.

\bibitem[{\citenamefont{Akiyama et~al.}(2019)}]{Akiyama:2019cqa}
\bibinfo{author}{\bibfnamefont{K.}~\bibnamefont{Akiyama}} \bibnamefont{et~al.}
  (\bibinfo{collaboration}{Event Horizon Telescope}),
  \bibinfo{journal}{Astrophys. J.} \textbf{\bibinfo{volume}{875}},
  \bibinfo{pages}{L1} (\bibinfo{year}{2019}), \eprint{1906.11238}.

\bibitem[{\citenamefont{Maldacena}(1996)}]{Maldacena:1996ky}
\bibinfo{author}{\bibfnamefont{J.~M.} \bibnamefont{Maldacena}}, Ph.D. thesis,
  \bibinfo{school}{Princeton U.} (\bibinfo{year}{1996}),
  \eprint{hep-th/9607235},
  \urlprefix\url{http://wwwlib.umi.com/dissertations/fullcit?p9627605}.

\bibitem[{\citenamefont{Padmanabhan}(2003)}]{Padmanabhan:2002ji}
\bibinfo{author}{\bibfnamefont{T.}~\bibnamefont{Padmanabhan}},
  \bibinfo{journal}{Phys. Rept.} \textbf{\bibinfo{volume}{380}},
  \bibinfo{pages}{235} (\bibinfo{year}{2003}), \eprint{hep-th/0212290}.

\bibitem[{\citenamefont{Weinberg}(1989)}]{Weinberg:1988cp}
\bibinfo{author}{\bibfnamefont{S.}~\bibnamefont{Weinberg}},
  \bibinfo{journal}{Rev. Mod. Phys.} \textbf{\bibinfo{volume}{61}},
  \bibinfo{pages}{1} (\bibinfo{year}{1989}), \bibinfo{note}{[,569(1988)]}.

\bibitem[{\citenamefont{Carroll}(1998)}]{Carroll:1998zi}
\bibinfo{author}{\bibfnamefont{S.~M.} \bibnamefont{Carroll}},
  \bibinfo{journal}{Phys. Rev. Lett.} \textbf{\bibinfo{volume}{81}},
  \bibinfo{pages}{3067} (\bibinfo{year}{1998}), \eprint{astro-ph/9806099}.

\bibitem[{\citenamefont{Khoury and Weltman}(2004)}]{Khoury:2003aq}
\bibinfo{author}{\bibfnamefont{J.}~\bibnamefont{Khoury}} \bibnamefont{and}
  \bibinfo{author}{\bibfnamefont{A.}~\bibnamefont{Weltman}},
  \bibinfo{journal}{Phys. Rev. Lett.} \textbf{\bibinfo{volume}{93}},
  \bibinfo{pages}{171104} (\bibinfo{year}{2004}), \eprint{astro-ph/0309300}.

\bibitem[{\citenamefont{Armendariz-Picon
  et~al.}(2000)\citenamefont{Armendariz-Picon, Mukhanov, and
  Steinhardt}}]{ArmendarizPicon:2000dh}
\bibinfo{author}{\bibfnamefont{C.}~\bibnamefont{Armendariz-Picon}},
  \bibinfo{author}{\bibfnamefont{V.~F.} \bibnamefont{Mukhanov}},
  \bibnamefont{and} \bibinfo{author}{\bibfnamefont{P.~J.}
  \bibnamefont{Steinhardt}}, \bibinfo{journal}{Phys. Rev. Lett.}
  \textbf{\bibinfo{volume}{85}}, \bibinfo{pages}{4438} (\bibinfo{year}{2000}),
  \eprint{astro-ph/0004134}.

\bibitem[{\citenamefont{Padmanabhan}(2002)}]{Padmanabhan:2002cp}
\bibinfo{author}{\bibfnamefont{T.}~\bibnamefont{Padmanabhan}},
  \bibinfo{journal}{Phys. Rev.} \textbf{\bibinfo{volume}{D66}},
  \bibinfo{pages}{021301} (\bibinfo{year}{2002}), \eprint{hep-th/0204150}.

\bibitem[{\citenamefont{Caldwell}(2002)}]{Caldwell:1999ew}
\bibinfo{author}{\bibfnamefont{R.~R.} \bibnamefont{Caldwell}},
  \bibinfo{journal}{Phys. Lett.} \textbf{\bibinfo{volume}{B545}},
  \bibinfo{pages}{23} (\bibinfo{year}{2002}), \eprint{astro-ph/9908168}.

\bibitem[{\citenamefont{Gasperini et~al.}(2002)\citenamefont{Gasperini, Piazza,
  and Veneziano}}]{Gasperini:2001pc}
\bibinfo{author}{\bibfnamefont{M.}~\bibnamefont{Gasperini}},
  \bibinfo{author}{\bibfnamefont{F.}~\bibnamefont{Piazza}}, \bibnamefont{and}
  \bibinfo{author}{\bibfnamefont{G.}~\bibnamefont{Veneziano}},
  \bibinfo{journal}{Phys. Rev.} \textbf{\bibinfo{volume}{D65}},
  \bibinfo{pages}{023508} (\bibinfo{year}{2002}), \eprint{gr-qc/0108016}.

\bibitem[{\citenamefont{Copeland et~al.}(2006)\citenamefont{Copeland, Sami, and
  Tsujikawa}}]{Copeland:2006wr}
\bibinfo{author}{\bibfnamefont{E.~J.} \bibnamefont{Copeland}},
  \bibinfo{author}{\bibfnamefont{M.}~\bibnamefont{Sami}}, \bibnamefont{and}
  \bibinfo{author}{\bibfnamefont{S.}~\bibnamefont{Tsujikawa}},
  \bibinfo{journal}{Int. J. Mod. Phys.} \textbf{\bibinfo{volume}{D15}},
  \bibinfo{pages}{1753} (\bibinfo{year}{2006}), \eprint{hep-th/0603057}.

\bibitem[{\citenamefont{Kiselev}(2003)}]{Kiselev:2002dx}
\bibinfo{author}{\bibfnamefont{V.~V.} \bibnamefont{Kiselev}},
  \bibinfo{journal}{Class. Quant. Grav.} \textbf{\bibinfo{volume}{20}},
  \bibinfo{pages}{1187} (\bibinfo{year}{2003}), \eprint{gr-qc/0210040}.

\bibitem[{\citenamefont{Uniyal et~al.}(2015)\citenamefont{Uniyal,
  Chandrachani~Devi, Nandan, and Purohit}}]{Uniyal:2014paa}
\bibinfo{author}{\bibfnamefont{R.}~\bibnamefont{Uniyal}},
  \bibinfo{author}{\bibfnamefont{N.}~\bibnamefont{Chandrachani~Devi}},
  \bibinfo{author}{\bibfnamefont{H.}~\bibnamefont{Nandan}}, \bibnamefont{and}
  \bibinfo{author}{\bibfnamefont{K.~D.} \bibnamefont{Purohit}},
  \bibinfo{journal}{Gen. Rel. Grav.} \textbf{\bibinfo{volume}{47}},
  \bibinfo{pages}{16} (\bibinfo{year}{2015}), \eprint{1406.3931}.

\bibitem[{\citenamefont{Banados et~al.}(2009)\citenamefont{Banados, Silk, and
  West}}]{Banados:2009pr}
\bibinfo{author}{\bibfnamefont{M.}~\bibnamefont{Banados}},
  \bibinfo{author}{\bibfnamefont{J.}~\bibnamefont{Silk}}, \bibnamefont{and}
  \bibinfo{author}{\bibfnamefont{S.~M.} \bibnamefont{West}},
  \bibinfo{journal}{Phys. Rev. Lett.} \textbf{\bibinfo{volume}{103}},
  \bibinfo{pages}{111102} (\bibinfo{year}{2009}), \eprint{0909.0169}.

\bibitem[{\citenamefont{Jacobson and Sotiriou}(2010)}]{Jacobson:2009zg}
\bibinfo{author}{\bibfnamefont{T.}~\bibnamefont{Jacobson}} \bibnamefont{and}
  \bibinfo{author}{\bibfnamefont{T.~P.} \bibnamefont{Sotiriou}},
  \bibinfo{journal}{Phys. Rev. Lett.} \textbf{\bibinfo{volume}{104}},
  \bibinfo{pages}{021101} (\bibinfo{year}{2010}), \eprint{0911.3363}.

\bibitem[{\citenamefont{Grib and Pavlov}(2011{\natexlab{a}})}]{Grib:2010dz}
\bibinfo{author}{\bibfnamefont{A.~A.} \bibnamefont{Grib}} \bibnamefont{and}
  \bibinfo{author}{\bibfnamefont{{\relax Yu}.~V.} \bibnamefont{Pavlov}},
  \bibinfo{journal}{Astropart. Phys.} \textbf{\bibinfo{volume}{34}},
  \bibinfo{pages}{581} (\bibinfo{year}{2011}{\natexlab{a}}),
  \eprint{1001.0756}.

\bibitem[{\citenamefont{Lake}(2010)}]{Lake:2010bq}
\bibinfo{author}{\bibfnamefont{K.}~\bibnamefont{Lake}}, \bibinfo{journal}{Phys.
  Rev. Lett.} \textbf{\bibinfo{volume}{104}}, \bibinfo{pages}{211102}
  (\bibinfo{year}{2010}), \bibinfo{note}{[Erratum: Phys. Rev.
  Lett.104,259903(2010)]}, \eprint{1001.5463}.

\bibitem[{\citenamefont{Wei et~al.}(2010{\natexlab{a}})\citenamefont{Wei, Liu,
  Guo, and Fu}}]{Wei:2010vca}
\bibinfo{author}{\bibfnamefont{S.-W.} \bibnamefont{Wei}},
  \bibinfo{author}{\bibfnamefont{Y.-X.} \bibnamefont{Liu}},
  \bibinfo{author}{\bibfnamefont{H.}~\bibnamefont{Guo}}, \bibnamefont{and}
  \bibinfo{author}{\bibfnamefont{C.-E.} \bibnamefont{Fu}},
  \bibinfo{journal}{Phys. Rev.} \textbf{\bibinfo{volume}{D82}},
  \bibinfo{pages}{103005} (\bibinfo{year}{2010}{\natexlab{a}}),
  \eprint{1006.1056}.

\bibitem[{\citenamefont{Grib and Pavlov}(2010)}]{Grib:2010zs}
\bibinfo{author}{\bibfnamefont{A.~A.} \bibnamefont{Grib}} \bibnamefont{and}
  \bibinfo{author}{\bibfnamefont{Y.~V.} \bibnamefont{Pavlov}}
  (\bibinfo{year}{2010}), \eprint{1007.3222}.

\bibitem[{\citenamefont{Zaslavskii}(2010)}]{Zaslavskii:2010aw}
\bibinfo{author}{\bibfnamefont{O.~B.} \bibnamefont{Zaslavskii}},
  \bibinfo{journal}{JETP Lett.} \textbf{\bibinfo{volume}{92}},
  \bibinfo{pages}{571} (\bibinfo{year}{2010}), \bibinfo{note}{[Pisma Zh. Eksp.
  Teor. Fiz.92,635(2010)]}, \eprint{1007.4598}.

\bibitem[{\citenamefont{Harada and Kimura}(2011{\natexlab{a}})}]{Harada:2010yv}
\bibinfo{author}{\bibfnamefont{T.}~\bibnamefont{Harada}} \bibnamefont{and}
  \bibinfo{author}{\bibfnamefont{M.}~\bibnamefont{Kimura}},
  \bibinfo{journal}{Phys. Rev.} \textbf{\bibinfo{volume}{D83}},
  \bibinfo{pages}{024002} (\bibinfo{year}{2011}{\natexlab{a}}),
  \eprint{1010.0962}.

\bibitem[{\citenamefont{Grib and Pavlov}(2011{\natexlab{b}})}]{Grib:2010xj}
\bibinfo{author}{\bibfnamefont{A.~A.} \bibnamefont{Grib}} \bibnamefont{and}
  \bibinfo{author}{\bibfnamefont{{\relax Yu}.~V.} \bibnamefont{Pavlov}},
  \bibinfo{journal}{Grav. Cosmol.} \textbf{\bibinfo{volume}{17}},
  \bibinfo{pages}{42} (\bibinfo{year}{2011}{\natexlab{b}}), \eprint{1010.2052}.

\bibitem[{\citenamefont{Banados et~al.}(2011)\citenamefont{Banados, Hassanain,
  Silk, and West}}]{Banados:2010kn}
\bibinfo{author}{\bibfnamefont{M.}~\bibnamefont{Banados}},
  \bibinfo{author}{\bibfnamefont{B.}~\bibnamefont{Hassanain}},
  \bibinfo{author}{\bibfnamefont{J.}~\bibnamefont{Silk}}, \bibnamefont{and}
  \bibinfo{author}{\bibfnamefont{S.~M.} \bibnamefont{West}},
  \bibinfo{journal}{Phys. Rev.} \textbf{\bibinfo{volume}{D83}},
  \bibinfo{pages}{023004} (\bibinfo{year}{2011}), \eprint{1010.2724}.

\bibitem[{\citenamefont{Williams}(2011)}]{Williams:2011uz}
\bibinfo{author}{\bibfnamefont{A.~J.} \bibnamefont{Williams}},
  \bibinfo{journal}{Phys. Rev.} \textbf{\bibinfo{volume}{D83}},
  \bibinfo{pages}{123004} (\bibinfo{year}{2011}), \eprint{1101.4819}.

\bibitem[{\citenamefont{Harada and Kimura}(2011{\natexlab{b}})}]{Harada:2011xz}
\bibinfo{author}{\bibfnamefont{T.}~\bibnamefont{Harada}} \bibnamefont{and}
  \bibinfo{author}{\bibfnamefont{M.}~\bibnamefont{Kimura}},
  \bibinfo{journal}{Phys. Rev.} \textbf{\bibinfo{volume}{D83}},
  \bibinfo{pages}{084041} (\bibinfo{year}{2011}{\natexlab{b}}),
  \eprint{1102.3316}.

\bibitem[{\citenamefont{Patil and Joshi}(2011)}]{Patil:2011yb}
\bibinfo{author}{\bibfnamefont{M.}~\bibnamefont{Patil}} \bibnamefont{and}
  \bibinfo{author}{\bibfnamefont{P.~S.} \bibnamefont{Joshi}},
  \bibinfo{journal}{Phys. Rev.} \textbf{\bibinfo{volume}{D84}},
  \bibinfo{pages}{104001} (\bibinfo{year}{2011}), \eprint{1103.1083}.

\bibitem[{\citenamefont{Zhu et~al.}(2011{\natexlab{a}})\citenamefont{Zhu, Wu,
  Liu, and Jiang}}]{Zhu:2011ae}
\bibinfo{author}{\bibfnamefont{Y.}~\bibnamefont{Zhu}},
  \bibinfo{author}{\bibfnamefont{S.-F.} \bibnamefont{Wu}},
  \bibinfo{author}{\bibfnamefont{Y.-X.} \bibnamefont{Liu}}, \bibnamefont{and}
  \bibinfo{author}{\bibfnamefont{Y.}~\bibnamefont{Jiang}},
  \bibinfo{journal}{Phys. Rev.} \textbf{\bibinfo{volume}{D84}},
  \bibinfo{pages}{043006} (\bibinfo{year}{2011}{\natexlab{a}}),
  \eprint{1103.3848}.

\bibitem[{\citenamefont{Liu et~al.}(2013)\citenamefont{Liu, Chen, and
  Jing}}]{Liu:2011wv}
\bibinfo{author}{\bibfnamefont{C.}~\bibnamefont{Liu}},
  \bibinfo{author}{\bibfnamefont{S.}~\bibnamefont{Chen}}, \bibnamefont{and}
  \bibinfo{author}{\bibfnamefont{J.}~\bibnamefont{Jing}},
  \bibinfo{journal}{Chin. Phys. Lett.} \textbf{\bibinfo{volume}{30}},
  \bibinfo{pages}{100401} (\bibinfo{year}{2013}), \eprint{1104.3225}.

\bibitem[{\citenamefont{Zaslavskii}(2011{\natexlab{a}})}]{Zaslavskii:2011dz}
\bibinfo{author}{\bibfnamefont{O.~B.} \bibnamefont{Zaslavskii}},
  \bibinfo{journal}{Phys. Rev.} \textbf{\bibinfo{volume}{D84}},
  \bibinfo{pages}{024007} (\bibinfo{year}{2011}{\natexlab{a}}),
  \eprint{1104.4802}.

\bibitem[{\citenamefont{Zaslavskii}(2011{\natexlab{b}})}]{Zaslavskii:2011ex}
\bibinfo{author}{\bibfnamefont{O.~B.} \bibnamefont{Zaslavskii}}
  (\bibinfo{year}{2011}{\natexlab{b}}), \eprint{1105.0303}.

\bibitem[{\citenamefont{Grib et~al.}(2011)\citenamefont{Grib, Pavlov,
  Piattella, and Piattella}}]{Grib:2011ph}
\bibinfo{author}{\bibfnamefont{A.~A.} \bibnamefont{Grib}},
  \bibinfo{author}{\bibfnamefont{{\relax Yu}.~V.} \bibnamefont{Pavlov}},
  \bibinfo{author}{\bibfnamefont{O.~F.} \bibnamefont{Piattella}},
  \bibnamefont{and} \bibinfo{author}{\bibfnamefont{O.~F.}
  \bibnamefont{Piattella}}, \bibinfo{journal}{Int. J. Mod. Phys.}
  \textbf{\bibinfo{volume}{A26}}, \bibinfo{pages}{3856} (\bibinfo{year}{2011}),
  \bibinfo{note}{[Int. J. Mod. Phys. Conf. Ser.03,342(2011)]},
  \eprint{1105.1540}.

\bibitem[{\citenamefont{Yao et~al.}(2012)\citenamefont{Yao, Chen, Liu, and
  Jing}}]{Yao:2011ai}
\bibinfo{author}{\bibfnamefont{W.-P.} \bibnamefont{Yao}},
  \bibinfo{author}{\bibfnamefont{S.}~\bibnamefont{Chen}},
  \bibinfo{author}{\bibfnamefont{C.}~\bibnamefont{Liu}}, \bibnamefont{and}
  \bibinfo{author}{\bibfnamefont{J.}~\bibnamefont{Jing}},
  \bibinfo{journal}{Eur. Phys. J.} \textbf{\bibinfo{volume}{C72}},
  \bibinfo{pages}{1898} (\bibinfo{year}{2012}), \eprint{1105.6156}.

\bibitem[{\citenamefont{Gao and Zhong}(2011)}]{Gao:2011sv}
\bibinfo{author}{\bibfnamefont{S.}~\bibnamefont{Gao}} \bibnamefont{and}
  \bibinfo{author}{\bibfnamefont{C.}~\bibnamefont{Zhong}},
  \bibinfo{journal}{Phys. Rev.} \textbf{\bibinfo{volume}{D84}},
  \bibinfo{pages}{044006} (\bibinfo{year}{2011}), \eprint{1106.2852}.

\bibitem[{\citenamefont{Zaslavsky}(2012)}]{Zaslavskii:2011uu}
\bibinfo{author}{\bibfnamefont{O.~B.} \bibnamefont{Zaslavsky}},
  \bibinfo{journal}{Grav. Cosmol.} \textbf{\bibinfo{volume}{18}},
  \bibinfo{pages}{139} (\bibinfo{year}{2012}), \eprint{1107.3964}.

\bibitem[{\citenamefont{Patil et~al.}(2012)\citenamefont{Patil, Joshi, Kimura,
  and Nakao}}]{Patil:2011uf}
\bibinfo{author}{\bibfnamefont{M.}~\bibnamefont{Patil}},
  \bibinfo{author}{\bibfnamefont{P.~S.} \bibnamefont{Joshi}},
  \bibinfo{author}{\bibfnamefont{M.}~\bibnamefont{Kimura}}, \bibnamefont{and}
  \bibinfo{author}{\bibfnamefont{K.-i.} \bibnamefont{Nakao}},
  \bibinfo{journal}{Phys. Rev.} \textbf{\bibinfo{volume}{D86}},
  \bibinfo{pages}{084023} (\bibinfo{year}{2012}), \eprint{1108.0288}.

\bibitem[{\citenamefont{Zhu et~al.}(2011{\natexlab{b}})\citenamefont{Zhu, Wu,
  Jiang, and Yang}}]{Zhu:2011ja}
\bibinfo{author}{\bibfnamefont{Y.}~\bibnamefont{Zhu}},
  \bibinfo{author}{\bibfnamefont{S.-F.} \bibnamefont{Wu}},
  \bibinfo{author}{\bibfnamefont{Y.}~\bibnamefont{Jiang}}, \bibnamefont{and}
  \bibinfo{author}{\bibfnamefont{G.-H.} \bibnamefont{Yang}},
  \bibinfo{journal}{Phys. Rev.} \textbf{\bibinfo{volume}{D84}},
  \bibinfo{pages}{123002} (\bibinfo{year}{2011}{\natexlab{b}}),
  \eprint{1108.1843}.

\bibitem[{\citenamefont{Harada and Kimura}(2011{\natexlab{c}})}]{Harada:2011pg}
\bibinfo{author}{\bibfnamefont{T.}~\bibnamefont{Harada}} \bibnamefont{and}
  \bibinfo{author}{\bibfnamefont{M.}~\bibnamefont{Kimura}},
  \bibinfo{journal}{Phys. Rev.} \textbf{\bibinfo{volume}{D84}},
  \bibinfo{pages}{124032} (\bibinfo{year}{2011}{\natexlab{c}}),
  \eprint{1109.6722}.

\bibitem[{\citenamefont{Zaslavsky}(2011)}]{Zaslavsky:2011zz}
\bibinfo{author}{\bibfnamefont{O.~B.} \bibnamefont{Zaslavsky}},
  \bibinfo{journal}{Int. J. Mod. Phys.} \textbf{\bibinfo{volume}{A26}},
  \bibinfo{pages}{3845} (\bibinfo{year}{2011}).

\bibitem[{\citenamefont{Zaslavskii}(2012{\natexlab{a}})}]{Zaslavskii:2011tj}
\bibinfo{author}{\bibfnamefont{O.~B.} \bibnamefont{Zaslavskii}},
  \bibinfo{journal}{Phys. Rev.} \textbf{\bibinfo{volume}{D85}},
  \bibinfo{pages}{024029} (\bibinfo{year}{2012}{\natexlab{a}}),
  \eprint{1110.5838}.

\bibitem[{\citenamefont{Frolov}(2012)}]{Frolov:2011ea}
\bibinfo{author}{\bibfnamefont{V.~P.} \bibnamefont{Frolov}},
  \bibinfo{journal}{Phys. Rev.} \textbf{\bibinfo{volume}{D85}},
  \bibinfo{pages}{024020} (\bibinfo{year}{2012}), \eprint{1110.6274}.

\bibitem[{\citenamefont{Zaslavskii}(2012{\natexlab{b}})}]{Zaslavskii:2012ua}
\bibinfo{author}{\bibfnamefont{O.~B.} \bibnamefont{Zaslavskii}},
  \bibinfo{journal}{Class. Quant. Grav.} \textbf{\bibinfo{volume}{29}},
  \bibinfo{pages}{205004} (\bibinfo{year}{2012}{\natexlab{b}}),
  \eprint{1201.5351}.

\bibitem[{\citenamefont{Grib et~al.}(2012)\citenamefont{Grib, Pavlov, and
  Piattella}}]{Grib:2012iq}
\bibinfo{author}{\bibfnamefont{A.~A.} \bibnamefont{Grib}},
  \bibinfo{author}{\bibfnamefont{{\relax Yu}.~V.} \bibnamefont{Pavlov}},
  \bibnamefont{and} \bibinfo{author}{\bibfnamefont{O.~F.}
  \bibnamefont{Piattella}}, \bibinfo{journal}{Grav. Cosmol.}
  \textbf{\bibinfo{volume}{18}}, \bibinfo{pages}{70} (\bibinfo{year}{2012}),
  \eprint{1203.4952}.

\bibitem[{\citenamefont{Hussain}(2012)}]{Hussain:2012zza}
\bibinfo{author}{\bibfnamefont{I.}~\bibnamefont{Hussain}},
  \bibinfo{journal}{Mod. Phys. Lett.} \textbf{\bibinfo{volume}{A27}},
  \bibinfo{pages}{1250017} (\bibinfo{year}{2012}).

\bibitem[{\citenamefont{Harada et~al.}(2012)\citenamefont{Harada, Nemoto, and
  Miyamoto}}]{Harada:2012ap}
\bibinfo{author}{\bibfnamefont{T.}~\bibnamefont{Harada}},
  \bibinfo{author}{\bibfnamefont{H.}~\bibnamefont{Nemoto}}, \bibnamefont{and}
  \bibinfo{author}{\bibfnamefont{U.}~\bibnamefont{Miyamoto}},
  \bibinfo{journal}{Phys. Rev.} \textbf{\bibinfo{volume}{D86}},
  \bibinfo{pages}{024027} (\bibinfo{year}{2012}), \bibinfo{note}{[Erratum:
  Phys. Rev.D86,069902(2012)]}, \eprint{1205.7088}.

\bibitem[{\citenamefont{Tanatarov and Zaslavskii}(2012)}]{Tanatarov:2012xj}
\bibinfo{author}{\bibfnamefont{I.~V.} \bibnamefont{Tanatarov}}
  \bibnamefont{and} \bibinfo{author}{\bibfnamefont{O.~B.}
  \bibnamefont{Zaslavskii}}, \bibinfo{journal}{Phys. Rev.}
  \textbf{\bibinfo{volume}{D86}}, \bibinfo{pages}{044019}
  (\bibinfo{year}{2012}), \eprint{1206.2580}.

\bibitem[{\citenamefont{Nemoto et~al.}(2013)\citenamefont{Nemoto, Miyamoto,
  Harada, and Kokubu}}]{Nemoto:2012cq}
\bibinfo{author}{\bibfnamefont{H.}~\bibnamefont{Nemoto}},
  \bibinfo{author}{\bibfnamefont{U.}~\bibnamefont{Miyamoto}},
  \bibinfo{author}{\bibfnamefont{T.}~\bibnamefont{Harada}}, \bibnamefont{and}
  \bibinfo{author}{\bibfnamefont{T.}~\bibnamefont{Kokubu}},
  \bibinfo{journal}{Phys. Rev.} \textbf{\bibinfo{volume}{D87}},
  \bibinfo{pages}{127502} (\bibinfo{year}{2013}), \eprint{1212.6701}.

\bibitem[{\citenamefont{Grib and Pavlov}(2013)}]{Grib:2013vc}
\bibinfo{author}{\bibfnamefont{A.~A.} \bibnamefont{Grib}} \bibnamefont{and}
  \bibinfo{author}{\bibfnamefont{{\relax Yu}.~V.} \bibnamefont{Pavlov}},
  \bibinfo{journal}{EPL} \textbf{\bibinfo{volume}{101}}, \bibinfo{pages}{20004}
  (\bibinfo{year}{2013}), \eprint{1301.0698}.

\bibitem[{\citenamefont{Galajinsky}(2013)}]{Galajinsky:2013as}
\bibinfo{author}{\bibfnamefont{A.}~\bibnamefont{Galajinsky}},
  \bibinfo{journal}{Phys. Rev.} \textbf{\bibinfo{volume}{D88}},
  \bibinfo{pages}{027505} (\bibinfo{year}{2013}), \eprint{1301.1159}.

\bibitem[{\citenamefont{Zaslavsky}(2013)}]{Zaslavsky:2013dra}
\bibinfo{author}{\bibfnamefont{O.~B.} \bibnamefont{Zaslavsky}},
  \bibinfo{journal}{Mod. Phys. Lett.} \textbf{\bibinfo{volume}{A28}},
  \bibinfo{pages}{1350037} (\bibinfo{year}{2013}), \eprint{1301.4699}.

\bibitem[{\citenamefont{Stuchlik and Schee}(2013)}]{Stuchlik:2013yca}
\bibinfo{author}{\bibfnamefont{Z.}~\bibnamefont{Stuchlik}} \bibnamefont{and}
  \bibinfo{author}{\bibfnamefont{J.}~\bibnamefont{Schee}},
  \bibinfo{journal}{Class. Quant. Grav.} \textbf{\bibinfo{volume}{30}},
  \bibinfo{pages}{075012} (\bibinfo{year}{2013}).

\bibitem[{\citenamefont{Abdujabbarov et~al.}(2013)\citenamefont{Abdujabbarov,
  Dadhich, Ahmedov, and Eshkuvatov}}]{Abdujabbarov:2013qka}
\bibinfo{author}{\bibfnamefont{A.}~\bibnamefont{Abdujabbarov}},
  \bibinfo{author}{\bibfnamefont{N.}~\bibnamefont{Dadhich}},
  \bibinfo{author}{\bibfnamefont{B.}~\bibnamefont{Ahmedov}}, \bibnamefont{and}
  \bibinfo{author}{\bibfnamefont{H.}~\bibnamefont{Eshkuvatov}},
  \bibinfo{journal}{Phys. Rev.} \textbf{\bibinfo{volume}{D88}},
  \bibinfo{pages}{084036} (\bibinfo{year}{2013}), \eprint{1310.4494}.

\bibitem[{\citenamefont{Tsukamoto et~al.}(2014)\citenamefont{Tsukamoto, Kimura,
  and Harada}}]{Tsukamoto:2013dna}
\bibinfo{author}{\bibfnamefont{N.}~\bibnamefont{Tsukamoto}},
  \bibinfo{author}{\bibfnamefont{M.}~\bibnamefont{Kimura}}, \bibnamefont{and}
  \bibinfo{author}{\bibfnamefont{T.}~\bibnamefont{Harada}},
  \bibinfo{journal}{Phys. Rev.} \textbf{\bibinfo{volume}{D89}},
  \bibinfo{pages}{024020} (\bibinfo{year}{2014}), \eprint{1310.5716}.

\bibitem[{\citenamefont{Zaslavskii}(2014)}]{Zaslavskii:2014dxa}
\bibinfo{author}{\bibfnamefont{O.~B.} \bibnamefont{Zaslavskii}},
  \bibinfo{journal}{Mod. Phys. Lett.} \textbf{\bibinfo{volume}{A29}},
  \bibinfo{pages}{1450112} (\bibinfo{year}{2014}), \eprint{1403.6286}.

\bibitem[{\citenamefont{Yumisaki}(2017)}]{Yumisaki:2016biz}
\bibinfo{author}{\bibfnamefont{H.}~\bibnamefont{Yumisaki}},
  \bibinfo{journal}{PTEP} \textbf{\bibinfo{volume}{2017}},
  \bibinfo{pages}{063B04} (\bibinfo{year}{2017}), \eprint{1606.09626}.

\bibitem[{\citenamefont{Sadeghi et~al.}(2014)\citenamefont{Sadeghi, Pourhassan,
  and Farahani}}]{Sadeghi:2013gmf}
\bibinfo{author}{\bibfnamefont{J.}~\bibnamefont{Sadeghi}},
  \bibinfo{author}{\bibfnamefont{B.}~\bibnamefont{Pourhassan}},
  \bibnamefont{and} \bibinfo{author}{\bibfnamefont{H.}~\bibnamefont{Farahani}},
  \bibinfo{journal}{Commun. Theor. Phys.} \textbf{\bibinfo{volume}{62}},
  \bibinfo{pages}{358} (\bibinfo{year}{2014}), \eprint{1310.7142}.

\bibitem[{\citenamefont{Pradhan}(2014)}]{Pradhan:2014oaa}
\bibinfo{author}{\bibfnamefont{P.}~\bibnamefont{Pradhan}}
  (\bibinfo{year}{2014}), \eprint{1402.2748}.

\bibitem[{\citenamefont{Pradhan}(2015)}]{Pradhan:2014eza}
\bibinfo{author}{\bibfnamefont{P.}~\bibnamefont{Pradhan}},
  \bibinfo{journal}{Astropart. Phys.} \textbf{\bibinfo{volume}{62}},
  \bibinfo{pages}{217} (\bibinfo{year}{2015}), \eprint{1407.0877}.

\bibitem[{\citenamefont{Ghosh et~al.}(2014)\citenamefont{Ghosh, Sheoran, and
  Amir}}]{Ghosh:2014mea}
\bibinfo{author}{\bibfnamefont{S.~G.} \bibnamefont{Ghosh}},
  \bibinfo{author}{\bibfnamefont{P.}~\bibnamefont{Sheoran}}, \bibnamefont{and}
  \bibinfo{author}{\bibfnamefont{M.}~\bibnamefont{Amir}},
  \bibinfo{journal}{Phys. Rev.} \textbf{\bibinfo{volume}{D90}},
  \bibinfo{pages}{103006} (\bibinfo{year}{2014}), \eprint{1410.5588}.

\bibitem[{\citenamefont{Hussain et~al.}(2015)\citenamefont{Hussain, Jamil, and
  Majeed}}]{Hussain:2014aea}
\bibinfo{author}{\bibfnamefont{I.}~\bibnamefont{Hussain}},
  \bibinfo{author}{\bibfnamefont{M.}~\bibnamefont{Jamil}}, \bibnamefont{and}
  \bibinfo{author}{\bibfnamefont{B.}~\bibnamefont{Majeed}},
  \bibinfo{journal}{Int. J. Theor. Phys.} \textbf{\bibinfo{volume}{54}},
  \bibinfo{pages}{1567} (\bibinfo{year}{2015}), \eprint{1412.1112}.

\bibitem[{\citenamefont{Zakria and Jamil}(2015)}]{Zakria:2015eua}
\bibinfo{author}{\bibfnamefont{A.}~\bibnamefont{Zakria}} \bibnamefont{and}
  \bibinfo{author}{\bibfnamefont{M.}~\bibnamefont{Jamil}},
  \bibinfo{journal}{JHEP} \textbf{\bibinfo{volume}{05}}, \bibinfo{pages}{147}
  (\bibinfo{year}{2015}), \eprint{1501.06306}.

\bibitem[{\citenamefont{Amir and Ghosh}(2015)}]{Amir:2015pja}
\bibinfo{author}{\bibfnamefont{M.}~\bibnamefont{Amir}} \bibnamefont{and}
  \bibinfo{author}{\bibfnamefont{S.~G.} \bibnamefont{Ghosh}},
  \bibinfo{journal}{JHEP} \textbf{\bibinfo{volume}{07}}, \bibinfo{pages}{015}
  (\bibinfo{year}{2015}), \eprint{1503.08553}.

\bibitem[{\citenamefont{Halilsoy and
  Ovgun}(2017{\natexlab{a}})}]{Halilsoy:2015rna}
\bibinfo{author}{\bibfnamefont{M.}~\bibnamefont{Halilsoy}} \bibnamefont{and}
  \bibinfo{author}{\bibfnamefont{A.}~\bibnamefont{Ovgun}},
  \bibinfo{journal}{Adv. High Energy Phys.} \textbf{\bibinfo{volume}{2017}},
  \bibinfo{pages}{4383617} (\bibinfo{year}{2017}{\natexlab{a}}),
  \eprint{1504.03840}.

\bibitem[{\citenamefont{Pourhassan and Debnath}(2019)}]{Pourhassan:2015lfa}
\bibinfo{author}{\bibfnamefont{B.}~\bibnamefont{Pourhassan}} \bibnamefont{and}
  \bibinfo{author}{\bibfnamefont{U.}~\bibnamefont{Debnath}},
  \bibinfo{journal}{Grav. Cosmol.} \textbf{\bibinfo{volume}{25}},
  \bibinfo{pages}{196} (\bibinfo{year}{2019}), \eprint{1506.03443}.

\bibitem[{\citenamefont{Ghosh and Amir}(2015)}]{Ghosh:2015pra}
\bibinfo{author}{\bibfnamefont{S.~G.} \bibnamefont{Ghosh}} \bibnamefont{and}
  \bibinfo{author}{\bibfnamefont{M.}~\bibnamefont{Amir}},
  \bibinfo{journal}{Eur. Phys. J.} \textbf{\bibinfo{volume}{C75}},
  \bibinfo{pages}{553} (\bibinfo{year}{2015}), \eprint{1506.04382}.

\bibitem[{\citenamefont{Saadat}(2014)}]{Saadat:2013iba}
\bibinfo{author}{\bibfnamefont{H.}~\bibnamefont{Saadat}},
  \bibinfo{journal}{Can. J. Phys.} \textbf{\bibinfo{volume}{92}},
  \bibinfo{pages}{1562} (\bibinfo{year}{2014}), \eprint{1306.4349}.

\bibitem[{\citenamefont{Abdujabbarov et~al.}(2011)\citenamefont{Abdujabbarov,
  Ahmedov, and Ahmedov}}]{Abdujabbarov:2011af}
\bibinfo{author}{\bibfnamefont{A.}~\bibnamefont{Abdujabbarov}},
  \bibinfo{author}{\bibfnamefont{B.}~\bibnamefont{Ahmedov}}, \bibnamefont{and}
  \bibinfo{author}{\bibfnamefont{B.}~\bibnamefont{Ahmedov}},
  \bibinfo{journal}{Phys. Rev.} \textbf{\bibinfo{volume}{D84}},
  \bibinfo{pages}{044044} (\bibinfo{year}{2011}), \eprint{1107.5389}.

\bibitem[{\citenamefont{Sadeghi and Pourhassan}(2012)}]{Sadeghi:2011qu}
\bibinfo{author}{\bibfnamefont{J.}~\bibnamefont{Sadeghi}} \bibnamefont{and}
  \bibinfo{author}{\bibfnamefont{B.}~\bibnamefont{Pourhassan}},
  \bibinfo{journal}{Eur. Phys. J.} \textbf{\bibinfo{volume}{C72}},
  \bibinfo{pages}{1984} (\bibinfo{year}{2012}), \eprint{1108.4530}.

\bibitem[{\citenamefont{Wei et~al.}(2010{\natexlab{b}})\citenamefont{Wei, Liu,
  Li, and Chen}}]{Wei:2010gq}
\bibinfo{author}{\bibfnamefont{S.-W.} \bibnamefont{Wei}},
  \bibinfo{author}{\bibfnamefont{Y.-X.} \bibnamefont{Liu}},
  \bibinfo{author}{\bibfnamefont{H.-T.} \bibnamefont{Li}}, \bibnamefont{and}
  \bibinfo{author}{\bibfnamefont{F.-W.} \bibnamefont{Chen}},
  \bibinfo{journal}{JHEP} \textbf{\bibinfo{volume}{12}}, \bibinfo{pages}{066}
  (\bibinfo{year}{2010}{\natexlab{b}}), \eprint{1007.4333}.

\bibitem[{\citenamefont{Mao et~al.}(2017)\citenamefont{Mao, Li, Jia, and
  Ren}}]{Mao:2010di}
\bibinfo{author}{\bibfnamefont{P.-J.} \bibnamefont{Mao}},
  \bibinfo{author}{\bibfnamefont{R.}~\bibnamefont{Li}},
  \bibinfo{author}{\bibfnamefont{L.-Y.} \bibnamefont{Jia}}, \bibnamefont{and}
  \bibinfo{author}{\bibfnamefont{J.-R.} \bibnamefont{Ren}},
  \bibinfo{journal}{Chin. Phys.} \textbf{\bibinfo{volume}{C41}},
  \bibinfo{pages}{065101} (\bibinfo{year}{2017}), \eprint{1008.2660}.

\bibitem[{\citenamefont{Li et~al.}(2011)\citenamefont{Li, Yang, Li, Wei, and
  Liu}}]{Li:2010ej}
\bibinfo{author}{\bibfnamefont{Y.}~\bibnamefont{Li}},
  \bibinfo{author}{\bibfnamefont{J.}~\bibnamefont{Yang}},
  \bibinfo{author}{\bibfnamefont{Y.-L.} \bibnamefont{Li}},
  \bibinfo{author}{\bibfnamefont{S.-W.} \bibnamefont{Wei}}, \bibnamefont{and}
  \bibinfo{author}{\bibfnamefont{Y.-X.} \bibnamefont{Liu}},
  \bibinfo{journal}{Class. Quant. Grav.} \textbf{\bibinfo{volume}{28}},
  \bibinfo{pages}{225006} (\bibinfo{year}{2011}), \eprint{1012.0748}.

\bibitem[{\citenamefont{Liu et~al.}(2011)\citenamefont{Liu, Chen, Ding, and
  Jing}}]{Liu:2010ja}
\bibinfo{author}{\bibfnamefont{C.}~\bibnamefont{Liu}},
  \bibinfo{author}{\bibfnamefont{S.}~\bibnamefont{Chen}},
  \bibinfo{author}{\bibfnamefont{C.}~\bibnamefont{Ding}}, \bibnamefont{and}
  \bibinfo{author}{\bibfnamefont{J.}~\bibnamefont{Jing}},
  \bibinfo{journal}{Phys. Lett.} \textbf{\bibinfo{volume}{B701}},
  \bibinfo{pages}{285} (\bibinfo{year}{2011}), \eprint{1012.5126}.

\bibitem[{\citenamefont{Halilsoy and
  Ovgun}(2017{\natexlab{b}})}]{Halilsoy:2015qta}
\bibinfo{author}{\bibfnamefont{M.}~\bibnamefont{Halilsoy}} \bibnamefont{and}
  \bibinfo{author}{\bibfnamefont{A.}~\bibnamefont{Ovgun}},
  \bibinfo{journal}{Can. J. Phys.} \textbf{\bibinfo{volume}{95}},
  \bibinfo{pages}{1037} (\bibinfo{year}{2017}{\natexlab{b}}),
  \eprint{1507.00633}.

\bibitem[{\citenamefont{Abdujabbarov et~al.}(2015)\citenamefont{Abdujabbarov,
  Atamurotov, Dadhich, Ahmedov, and Stuchlík}}]{Abdujabbarov:2015rqa}
\bibinfo{author}{\bibfnamefont{A.}~\bibnamefont{Abdujabbarov}},
  \bibinfo{author}{\bibfnamefont{F.}~\bibnamefont{Atamurotov}},
  \bibinfo{author}{\bibfnamefont{N.}~\bibnamefont{Dadhich}},
  \bibinfo{author}{\bibfnamefont{B.}~\bibnamefont{Ahmedov}}, \bibnamefont{and}
  \bibinfo{author}{\bibfnamefont{Z.}~\bibnamefont{Stuchlík}},
  \bibinfo{journal}{Eur. Phys. J.} \textbf{\bibinfo{volume}{C75}},
  \bibinfo{pages}{399} (\bibinfo{year}{2015}), \eprint{1508.00331}.

\bibitem[{\citenamefont{Debnath}(2015)}]{Debnath:2015bna}
\bibinfo{author}{\bibfnamefont{U.}~\bibnamefont{Debnath}}
  (\bibinfo{year}{2015}), \eprint{1508.02385}.

\bibitem[{\citenamefont{Toshmatov et~al.}(2015)\citenamefont{Toshmatov,
  Abdujabbarov, Ahmedov, and Stuchlík}}]{Toshmatov:2015gna}
\bibinfo{author}{\bibfnamefont{B.}~\bibnamefont{Toshmatov}},
  \bibinfo{author}{\bibfnamefont{A.}~\bibnamefont{Abdujabbarov}},
  \bibinfo{author}{\bibfnamefont{B.}~\bibnamefont{Ahmedov}}, \bibnamefont{and}
  \bibinfo{author}{\bibfnamefont{Z.}~\bibnamefont{Stuchlík}},
  \bibinfo{journal}{Astrophys. Space Sci.} \textbf{\bibinfo{volume}{360}},
  \bibinfo{pages}{19} (\bibinfo{year}{2015}).

\bibitem[{\citenamefont{Sultana and Bose}(2015)}]{Sultana:2015avz}
\bibinfo{author}{\bibfnamefont{J.}~\bibnamefont{Sultana}} \bibnamefont{and}
  \bibinfo{author}{\bibfnamefont{B.}~\bibnamefont{Bose}},
  \bibinfo{journal}{Phys. Rev.} \textbf{\bibinfo{volume}{D92}},
  \bibinfo{pages}{104022} (\bibinfo{year}{2015}).

\bibitem[{\citenamefont{Zaslavskii}(2017)}]{Zaslavskii:2016stw}
\bibinfo{author}{\bibfnamefont{O.~B.} \bibnamefont{Zaslavskii}},
  \bibinfo{journal}{Int. J. Mod. Phys.} \textbf{\bibinfo{volume}{D26}},
  \bibinfo{pages}{1750108} (\bibinfo{year}{2017}), \eprint{1602.08779}.

\bibitem[{\citenamefont{Amir et~al.}(2016)\citenamefont{Amir, Ahmed, and
  Ghosh}}]{Amir:2016nti}
\bibinfo{author}{\bibfnamefont{M.}~\bibnamefont{Amir}},
  \bibinfo{author}{\bibfnamefont{F.}~\bibnamefont{Ahmed}}, \bibnamefont{and}
  \bibinfo{author}{\bibfnamefont{S.~G.} \bibnamefont{Ghosh}},
  \bibinfo{journal}{Eur. Phys. J.} \textbf{\bibinfo{volume}{C76}},
  \bibinfo{pages}{532} (\bibinfo{year}{2016}), \eprint{1607.05063}.

\bibitem[{\citenamefont{Oteev et~al.}(2016)\citenamefont{Oteev, Abdujabbarov,
  Stuchlík, and Ahmedov}}]{Oteev:2016fbp}
\bibinfo{author}{\bibfnamefont{T.}~\bibnamefont{Oteev}},
  \bibinfo{author}{\bibfnamefont{A.}~\bibnamefont{Abdujabbarov}},
  \bibinfo{author}{\bibfnamefont{Z.}~\bibnamefont{Stuchlík}},
  \bibnamefont{and} \bibinfo{author}{\bibfnamefont{B.}~\bibnamefont{Ahmedov}},
  \bibinfo{journal}{Astrophys. Space Sci.} \textbf{\bibinfo{volume}{361}},
  \bibinfo{pages}{269} (\bibinfo{year}{2016}).

\bibitem[{\citenamefont{Jawad et~al.}(2016)\citenamefont{Jawad, Ali, Jamil, and
  Debnath}}]{Jawad:2016ccw}
\bibinfo{author}{\bibfnamefont{A.}~\bibnamefont{Jawad}},
  \bibinfo{author}{\bibfnamefont{F.}~\bibnamefont{Ali}},
  \bibinfo{author}{\bibfnamefont{M.}~\bibnamefont{Jamil}}, \bibnamefont{and}
  \bibinfo{author}{\bibfnamefont{U.}~\bibnamefont{Debnath}},
  \bibinfo{journal}{Commun. Theor. Phys.} \textbf{\bibinfo{volume}{66}},
  \bibinfo{pages}{509} (\bibinfo{year}{2016}), \eprint{1610.07411}.

\bibitem[{\citenamefont{Fernando}(2017{\natexlab{a}})}]{Fernando:2017kut}
\bibinfo{author}{\bibfnamefont{S.}~\bibnamefont{Fernando}},
  \bibinfo{journal}{Mod. Phys. Lett.} \textbf{\bibinfo{volume}{A32}},
  \bibinfo{pages}{1750074} (\bibinfo{year}{2017}{\natexlab{a}}),
  \eprint{1703.00373}.

\bibitem[{\citenamefont{Fernando}(2017{\natexlab{b}})}]{Fernando:2017qrq}
\bibinfo{author}{\bibfnamefont{S.}~\bibnamefont{Fernando}},
  \bibinfo{journal}{Mod. Phys. Lett.} \textbf{\bibinfo{volume}{A32}},
  \bibinfo{pages}{1750088} (\bibinfo{year}{2017}{\natexlab{b}}),
  \eprint{1703.10072}.

\bibitem[{\citenamefont{Majeed and Jamil}(2017)}]{Majeed:2017txa}
\bibinfo{author}{\bibfnamefont{B.}~\bibnamefont{Majeed}} \bibnamefont{and}
  \bibinfo{author}{\bibfnamefont{M.}~\bibnamefont{Jamil}},
  \bibinfo{journal}{Int. J. Mod. Phys.} \textbf{\bibinfo{volume}{D26}},
  \bibinfo{pages}{1741017} (\bibinfo{year}{2017}), \eprint{1705.04167}.

\bibitem[{\citenamefont{Sharif and Shahzadi}(2017)}]{Sharif:2017owq}
\bibinfo{author}{\bibfnamefont{M.}~\bibnamefont{Sharif}} \bibnamefont{and}
  \bibinfo{author}{\bibfnamefont{M.}~\bibnamefont{Shahzadi}},
  \bibinfo{journal}{Eur. Phys. J.} \textbf{\bibinfo{volume}{C77}},
  \bibinfo{pages}{363} (\bibinfo{year}{2017}), \eprint{1705.03058}.

\bibitem[{\citenamefont{Tsukamoto et~al.}(2017)\citenamefont{Tsukamoto,
  Ogasawara, and Gong}}]{Tsukamoto:2017rrl}
\bibinfo{author}{\bibfnamefont{N.}~\bibnamefont{Tsukamoto}},
  \bibinfo{author}{\bibfnamefont{K.}~\bibnamefont{Ogasawara}},
  \bibnamefont{and} \bibinfo{author}{\bibfnamefont{Y.}~\bibnamefont{Gong}},
  \bibinfo{journal}{Phys. Rev.} \textbf{\bibinfo{volume}{D96}},
  \bibinfo{pages}{024042} (\bibinfo{year}{2017}), \eprint{1705.10477}.

\bibitem[{\citenamefont{An and Gao}(2017)}]{An:2017tlp}
\bibinfo{author}{\bibfnamefont{J.}~\bibnamefont{An}} \bibnamefont{and}
  \bibinfo{author}{\bibfnamefont{S.}~\bibnamefont{Gao}} (\bibinfo{year}{2017}),
  \eprint{1708.09576}.

\bibitem[{\citenamefont{An et~al.}(2018)\citenamefont{An, Peng, Liu, and
  Feng}}]{An:2017hby}
\bibinfo{author}{\bibfnamefont{J.}~\bibnamefont{An}},
  \bibinfo{author}{\bibfnamefont{J.}~\bibnamefont{Peng}},
  \bibinfo{author}{\bibfnamefont{Y.}~\bibnamefont{Liu}}, \bibnamefont{and}
  \bibinfo{author}{\bibfnamefont{X.-H.} \bibnamefont{Feng}},
  \bibinfo{journal}{Phys. Rev.} \textbf{\bibinfo{volume}{D97}},
  \bibinfo{pages}{024003} (\bibinfo{year}{2018}), \eprint{1710.08630}.

\bibitem[{\citenamefont{Bécar et~al.}(2018)\citenamefont{Bécar, González,
  and Vásquez}}]{Becar:2017aag}
\bibinfo{author}{\bibfnamefont{R.}~\bibnamefont{Bécar}},
  \bibinfo{author}{\bibfnamefont{P.~A.} \bibnamefont{González}},
  \bibnamefont{and} \bibinfo{author}{\bibfnamefont{Y.}~\bibnamefont{Vásquez}},
  \bibinfo{journal}{Eur. Phys. J.} \textbf{\bibinfo{volume}{C78}},
  \bibinfo{pages}{335} (\bibinfo{year}{2018}), \eprint{1712.00868}.

\bibitem[{\citenamefont{González et~al.}(2018)\citenamefont{González,
  Olivares, Papantonopoulos, and Vásquez}}]{Gonzalez:2018lfs}
\bibinfo{author}{\bibfnamefont{P.~A.} \bibnamefont{González}},
  \bibinfo{author}{\bibfnamefont{M.}~\bibnamefont{Olivares}},
  \bibinfo{author}{\bibfnamefont{E.}~\bibnamefont{Papantonopoulos}},
  \bibnamefont{and} \bibinfo{author}{\bibfnamefont{Y.}~\bibnamefont{Vásquez}},
  \bibinfo{journal}{Phys. Rev.} \textbf{\bibinfo{volume}{D97}},
  \bibinfo{pages}{064034} (\bibinfo{year}{2018}), \eprint{1802.01760}.

\bibitem[{\citenamefont{Ahmed et~al.}(2019)\citenamefont{Ahmed, Amir, and
  Ghosh}}]{Ahmed:2018fge}
\bibinfo{author}{\bibfnamefont{F.}~\bibnamefont{Ahmed}},
  \bibinfo{author}{\bibfnamefont{M.}~\bibnamefont{Amir}}, \bibnamefont{and}
  \bibinfo{author}{\bibfnamefont{S.~G.} \bibnamefont{Ghosh}},
  \bibinfo{journal}{Astrophys. Space Sci.} \textbf{\bibinfo{volume}{364}},
  \bibinfo{pages}{10} (\bibinfo{year}{2019}), \eprint{1805.00804}.

\bibitem[{\citenamefont{Shaymatov et~al.}(2018)\citenamefont{Shaymatov,
  Ahmedov, Stuchlík, and Abdujabbarov}}]{Shaymatov:2018azq}
\bibinfo{author}{\bibfnamefont{S.}~\bibnamefont{Shaymatov}},
  \bibinfo{author}{\bibfnamefont{B.}~\bibnamefont{Ahmedov}},
  \bibinfo{author}{\bibfnamefont{Z.}~\bibnamefont{Stuchlík}},
  \bibnamefont{and}
  \bibinfo{author}{\bibfnamefont{A.}~\bibnamefont{Abdujabbarov}},
  \bibinfo{journal}{Int. J. Mod. Phys.} \textbf{\bibinfo{volume}{D27}},
  \bibinfo{pages}{1850088} (\bibinfo{year}{2018}).

\bibitem[{\citenamefont{Ogasawara and Tsukamoto}(2019)}]{Ogasawara:2018gni}
\bibinfo{author}{\bibfnamefont{K.}~\bibnamefont{Ogasawara}} \bibnamefont{and}
  \bibinfo{author}{\bibfnamefont{N.}~\bibnamefont{Tsukamoto}},
  \bibinfo{journal}{Phys. Rev.} \textbf{\bibinfo{volume}{D99}},
  \bibinfo{pages}{024016} (\bibinfo{year}{2019}), \eprint{1810.03294}.

\bibitem[{\citenamefont{González et~al.}(2019)\citenamefont{González,
  Olivares, Vásquez, Saavedra, and Övgün}}]{Gonzalez:2018zuu}
\bibinfo{author}{\bibfnamefont{P.~A.} \bibnamefont{González}},
  \bibinfo{author}{\bibfnamefont{M.}~\bibnamefont{Olivares}},
  \bibinfo{author}{\bibfnamefont{Y.}~\bibnamefont{Vásquez}},
  \bibinfo{author}{\bibfnamefont{J.}~\bibnamefont{Saavedra}}, \bibnamefont{and}
  \bibinfo{author}{\bibfnamefont{A.}~\bibnamefont{Övgün}},
  \bibinfo{journal}{Eur. Phys. J.} \textbf{\bibinfo{volume}{C79}},
  \bibinfo{pages}{528} (\bibinfo{year}{2019}), \eprint{1811.08551}.

\bibitem[{\citenamefont{Saha and Debnath}(2019)}]{Saha:2019xql}
\bibinfo{author}{\bibfnamefont{P.}~\bibnamefont{Saha}} \bibnamefont{and}
  \bibinfo{author}{\bibfnamefont{U.}~\bibnamefont{Debnath}},
  \bibinfo{journal}{Mod. Phys. Lett.} \textbf{\bibinfo{volume}{A34}},
  \bibinfo{pages}{1950127} (\bibinfo{year}{2019}), \eprint{1905.05612}.

\bibitem[{\citenamefont{Rudra et~al.}(2019)\citenamefont{Rudra, Nandan,
  Gannouji, Chakraborty, and Chatterjee}}]{Rudra:2019ssz}
\bibinfo{author}{\bibfnamefont{A.}~\bibnamefont{Rudra}},
  \bibinfo{author}{\bibfnamefont{H.}~\bibnamefont{Nandan}},
  \bibinfo{author}{\bibfnamefont{R.}~\bibnamefont{Gannouji}},
  \bibinfo{author}{\bibfnamefont{S.}~\bibnamefont{Chakraborty}},
  \bibnamefont{and} \bibinfo{author}{\bibfnamefont{A.~K.}
  \bibnamefont{Chatterjee}} (\bibinfo{year}{2019}), \eprint{1906.03566}.

\bibitem[{\citenamefont{Rahim and Saifullah}(2019)}]{Rahim:2019lip}
\bibinfo{author}{\bibfnamefont{R.}~\bibnamefont{Rahim}} \bibnamefont{and}
  \bibinfo{author}{\bibfnamefont{K.}~\bibnamefont{Saifullah}}
  (\bibinfo{year}{2019}), \eprint{1906.05632}.

\bibitem[{\citenamefont{Harada and Kimura}(2014)}]{Harada:2014vka}
\bibinfo{author}{\bibfnamefont{T.}~\bibnamefont{Harada}} \bibnamefont{and}
  \bibinfo{author}{\bibfnamefont{M.}~\bibnamefont{Kimura}},
  \bibinfo{journal}{Class. Quant. Grav.} \textbf{\bibinfo{volume}{31}},
  \bibinfo{pages}{243001} (\bibinfo{year}{2014}), \eprint{1409.7502}.

\bibitem[{\citenamefont{Mathisson}(1937)}]{Mathisson:1937zz}
\bibinfo{author}{\bibfnamefont{M.}~\bibnamefont{Mathisson}},
  \bibinfo{journal}{Acta Phys. Polon.} \textbf{\bibinfo{volume}{6}},
  \bibinfo{pages}{163} (\bibinfo{year}{1937}).

\bibitem[{\citenamefont{Papapetrou}(1951)}]{Papapetrou:1951pa}
\bibinfo{author}{\bibfnamefont{A.}~\bibnamefont{Papapetrou}},
  \bibinfo{journal}{Proc. Roy. Soc. Lond.} \textbf{\bibinfo{volume}{A209}},
  \bibinfo{pages}{248} (\bibinfo{year}{1951}).

\bibitem[{\citenamefont{{Dixon}}(1964)}]{Dixon:1964}
\bibinfo{author}{\bibfnamefont{W.~G.} \bibnamefont{{Dixon}}},
  \bibinfo{journal}{Il Nuovo Cimento} \textbf{\bibinfo{volume}{34}},
  \bibinfo{pages}{317} (\bibinfo{year}{1964}).

\bibitem[{\citenamefont{Armaza et~al.}(2016)\citenamefont{Armaza, Bañados, and
  Koch}}]{Armaza:2015eha}
\bibinfo{author}{\bibfnamefont{C.}~\bibnamefont{Armaza}},
  \bibinfo{author}{\bibfnamefont{M.}~\bibnamefont{Bañados}}, \bibnamefont{and}
  \bibinfo{author}{\bibfnamefont{B.}~\bibnamefont{Koch}},
  \bibinfo{journal}{Class. Quant. Grav.} \textbf{\bibinfo{volume}{33}},
  \bibinfo{pages}{105014} (\bibinfo{year}{2016}), \eprint{1510.01223}.

\bibitem[{\citenamefont{Zhang et~al.}(2016)\citenamefont{Zhang, Gu, Wei, Yang,
  and Liu}}]{Zhang:2016btg}
\bibinfo{author}{\bibfnamefont{Y.-P.} \bibnamefont{Zhang}},
  \bibinfo{author}{\bibfnamefont{B.-M.} \bibnamefont{Gu}},
  \bibinfo{author}{\bibfnamefont{S.-W.} \bibnamefont{Wei}},
  \bibinfo{author}{\bibfnamefont{J.}~\bibnamefont{Yang}}, \bibnamefont{and}
  \bibinfo{author}{\bibfnamefont{Y.-X.} \bibnamefont{Liu}},
  \bibinfo{journal}{Phys. Rev.} \textbf{\bibinfo{volume}{D94}},
  \bibinfo{pages}{124017} (\bibinfo{year}{2016}), \eprint{1608.08705}.

\bibitem[{\citenamefont{Jiang and Gao}(2019)}]{Jiang:2019cuc}
\bibinfo{author}{\bibfnamefont{J.}~\bibnamefont{Jiang}} \bibnamefont{and}
  \bibinfo{author}{\bibfnamefont{S.}~\bibnamefont{Gao}}, \bibinfo{journal}{Eur.
  Phys. J.} \textbf{\bibinfo{volume}{C79}}, \bibinfo{pages}{378}
  (\bibinfo{year}{2019}), \eprint{1905.02491}.

\bibitem[{\citenamefont{Hojman and Hojman}(1977)}]{Hojman:1976kn}
\bibinfo{author}{\bibfnamefont{R.}~\bibnamefont{Hojman}} \bibnamefont{and}
  \bibinfo{author}{\bibfnamefont{S.}~\bibnamefont{Hojman}},
  \bibinfo{journal}{Phys. Rev.} \textbf{\bibinfo{volume}{D15}},
  \bibinfo{pages}{2724} (\bibinfo{year}{1977}).

\bibitem[{\citenamefont{Zalaquett et~al.}(2014)\citenamefont{Zalaquett, Hojman,
  and Asenjo}}]{Zalaquett:2014eia}
\bibinfo{author}{\bibfnamefont{N.}~\bibnamefont{Zalaquett}},
  \bibinfo{author}{\bibfnamefont{S.~A.} \bibnamefont{Hojman}},
  \bibnamefont{and} \bibinfo{author}{\bibfnamefont{F.~A.}
  \bibnamefont{Asenjo}}, \bibinfo{journal}{Class. Quant. Grav.}
  \textbf{\bibinfo{volume}{31}}, \bibinfo{pages}{085011}
  (\bibinfo{year}{2014}), \eprint{1308.4435}.

\bibitem[{\citenamefont{Saijo et~al.}(1998)\citenamefont{Saijo, Maeda, Shibata,
  and Mino}}]{Saijo:1998mn}
\bibinfo{author}{\bibfnamefont{M.}~\bibnamefont{Saijo}},
  \bibinfo{author}{\bibfnamefont{K.-i.} \bibnamefont{Maeda}},
  \bibinfo{author}{\bibfnamefont{M.}~\bibnamefont{Shibata}}, \bibnamefont{and}
  \bibinfo{author}{\bibfnamefont{Y.}~\bibnamefont{Mino}},
  \bibinfo{journal}{Phys. Rev.} \textbf{\bibinfo{volume}{D58}},
  \bibinfo{pages}{064005} (\bibinfo{year}{1998}).

\bibitem[{\citenamefont{Zaslavskii}(2016)}]{Zaslavskii:2016dfh}
\bibinfo{author}{\bibfnamefont{O.~B.} \bibnamefont{Zaslavskii}},
  \bibinfo{journal}{EPL} \textbf{\bibinfo{volume}{114}}, \bibinfo{pages}{30003}
  (\bibinfo{year}{2016}), \eprint{1603.09353}.

\bibitem[{\citenamefont{M\o~ller}(1949)}]{Moller:1949}
\bibinfo{author}{\bibfnamefont{C.}~\bibnamefont{M\o~ller}},
  \bibinfo{journal}{Commun. Dublin Inst. Advan. Stud.}
  \textbf{\bibinfo{volume}{A5}} (\bibinfo{year}{1949}).

\bibitem[{\citenamefont{Wald}(1972)}]{PhysRevD.6.406}
\bibinfo{author}{\bibfnamefont{R.}~\bibnamefont{Wald}}, \bibinfo{journal}{Phys.
  Rev. D} \textbf{\bibinfo{volume}{6}}, \bibinfo{pages}{406}
  (\bibinfo{year}{1972}),
  \urlprefix\url{https://link.aps.org/doi/10.1103/PhysRevD.6.406}.

\bibitem[{\citenamefont{Zhang et~al.}(2018)\citenamefont{Zhang, Jiang, Liu, and
  Liu}}]{Zhang:2018gpn}
\bibinfo{author}{\bibfnamefont{M.}~\bibnamefont{Zhang}},
  \bibinfo{author}{\bibfnamefont{J.}~\bibnamefont{Jiang}},
  \bibinfo{author}{\bibfnamefont{Y.}~\bibnamefont{Liu}}, \bibnamefont{and}
  \bibinfo{author}{\bibfnamefont{W.-B.} \bibnamefont{Liu}},
  \bibinfo{journal}{Phys. Rev.} \textbf{\bibinfo{volume}{D98}},
  \bibinfo{pages}{044006} (\bibinfo{year}{2018}).

\bibitem[{\citenamefont{Hojman and Asenjo}(2016)}]{Hojman:2016gep}
\bibinfo{author}{\bibfnamefont{S.~A.} \bibnamefont{Hojman}} \bibnamefont{and}
  \bibinfo{author}{\bibfnamefont{F.~A.} \bibnamefont{Asenjo}},
  \bibinfo{journal}{Phys. Rev.} \textbf{\bibinfo{volume}{D93}},
  \bibinfo{pages}{028501} (\bibinfo{year}{2016}).

\bibitem[{\citenamefont{Suzuki and Maeda}(1998)}]{Suzuki:1998}
\bibinfo{author}{\bibfnamefont{S.}~\bibnamefont{Suzuki}} \bibnamefont{and}
  \bibinfo{author}{\bibfnamefont{K.-i.} \bibnamefont{Maeda}},
  \bibinfo{journal}{Phys. Rev. D} \textbf{\bibinfo{volume}{58}},
  \bibinfo{pages}{023005} (\bibinfo{year}{1998}),
  \urlprefix\url{https://link.aps.org/doi/10.1103/PhysRevD.58.023005}.

\bibitem[{\citenamefont{Hojman and Asenjo}(2013)}]{Hojman:2012me}
\bibinfo{author}{\bibfnamefont{S.~A.} \bibnamefont{Hojman}} \bibnamefont{and}
  \bibinfo{author}{\bibfnamefont{F.~A.} \bibnamefont{Asenjo}},
  \bibinfo{journal}{Class. Quant. Grav.} \textbf{\bibinfo{volume}{30}},
  \bibinfo{pages}{025008} (\bibinfo{year}{2013}), \eprint{1203.5008}.

\bibitem[{\citenamefont{Nucamendi and Sudarsky}(1997)}]{Nucamendi:1996ac}
\bibinfo{author}{\bibfnamefont{U.}~\bibnamefont{Nucamendi}} \bibnamefont{and}
  \bibinfo{author}{\bibfnamefont{D.}~\bibnamefont{Sudarsky}},
  \bibinfo{journal}{Class. Quant. Grav.} \textbf{\bibinfo{volume}{14}},
  \bibinfo{pages}{1309} (\bibinfo{year}{1997}), \eprint{gr-qc/9611043}.

\bibitem[{\citenamefont{Nucamendi and Sudarsky}(2000)}]{Nucamendi:2000af}
\bibinfo{author}{\bibfnamefont{U.}~\bibnamefont{Nucamendi}} \bibnamefont{and}
  \bibinfo{author}{\bibfnamefont{D.}~\bibnamefont{Sudarsky}},
  \bibinfo{journal}{Class. Quant. Grav.} \textbf{\bibinfo{volume}{17}},
  \bibinfo{pages}{4051} (\bibinfo{year}{2000}), \eprint{gr-qc/0004068}.

\bibitem[{\citenamefont{Hackmann et~al.}(2014)\citenamefont{Hackmann,
  L\"ammerzahl, Obukhov, Puetzfeld, and Schaffer}}]{Hackmann2014}
\bibinfo{author}{\bibfnamefont{E.}~\bibnamefont{Hackmann}},
  \bibinfo{author}{\bibfnamefont{C.}~\bibnamefont{L\"ammerzahl}},
  \bibinfo{author}{\bibfnamefont{Y.~N.} \bibnamefont{Obukhov}},
  \bibinfo{author}{\bibfnamefont{D.}~\bibnamefont{Puetzfeld}},
  \bibnamefont{and} \bibinfo{author}{\bibfnamefont{I.}~\bibnamefont{Schaffer}},
  \bibinfo{journal}{Phys. Rev. D} \textbf{\bibinfo{volume}{90}},
  \bibinfo{pages}{064035} (\bibinfo{year}{2014}),
  \urlprefix\url{http://link.aps.org/doi/10.1103/PhysRevD.90.064035}.

\bibitem[{\citenamefont{{Stuchlik}}(1983)}]{Stuchlik:1983}
\bibinfo{author}{\bibfnamefont{Z.}~\bibnamefont{{Stuchlik}}},
  \bibinfo{journal}{Bulletin of the Astronomical Institutes of Czechoslovakia}
  \textbf{\bibinfo{volume}{34}}, \bibinfo{pages}{129} (\bibinfo{year}{1983}).

\bibitem[{\citenamefont{{Toshmatov} et~al.}(2017)\citenamefont{{Toshmatov},
  {Stuchl{\'\i}k}, and {Ahmedov}}}]{Toshmatov:2017}
\bibinfo{author}{\bibfnamefont{B.}~\bibnamefont{{Toshmatov}}},
  \bibinfo{author}{\bibfnamefont{Z.}~\bibnamefont{{Stuchl{\'\i}k}}},
  \bibnamefont{and}
  \bibinfo{author}{\bibfnamefont{B.}~\bibnamefont{{Ahmedov}}},
  \bibinfo{journal}{European Physical Journal Plus}
  \textbf{\bibinfo{volume}{132}}, \bibinfo{eid}{98} (\bibinfo{year}{2017}).

\bibitem[{\citenamefont{Hojman}(1975)}]{Hojman_thesis:1975}
\bibinfo{author}{\bibfnamefont{S.~A.} \bibnamefont{Hojman}},
  \bibinfo{journal}{(unpublished)}  (\bibinfo{year}{1975}),
  \eprint{(unpublished)}.

\bibitem[{\citenamefont{Deriglazov and
  Ramírez}(2016{\natexlab{a}})}]{Deriglazov:2015wde}
\bibinfo{author}{\bibfnamefont{A.~A.} \bibnamefont{Deriglazov}}
  \bibnamefont{and} \bibinfo{author}{\bibfnamefont{W.~G.}
  \bibnamefont{Ramírez}}, \bibinfo{journal}{Adv. High Energy Phys.}
  \textbf{\bibinfo{volume}{2016}}, \bibinfo{pages}{1376016}
  (\bibinfo{year}{2016}{\natexlab{a}}), \eprint{1511.00645}.

\bibitem[{\citenamefont{Deriglazov and
  Ramírez}(2016{\natexlab{b}})}]{Deriglazov:2015zta}
\bibinfo{author}{\bibfnamefont{A.~A.} \bibnamefont{Deriglazov}}
  \bibnamefont{and} \bibinfo{author}{\bibfnamefont{W.~G.}
  \bibnamefont{Ramírez}}, \bibinfo{journal}{Int. J. Mod. Phys.}
  \textbf{\bibinfo{volume}{D26}}, \bibinfo{pages}{1750047}
  (\bibinfo{year}{2016}{\natexlab{b}}), \eprint{1509.05357}.

\bibitem[{\citenamefont{{Hackmann} et~al.}(2020)\citenamefont{{Hackmann},
  {Nandan}, and {Sheoran}}}]{Hackmann2020}
\bibinfo{author}{\bibfnamefont{E.}~\bibnamefont{{Hackmann}}},
  \bibinfo{author}{\bibfnamefont{H.}~\bibnamefont{{Nandan}}}, \bibnamefont{and}
  \bibinfo{author}{\bibfnamefont{P.}~\bibnamefont{{Sheoran}}},
  \bibinfo{journal}{arXiv e-prints} \bibinfo{eid}{arXiv:2006.05045}
  (\bibinfo{year}{2020}), \eprint{2006.05045}.

\bibitem[{\citenamefont{{Zaslavskii}}(2020)}]{Zaslavskii2020}
\bibinfo{author}{\bibfnamefont{O.~B.} \bibnamefont{{Zaslavskii}}},
  \bibinfo{journal}{arXiv e-prints} \bibinfo{eid}{arXiv:2006.11552}
  (\bibinfo{year}{2020}), \eprint{2006.11552}.

\end{thebibliography}
%\bibliographystyle{spbasic}
%\renewcommand{\bibname}{References}
%\nocite{*}%\bibliographystyle{revcompchem}
%\bibliographystyle{naturemag}
%\bibliographystyle{amsalpha} %% acm, naturemag, revcompchem
%\bibliographystyle{alpha}
%\bibliographystyle{unsrt}
%\bibliographystyle{apalike}
%\bibliographystyle{amsplain}
%\bibliographystyle{plain}
%\bibliographystyle{h-physrev3.bst}
%\bibliographystyle{amsplain}
%\bibliographystyle{abbrv}
\bibliographystyle{apsrev}
\end{document}